%% file: Paper2.tex
\documentclass[11pt,a4paper,final]{article}
\input{paquetes.tex}
\title{\vspace*{-1.35cm}\textbf{Transverse expansion of the metric at null hypersurfaces II. Existence results and application to Killing horizons}}
\author{Marc Mars\footnote{\href{mailto:marc@usal.es}{marc@usal.es}}\,\, and Gabriel Sánchez-Pérez\footnote{\href{mailto:gasape21@usal.es}{gasape21@usal.es}} \\
	Departamento de Física Fundamental, Universidad de Salamanca\\
	Plaza de la Merced s/n, 37008 Salamanca, Spain}
\date{\today}

\begin{document}
	\maketitle
	
\input{Intro2}

\input{Notation}
\input{Datashort}
\input{reviewtransverse}
\input{Existence}
\input{Killing2}

\input{Ack}

	\begingroup
	\let\itshape\upshape
	
	\renewcommand{\bibname}{References}
	\bibliographystyle{acm}
	\bibliography{biblio} 
	
\end{document}

%% file: paquetes.tex
\usepackage[utf8]{inputenc}
\usepackage[english]{babel}
\usepackage{amsmath}
\usepackage{amsfonts}
\usepackage{amssymb}
\usepackage{appendix}
\usepackage{faktor}
\usepackage{yfonts}
\usepackage{upgreek}
\usepackage{lipsum}
\usepackage[hidelinks]{hyperref}
\hypersetup{
    colorlinks=true,
    urlcolor=blue,
    linkcolor=blue,
    citecolor=red,
    }
\usepackage{amsthm}
\usepackage{bbold}
\usepackage{mathrsfs}
\usepackage{wasysym}
\usepackage{mathtools}
\usepackage{bm}
\usepackage{cancel}
\usepackage{color}
\usepackage{enumitem,multicol}
\usepackage{multicol}
\usepackage{tikz}
\usepackage{tikz-cd}
\usepackage[Glenn]{fncychap}
\usepackage{subfig}
\usepackage{accents}
\usepackage{slashed}
\usepackage{psfrag}
\usepackage{todonotes}

\usepackage{stackengine}
	
\usetikzlibrary{babel}
\makeatletter
\newcommand*\owedge{\mathpalette\@owedge\relax}
\newcommand*\@owedge[1]{%
	\mathbin{%
		\ooalign{%
			$#1\m@th\bigcirc$\cr
			\hidewidth$#1\m@th\wedge$\hidewidth\cr
		}%
	}%
}
\makeatother

\usepackage{cjhebrew}

\newcounter{mnotecount}

\newcommand{\mnote}[1]
{\protect{\stepcounter{mnotecount}}$^{\mbox{\tiny
			$\,\bullet$\themnotecount}}$ \marginpar{
		\raggedright\tiny\em
		$\,\bullet$\themnotecount: #1} }

\newtheorem{teo}{Theorem}[section]
\newtheorem{cor}[teo]{Corollary}
\newtheorem{prop}[teo]{Proposition}
\newtheorem{lema}[teo]{Lemma}
\newtheorem{defi}[teo]{Definition}

\newtheorem{rmk}[teo]{Remark}

\usepackage{makeidx}
\usepackage{graphicx}
\usepackage[left=2.60cm, right=2.60cm, top=2.45cm, bottom=2.60cm]{geometry}

\usepackage{scalerel}
\usepackage{stackengine,wasysym}

\makeatletter
\newcommand{\douwidehat}[2]{%
	\sbox0{$\m@th#1\widehat{\hphantom{#2}}$}%
	\sbox2{$\m@th#1x$}
	\sbox4{$\m@th#1#2$}
	\dimen0=\ht0
	\advance\dimen0 -.8\ht2
	\dimen2=\dp4
	\rlap{%
		\raisebox{\dimexpr\dimen0-\dimen2}{%
			\scalebox{1}[-1]{\box0}%
		}%
	}%
	{#2}%
}
\makeatother

\usepackage{color}
    \makeatletter
\renewcommand\part{%
	\if@openright
	\cleardoublepage
	\else
	\clearpage
	\fi
	\thispagestyle{empty}%
	\if@twocolumn
	\onecolumn
	\@tempswatrue
	\else
	\@tempswafalse
	\fi
	\null\vfil
	\secdef\@part\@spart}
\makeatother

\newcommand{\wt}{\widetilde}
\newcommand{\wh}{\widehat}

\newcommand{\elltwo}{\ell^{(2)}}
\newcommand{\ntwo}{n^{(2)}}
\newcommand{\nablacero}{\accentset{\circ}{\nabla}}

\newcommand{\bY}{\textup{\textbf Y}}
\newcommand{\Y}{\textup{Y}}
\newcommand{\bcY}{\mathbb Y}
\newcommand{\bcr}{\mathbb r}
\newcommand{\bck}{\mathbb k}
\renewcommand{\k}{\mathscr{K}}
\newcommand{\bT}{\textup{\textbf T}}
\newcommand{\T}{\textup{T}}
\newcommand{\bS}{\textup{\textbf S}}
\renewcommand{\S}{\textup{S}}
\newcommand{\Sigmacero}{\accentset{\circ}{\Sigma}}

\newcommand{\bU}{\textup{\textbf U}}
\newcommand{\U}{\textup{U}}

\newcommand{\bF}{\textup{\textbf F}}
\newcommand{\F}{\textup{F}}
\newcommand{\br}{\textup{\textbf r}}
\renewcommand{\r}{\textup{r}}

\renewcommand{\H}{\mc{H}}

\newcommand{\bs}{\textup{\textbf s}}
\newcommand{\s}{\textup{s}}

\newcommand{\bg}{\bm{\gamma}}

\newcommand{\mf}{\mathfrak}
\newcommand{\real}{\mathbb R}
\newcommand{\tr}{\operatorname{tr}}

\renewcommand{\div}{\operatorname{div}}

\newcommand{\R}{\bm{\mc R}}
\newcommand{\st}{\stackrel}

\newcommand{\para}{\parallel}
\newcommand{\Rcero}{\accentset{\circ}{R}}

\newcommand{\eps}{\varepsilon}
\newcommand{\X}{\mathfrak{X}}
\renewcommand{\d}{\coloneqq}

\newcommand{\ric}{\operatorname{Ric}}

\newcommand{\lie}{\mathcal{L}}
\newcommand{\mc}{\mathcal}

\renewcommand{\to}{\longrightarrow}

\renewcommand{\mapsto}{\longmapsto}

%% file: Intro2.tex
\begin{abstract}
This paper finishes the series of two papers that we started with \cite{Mio3}, where we analyzed the transverse expansion of the metric at a general null hypersurface. While \cite{Mio3} focused on uniqueness results, here we show existence of ambient manifolds given the full asymptotic expansion at the null hypersurface. When such expansion fulfills a set of ``constraint equations'' we prove that the ambient manifold solves the Einstein equations to infinite order at the hypersurface. Our approach does not make any assumptions regarding the dimension or topology of the null hypersurface and is entirely covariant. Furthermore, when the hypersurface exhibits a product topology we find the minimum amount of data on a cross-section that ensures the existence of an ambient space solving the Einstein equations to infinite order on the hypersurface. As an application we recall the notion of \textit{abstract Killing horizon data} (AKH) introduced in \cite{Mio3}, namely the minimal data needed to define a non-degenerate Killing horizon from a detached viewpoint, and we prove that every AKH of arbitrary dimension and topology gives rise to an ambient space solving the $\Lambda$-vacuum equations to infinite order and with the given data as Killing horizon. Our result also includes the possibility of the Killing vector having zeroes at the horizon.
\end{abstract}

\section{Introduction}

An important problem in General Relativity is the study of solutions to the Einstein field equations given initial data. The fundamental result of Choquet-Buhat \cite{Choquet,choquet1969global,ringstrom} establishes that by prescribing an $n$-dimensional Riemannian manifold $(\Sigma,h)$ and a symmetric, two covariant tensor field $K$ subject to the so-called constraint equations, it exists an $(n+1)$-dimensional spacetime $(\mc M,g)$ solution of the Einstein equations where $(\Sigma,h,K)$ is embedded. Another notable initial value problem is the characteristic one, where the initial data is posed on a pair of intersecting null hypersurfaces. Rendall's result \cite{Rendall} shows that in this scenario the initial data consists of the full spacetime metric (and none of its transverse derivatives) on the initial hypersurfaces. This problem has been recently addressed from a more geometric point of view \cite{Mio1,Mio2}, where we also consider transverse derivatives of the metric as initial data (and thus we need to incorporate the corresponding constraint equations). Various other approaches to this problem can be found in \cite{Luk,CandP,cabet,chrusciel2023neighborhood}. A natural question that arises regarding the characteristic problem is whether initial data at a single null hypersurface can give rise to an ambient manifold solution of the Einstein equations. \\

This topic has been analyzed in different contexts over the last years. Of particular interest is the work of C. Fefferman and R. Graham \cite{fefferman2012ambient} where they show that given an arbitrary conformal class and a given $n$-dimensional manifold it is possible to construct a homothetic horizon and to embed it into an ambient manifold of dimension $n+2$ that satisfies the Einstein equations to infinite order at the horizon. Their approach involves working within a specific coordinate system where the metric coefficients are given by a formal series expansion in a coordinate transverse to the horizon. By imposing the Einstein equations order by order they are able to prove that in odd-dimensional ambient manifolds, all metric coefficients are determined solely by the conformal data. However, in even-dimensional manifolds, additional data is required to fully determine the expansion at the horizon. Furthermore, when the initial data is real analytic, Fefferman and Graham show that the metric expansion converges, resulting in an ambient space that satisfies the Einstein equations throughout. An alternative approach to this problem has been explored in \cite{rodnianski}, where the authors define initial data on a closed and oriented Riemannian manifold of arbitrary dimension $n\ge 2$ and prove that such data give rise to an $n+2$-dimensional spacetime that admits a homothetic vector. Unlike the ambient space of Fefferman and Graham, the spacetime in \cite{rodnianski} is \textit{exactly} Ricci flat, and not only \textit{asymptotically} Ricci flat at the horizon.\\

Another example where data on a single null hypersurface gives rise to an ambient manifold is the case of non-degenerate Killing horizons. In \cite{moncrief1982neighborhoods} Moncrief shows that by providing initial data on a three-dimensional null hypersurface with product topology there exists a four dimensional spacetime solving the Einstein equations to infinite order that admits a Killing vector for which the given null hypersurface is a non-degenerate Killing horizon. This result has been recently written in more geometric terms in \cite{oliver}, where the authors encode the initial data on a Killing horizon into a Riemannian manifold $(\H, h)$ endowed with a non-vanishing Killing field. \\

Motivated by these examples, this paper aims at determining the necessary and sufficient conditions for a null hypersurface to be embedded in an ambient manifold that satisfies the Einstein equations to infinite order. Our approach is entirely covariant and independent of the dimension or the topology of the null hypersurface. In addition, we analyze in deeper detail the case where the hypersurface admits a product topology, thereby characterizing the initial data in terms of geometric objects at any section. As an application of our findings, we recall the abstract definition of Killing horizon data (AKH) introduced in \cite{Mio3} and we prove that every AKH data gives rise to an ambient manifold that solves the $\Lambda$-vacuum equations to infinite order, with the initial data being embedded as a non-degenerate Killing horizon.\\

In order to deal with abstract null hypersurfaces it is convenient to employ the so-called hypersurface data formalism \cite{Marc3,Marc1,Marc2}, as it allows to work with hypersurfaces of arbitrary causal character in a detached way. This formalism relies on the notion of metric hypersurface data, which consists of a $(0,2)$ symmetric tensor field $\bg$, a one-form $\bm\ell$ and a scalar $\elltwo$ that encodes, at the abstract level, the restriction of the ambient metric at the hypersurface. This data is sufficient to define a torsion-free connection in the abstract hypersurface, even in the null case, which allows us to do geometry at null hypersurfaces in a detached manner. From $\{\bg,\bm\ell,\elltwo\}$ one can uniquely introduce the ``contravariant metric data'' that consists of a $(2,0)$ symmetric tensor $P$, a vector field $n$ and a scalar $n^{(2)}$. In the null case, $n$ is a null generator of the hypersurface. \\

In addition to the hypersurface metric data one can also encode transverse information of the metric, known as the \textit{transverse expansion}, into a collection of tensors denoted by $\{\bY^{(k)}\}$. In the embedded case, the idea is that for each $k$ the tensor $2\bY^{(k)}$ agrees with the pullback of the $k$-th derivative along the transverse direction of the ambient metric at the null hypersurface. Informally, the transverse expansion allows us to ``reconstruct'' the ambient metric order by order at the null hypersurface. To construct an ambient manifold from such expansion we adapt a result due to Borel to the tensorial case and we prove that every null metric hypersurface data $\{\H,\bg,\bm\ell,\elltwo\}$ along with a collection of tensors $\{\bcY^{(k)}\}$ leads to an ambient manifold where the data is embedded, in the sense that each $2\bcY^{(k)}$ agrees with $2\bY^{(k)}$, i.e. the $k$-th transverse derivative of the ambient metric at the null hypersurface (Theorem \ref{borel}). It is crucial to emphasize that even if the Taylor series associated to $\{\bcY^{(k)}\}$ may not converge, we are still able to construct a smooth ambient manifold.\\

A priori, such manifold does not satisfy any field equations. In order to analyze necessary and sufficient conditions for the ambient space to satisfy suitable field equations (such as the Einstein equations) it is key to know how to link the transverse derivatives of the ambient curvature with the transverse derivatives of the metric at a general null hypersurface. In our previous paper \cite{Mio3} we derived general identities in that regard and we also explored the case where the hypersurface exhibits some sort of symmetry. Our identities show that, given the $m$-th transverse derivative of the ambient Ricci tensor, (i) its completely transverse components depend algebraically on the trace of $\bY^{(m+2)}$ w.r.t. the tensor $P$, (ii) its transverse-tangent components are given by the contraction $\bY^{(m+2)}(n,\cdot)$, and (iii) its completely tangent components depend on $\bY^{(m+1)}$ via a transport equation along $n$, as well as on lower order terms. \\

Employing the previous identities and our existence result mentioned above, in Theorem \ref{existencia} we determine the necessary and sufficient conditions for a metric data $\{\H,\bg,\bm\ell,\elltwo\}$ and the abstract expansion $\{\bcY^{(k)}\}$ that guarantees the existence of an ambient manifold solving the $\Lambda$-vacuum equations to infinite order at the hypersurface, namely that the expansion satisfies the \textit{higher order constraint equations}. An informal version of Theorem \ref{existencia} can be stated as follows.
\begin{teo}
Given null metric hypersurface data satisfying the higher order constrain equations \eqref{xconstraint}-\eqref{xtt}, there exists a spacetime that satisfies the $\Lambda$-vacuum equations to infinite order where the data is embedded.
\end{teo} 
Our result holds regardless the dimension or topology of the hypersurface. In particular, when the null hypersurface has a product topology, the constraints can be reinterpreted as evolution equations along the null generator of the hypersurface. This reinterpretation allows us to solve the higher order constraints and restate our existence theorem in terms of geometric objects defined at some cross-section of the hypersurface (Theorem \ref{teoexistencia}). Specifically, given a one-form and a collection of symmetric, traceless tensors on any cross-section, as well as a function defined everywhere but restricted to satisfy the algebraic equation \eqref{equationf} (which in essence is the Raychaudhuri equation when expressed in terms of detached quantities), there exists a $\Lambda$-vacuum spacetime where the data is embedded. One can state this result informally as follows.
\begin{teo}
Given null metric hypersurface data together with suitable geometric data at a cross-section and a function defined everywhere and satisfying \eqref{equationf}, there exists a spacetime that satisfies the $\Lambda$-vacuum equations to infinite order where the data is embedded.
\end{teo}

The results of this paper can be applied to many contexts. For instance, in a recent work \cite{katona2024uniqueness} the authors find explicitly the transverse expansion of the metric at a static degenerate Killing horizon with maximally symmetric cross-sections and they show that either it agrees with that of Schwarzschild-de Sitter or with its near horizon limit (Nariai). In both cases the expansion is manifestly convergent and the spacetime they give rise to is either Schwarzschild-de Sitter or Nariai. However, there could be other situations (not necessarily Killing horizons) where one can find explicitly the transverse expansion but it may not converge (e.g. Fefferman-Graham metrics with non-analytic conformal class where the series does not converge). In these cases, our results may be applied to construct a smooth ambient manifold, and also to check whether it satisfies the Einstein equations to infinite order or not.\\

When the hypersurface $\H$ admits a symmetry generator $\eta$ that is null and tangent to $\H$ (which may have zeros on $\H$) it is possible to translate information from the deformation tensor of $\eta$ into the identities (iii) described two paragraphs above. This yields a new collection of identities, known as the \textit{generalized master equations}, in which the dependence on the tensors $\{\bY^{(k)}\}$ becomes algebraic rather than via transport equations along $n$. \\

The generalized master equations have a particularly interesting application in the Killing horizon case. As shown in \cite{Mio3}, for non-degenerate Killing horizons the generalized master equations allow us to obtain the full transverse expansion in terms of just abstract data at the horizon and the tower of derivatives of the ambient Ricci tensor at the horizon. This generalizes the recent work by O. Petersen and K. Kroencke \cite{oliver} because we allow for $\eta$ having zeros at $\H$ and other field equations besides vacuum with $\Lambda=0$. Such abstract data, known as \textit{abstract Killing horizon data}, was introduced in \cite{Mio3} and is given by metric hypersurface data, a one-form $\bm\tau$ and a scalar function $\alpha$. Geometrically, $\bm\tau$ can be interpreted as the torsion one-form of the null hypersurface, while $\alpha$ represents the proportionality function between the Killing field (which may have zeroes) and the null generator $n$. By applying our existence results to this scenario, we prove that every abstract Killing horizon data gives rise to an ambient manifold which solves the $\Lambda$-vacuum equations to infinite order and for which the initial data is a non-degenerate horizon. In informal terms, we prove the following (Theorem \ref{teorema2}). 
\begin{teo}
Given (non-degenerate) abstract Killing horizon data, there exists a spacetime satisfying the $\Lambda$-vacuum equations to infinite order and where the data is embedded as a non-degenerate Killing horizon.
\end{teo}
It is important to emphasize that the topology and dimension of the horizon are completely arbitrary, and that we allow for the Killing vector to have zeroes, thereby encompassing bifurcate horizons.\\

The structure of this manuscript is as follows. Section \ref{sec_hypersurfacedata} is mainly an overview of the main aspects of hypersurface data, but some new results needed below are also proved. Section \ref{section_higher} summarizes the results of our previous paper \cite{Mio3}, specifically focusing on the relation between the transverse derivatives of the ambient Ricci tensor with the transverse derivatives of the metric at a null hypersurface. Section \ref{sec_existence} is devoted to proving our general existence theorem, namely that metric hypersurface data $\{\H,\bg,\bm\ell,\elltwo\}$ together with a collection of tensors $\{\bcY^{(k)}\}$ satisfying a set of constraint equations give rise to an ambient manifold solving the Einstein equations to infinite order at $\H$. We also reformulate the theorem when $\H$ admits sections. Finally, in Section \ref{sec_KH2}, after revisiting the main results regarding Killing horizons from \cite{Mio3}, we prove that every abstract Killing horizon data gives rise to an ambient manifold solution of the $\Lambda$-vacuum equations to infinite order at $\H$, and admitting a Killing vector for which the initial data is embedded as a non-degenerate Killing horizon.

%% file: Notation.tex
\section*{Notation and conventions}

Throughout this paper $(\mc M,g)$ denotes an arbitrary smooth $\mf{n}$-dimensional semi-Riemannian manifold of any signature $(p,q)$ with both $p$ and $q$ different from zero. When we specifically need this signature to be Lorentzian, we will say so explicitly. We employ both index-free and abstract index notation at our convenience. Ambient indices are denoted with Greek letters, abstract indices on a hypersurface are written in lowercase Latin letters, and abstract indices at cross-sections of a hypersurface are expressed in uppercase Latin letters. As usual, square brackets enclosing indices denote antisymmetrization and parenthesis are for symmetrization. The symmetrized tensor product is denoted with $\otimes_s$. By $\mc F(\mc M)$, $\X(\mc M)$ and $\X^{\star}(\mc M)$ we denote, respectively, the set of smooth functions, vector fields and one-forms on $\mc M$. The subset $\mc F^{\star}(\mc M)\subset\mc F(\mc M)$ consists of the nowhere vanishing functions on $\mc M$. The pullback of a function $f$ via a diffeomorphism $\Phi$ will be denoted by $\Phi^{\star}f$ or simply by $f$ depending on the context. Given a diffeomorphism $\Phi$ and a vector field $X$, we define $\Phi^{\star}X\d (\Phi^{-1})_{\star}X$. A $(p,q)$-tensor refers to a tensor field $p$ times contravariant and $q$ times covariant. Given any pair of $(2,0)$ and $(0,2)$ tensors $A^{ab}$ and $B_{cd}$ we denote $\tr_A \bm B \d A^{ab}B_{ab}$. We employ the symbol $\nabla$ for the Levi-Civita connection of $g$. Throughout this paper we use the notation $\lie_X^{(m)}T$ to denote the $m$-th Lie derivative of the tensor $T$ along $X$, and $X^{(m)}(f)$ for the $m$-th directional derivative of the function $f$ along $X$. When $m=1$ we also write $\lie_X T$ and $X(f)$, respectively, and when $m=0$ they are just the identity operators. All manifolds are assumed to be connected and smooth.

%% file: Datashort.tex
\section{Review of hypersurface data formalism}
\label{sec_hypersurfacedata}

In this section we review the basic notions of the so-called \textit{hypersurface data formalism} that we shall need along the paper. Further details can be found in \cite{Marc1,Marc2,Marc3}. The usefulness of this formalism has been demonstrated in several different contexts such as matching of spacetimes \cite{miguel1,miguel2,miguel4,manzano2024embedded} or the characteristic problem of general relativity \cite{Mio1,Mio2}. We start by introducing the notion of hypersurface data.

\begin{defi}
	\label{def_hypersurfacedata}
	Let $\mc H$ be a $d$-dimensional manifold, $\bg$ a $(0,2)$ symmetric tensor field, $\bm\ell$ a one-form and $\elltwo$ a scalar function on $\mc H$. We say that $\{\mc H,\bg,\bm\ell,\elltwo\}$ defines a metric hypersurface data set provided that the 2-covariant, symmetric tensor $\bm{\mc A}|_p$ on $T_p\mc H\times\real$ defined by 
	\begin{equation}
		\label{def_A}
		\mc A|_p\left((W,a),(V,b)\right) \d \bg|_p (W,V) + a\bm\ell|_p(V)+b\bm\ell|_p(W)+ab\ell^{(2)}|_p
	\end{equation}
	is non-degenerate at every $p\in\mc H$. A five-tuple $\{\mc H,\bg,\bm\ell,\elltwo,\bY\}$, where $\bY$ is a $(0,2)$ symmetric tensor field on $\mc H$, is called hypersurface data.
\end{defi}
Since $\bm{\mc A}$ is non-degenerate one can introduce its inverse $\bm{\mc A}^{\sharp}$ by means of $\bm{\mc A}^{\sharp}\big(\bm{\mc A}((V,a),\cdot),\cdot\big)=(V,a)$ for every $(V,a)\in\X(\mc H)\otimes\mc{F}(\mc H)$. This defines a $(2,0)$ symmetric tensor field $P$, a vector $n$ and a scalar $n^{(2)}$ on $\mc H$ by the decomposition 
\begin{equation}
	\label{def_Asharp}
	\mc A^{\sharp}\left((\bm\alpha,a),(\bm\beta,b)\right) = P (\bm\alpha,\bm\beta)+a n(\bm\beta)+bn(\bm\alpha)+ab n^{(2)},
\end{equation} 
for all $\bm\alpha,\bm\beta\in\X^{\star}(\mc H)$ and $a,b\in\mc F(\mc H)$. Equivalently, $P$, $n$ and $n^{(2)}$ are defined by
\begin{multicols}{2}
	\noindent
	\begin{align}
		\gamma_{ab}n^b + n^{(2)}\ell_a&=0,\label{gamman}\\
		\ell_an^a+n^{(2)}\ell^{(2)}&=1,\label{ell(n)}
	\end{align}
	\begin{align}
		P^{ab}\ell_b+\ell^{(2)} n^a&=0,\label{Pell}\\
		P^{ac}\gamma_{cb} + \ell_b n^a &=\delta^a_b.\label{Pgamma}
	\end{align}
\end{multicols}
Despite its name, the notion of hypersurface data does not view $\mc H$ as embedded in another ambient manifold. The relation between Definition \ref{def_hypersurfacedata} and the common definition of embedded hypersurface is as follows.
\begin{defi}
	\label{defi_embedded}
	Metric hypersurface data $\{\mc H,\bg,\bm\ell,\ell^{(2)}\}$ is $(\Phi,\xi)$-embedded in a semi-Riemannian manifold $(\mc M,g)$ if there exists an embedding $\Phi:\mc H\hookrightarrow\mc M$ and a vector field $\xi$ along $\Phi(\mc H)$ everywhere transversal to $\Phi(\mc H)$, called rigging, such that
	\begin{equation}
		\label{embedded_equations}
		\Phi^{\star}(g)=\bg, \hspace{0.5cm} \Phi^{\star}\left(g(\xi,\cdot)\right) = \bm\ell,\hspace{0.5cm} \Phi^{\star}\left(g(\xi,\xi)\right) = \ell^{(2)}.
	\end{equation}
	Hypersurface data $\{\mc H,\bg,\bm\ell,\ell^{(2)},\bY\}$ is embedded provided that, in addition, 
	\begin{equation}
		\label{Yembedded}
		\dfrac{1}{2}\Phi^{\star}\left(\lie_{\xi} g\right) = \bY.
	\end{equation}
\end{defi}
Given metric hypersurface data, the condition of $\bg$ being degenerate is equivalent to $\Phi(\mc H)$ being an embedded null hypersurface, and it is also equivalent to $n^{(2)}=0$ \cite{Marc2}. Hence, (metric) hypersurface data satisfying $n^{(2)}=0$ are called \textbf{null (metric) hypersurface data}. We restrict to this case from now on. Note that $\bg(n,\cdot)=0$ in this case. A cross-section (or simply a section) $\mc S$ of $\mc H$ is an embedded hypersurface $\mc S\hookrightarrow\mc H$ satisfying that every integral curve of $n$ intersects $\mc S$ exactly once. \\

Definition \ref{defi_embedded} implies that the signature of $g$ is the same as the signature of the tensor $\mc A$ in Definition \ref{def_hypersurfacedata}. In particular, when $\bg$ has signature $\{0,+,\cdots,+\}$, then $(\mc M,g)$ is a spacetime. Although this is the most relevant situation in physical applications, the results in this paper hold for null metric hypersurface data of any signature.\\


Given null metric hypersurface data we define
\begin{equation}
	\label{defK}
	\bU\d \dfrac{1}{2}\lie_n\bg.
\end{equation} 
When the data is embedded $\bU$ coincides with the second fundamental form of $\Phi(\mc H)$ w.r.t. the unique normal one-form $\bm{\nu}$ satisfying $\bm{\nu}(\xi)=1$ (see \cite{Marc1}). It is also convenient to introduce $\bF\d \frac{1}{2}d\bm\ell$ and $\bs\d\bF(n,\cdot)$. Applying the Cartan identity $\lie_n = \iota_n \circ d + d\circ \iota_n$ to the one-form $\bm\ell$ and using $2\bF = d\bm\ell$ and $\bm\ell(n)=1$ one has 
\begin{equation}
	\label{lienell}
	\lie_n\bm\ell = 2\bs.
\end{equation}
We shall use repeatedly that $\bU(n,\cdot)=0$ and $\bs(n)=0$ without further notice.\\

Let $\{\mc H,\bg,\bm\ell,\elltwo\}$ be null metric hypersurface data $(\Phi,\xi)$-embedded in $(\mc M,g)$. By the transversality of $\xi$, given a (local) basis $\{e_a\}$ of $\mc H$, the set $\{\wh{e}_a\d\Phi_{\star}e_a,\xi\}$ is a (local) basis of $\Phi(\mc H)$ with dual basis $\{\bm\theta^a,\bm\nu\}$. From \eqref{def_Asharp} the inverse metric $g^{\alpha\beta}$ at $\Phi(\mc H)$ can be written in the basis $\{\xi, \wh{e}_a\}$ as
\begin{equation}
	\label{inversemetric}
	g^{\alpha\beta}\st{\mc H}{=}	P^{ab}\wh e_a^{\alpha}\wh e_b^{\beta} + n^{a}\wh e_a^{\alpha}\xi^{\beta} + n^{b}\wh e_b^{\beta}\xi^{\alpha} .
\end{equation}

Given embedded (metric) hypersurface data the notion of rigging vector is non-unique, since given a rigging $\xi$ any other vector $\xi' = z(\xi+\Phi_{\star}V)$ with $(z,V)\in\mc{F}^{\star}(\mc H)\times\X(\mc H)$ is also a rigging of $\Phi(\mc H)$. This translates into the abstract setting as a gauge freedom.
\begin{defi}
	\label{defi_gauge}
	Let $\{\mc H,\bg,\bm\ell,\ell^{(2)},\bY\}$ be hypersurface data and $(z,V)\in\mc{F}^{\star}(\mc H)\times\X(\mc H)$. We define the gauge transformed hypersurface data with gauge parameters $(z,V)$ by 
	\begin{align}
		\mc{G}_{(z,V)}\left(\bg \right)&\d \bg,\label{transgamma}\\
		\mc{G}_{(z,V)}\left( \bm{\ell}\right)&\d z\left(\bm{\ell}+\bg(V,\cdot)\right),\label{tranfell}\\
		\mc{G}_{(z,V)}\big( \ell^{(2)} \big)&\d z^2\big(\ell^{(2)}+2\bm\ell(V)+\bg(V,V)\big),\label{transell2}\\
		\mc{G}_{(z,V)}\left( \bY\right)&\d z \bY + \bm\ell\otimes_s d z +\dfrac{1}{2}\lie_{zV}\bg.\label{transY}
	\end{align}
\end{defi}
Transformations \eqref{transgamma}-\eqref{transY} induce the following \cite{Marc1}
\begin{multicols}{2}
	\noindent
	\begin{equation}
		\label{gaugeP}
		\mc{G}_{(z,V)}\left(P \right) = P -2V\otimes_s n,
	\end{equation}
	\begin{equation}
		\label{transn}
		\mc{G}_{(z,V)}\left( n \right)= z^{-1}n.
	\end{equation}
\end{multicols}
Given null metric hypersurface data $\{\mc H,\bg,\bm\ell,\elltwo\}$ there exists a unique torsion-free connection $\nablacero$ on $\mc H$ that satisfies \cite{Marc2}
\begin{multicols}{2}
	\noindent
	\begin{equation}
		\label{nablagamma}
		\nablacero_a\gamma_{bc} = -\ell_c\U_{ab} - \ell_b\U_{ac},
	\end{equation}
	\begin{equation}
		\label{nablaell}
		\nablacero_a\ell_b  = \F_{ab} - \elltwo\U_{ab}.
	\end{equation}
\end{multicols}
When the data is embedded in a semi-Riemannian manifold $(\mc M,g)$, the connection $\nablacero$ is related with the Levi-Civita connection $\nabla$ of $g$ by \cite{Marc2}
\begin{equation}
	\label{connections}
	\nabla_{\Phi_{\star}X}\Phi_{\star}Y \st{\mc H}{=} \Phi_{\star}\nablacero_X Y - \bY(X,Y)\nu - \bU(X,Y)\xi,
\end{equation}
for every $X,Y\in\X(\mc H)$. Unless otherwise indicated, scalar functions related by $\Phi^{\star}$ are denoted with the same symbol. The action of $\nablacero$ on the contravariant data $\{P,n\}$ is given by \cite{Marc2}
\begin{align}
	\nablacero_c n^b & =\s_c n^b + P^{ba}\U_{ca},\label{derivadannull}\\
	\nablacero_c P^{ab} & = -\big(n^aP^{bd}+n^bP^{ad}\big) \F_{cd} - n^an^b (d\elltwo)_c.\label{derivadaP}
\end{align}
From \eqref{derivadaP} it can be shown that \cite{MarcAbstract} 
\begin{equation}
	\label{lienP}
	\lie_n P^{ab} = -2 \s_c \big(P^{ac}n^b + P^{bc}n^a\big) -2P^{ac}P^{bd}\U_{cd} - n^an^bn\big(\elltwo\big).
\end{equation}
A direct consequence of \eqref{connections} and \eqref{derivadannull} is 
\begin{equation}
	\nabla_{\nu}\nu = -\bY(n,n)\nu.
\end{equation}
It is then natural to define the surface gravity of $n$ by $\kappa_n\d -\bY(n,n)$. Another consequence of \eqref{derivadannull} is that for any one-form $\bm\theta$ the following identity holds \cite{miguel3}
\begin{equation}
	\label{nnablacerotheta}
	2n^b \nablacero_{(a}\theta_{b)} = \lie_n\theta_a + \nablacero_a\big(\bm\theta(n)\big) - 2\big(\bm\theta(n)\s_a + P^{bc}\U_{ab}\theta_c\big).
\end{equation}
This expression will be mostly used in cases where $\bs=0$, $\bU=0$ and $\bm\theta$ is Lie-constant $\lie_n\bm\theta=0$, in which case it reduces to
\begin{equation}
\label{nnablacerotheta2}
2n^b \nablacero_{(a}\theta_{b)} =  \nablacero_a\big(\bm\theta(n)\big).
\end{equation}
As shown in \cite{Marc1}, the completely tangential components of the ambient Riemann tensor, as well as its 3-tangential, 1-transverse components can be written as
\begin{equation}
	\label{ABembedded}
	R_{\alpha\beta\mu\nu}\xi^{\alpha}e^{\beta}_be^{\mu}_ce^{\nu}_d \st{\mc H}{=} A_{bcd}, \hspace{1cm} R_{\alpha\beta\mu\nu} e^{\alpha}_ae^{\beta}_be^{\mu}_ce^{\nu}_d \st{\mc H}{=} B_{abcd}.
\end{equation}
where $A$ and $B$ are tensors on $\mc H$ defined solely in terms of hypersurface data as
\begin{align}
	A_{bcd}&\d 2\nablacero_{[d}\F_{c]b} + 2\nablacero_{[d}\Y_{c]b} + \U_{b[d}\nablacero_{c]}\elltwo + 2\Y_{b[d}(\r-\s)_{c]},\label{A} \\
	B_{abcd}&\d \gamma_{af}\Rcero^f{}_{bcd} + 2\ell_a\nablacero_{[d}\U_{c]b}+2\U_{a[d}\Y_{c]b} + 2\U_{b[c}(\Y+\F)_{d]a},\label{B}
\end{align}
where $\Rcero^f{}_{bcd}$ is the curvature of $\nablacero$ and $\br\d\bY(n,\cdot)$.
From \eqref{ABembedded} it follows that the pullback of the ambient Ricci tensor can be written in terms of hypersurface data as \cite{Mio1} $$g^{\alpha\beta} R_{\alpha \mu \beta \nu} e_a^{\mu} e_b^{\nu}\st{\mc H}{=}B_{acbd}P^{cd}- (A_{bac}+A_{abc})n^c.$$

The RHS defines an abstract tensor on $\mc H$, called \textit{constraint tensor} $\R$, whose explicit form is \cite{miguel3}
	\begin{equation}
		\label{constraint}
		\begin{aligned}
			\mc R_{ab}& = \accentset{\circ}{R}_{(ab)} -2\lie_n \Y_{ab} - (2\kappa_n+\tr_P\bU)\Y_{ab} + \nablacero_{(a}\left(\s_{b)}+2\r_{b)}\right)\\
			&\quad -2\r_a\r_b + 4\r_{(a}\s_{b)} - \s_a\s_b - (\tr_P\bY)\U_{ab} + 2P^{cd}\U_{d(a}\left(2\Y_{b)c}+\F_{b)c}\right),
		\end{aligned}
	\end{equation}
where $\Rcero_{ab}$ is the Ricci tensor of $\nablacero$. Note that all the dependence on the tensor $\bY$ in \eqref{constraint} is explicit. The contraction of $\mc{R}_{ab}$ with $n^a$ relies on the following general identity \cite{miguel3}
\begin{equation}
	\label{Rceron}
	\Rcero_{(ab)}n^a = \dfrac{1}{2}\lie_n\s_b - 2P^{ac}\U_{ab}\s_c + P^{ac}\nablacero_c\U_{ab} - \nablacero_b\big(\tr_P\bU\big) + \big(\tr_P\bU\big)\s_b
\end{equation}
and is 
\begin{equation}
	\label{constraintn}
	\begin{aligned}
		\mc R_{ab}n^a  = -\lie_n(\r_b-\s_b) - \nablacero_b \kappa_n - (\tr_P\bU) (\r_b-\s_b)  - \nablacero_b (\tr_P\bU) + P^{cd}\nablacero_c\U_{bd} - 2P^{cd}\U_{bd}s_c.
	\end{aligned}
\end{equation}
Another contraction with $n^b$ gives
\begin{equation}
	\label{constraintnn}
	\mc R_{ab}n^an^b = -n(\tr_P\bU) + (\tr_P\bU)\kappa_n - P^{ab}P^{cd}\U_{ac}\U_{bd}.
\end{equation}
The contraction of $\mc{R}_{ab}$ with $P^{ab}$ is \cite{Mio3}
\begin{equation}
	\label{trPR}
	\begin{aligned}
		\tr_P\bm{\mc R} &= \tr_P \Rcero -2\lie_n(\tr_P\bY)  - 2\big(\kappa_n+\tr_P\bU\big)\tr_P\bY+ \div_P(\bs + 2\br) \\
		&\quad -2P(\br,\br)-4P(\br,\bs)-P(\bs,\bs)+2\kappa_n n(\elltwo),
	\end{aligned}
\end{equation}
where for any one-form $\bm t$ we define $\div_P\bm{t}\d P^{ab}\nablacero_a t_b$.\\

For later use we analyze the behavior of the tensor $\Rcero_{ab}$ under gauge transformations of the form $\mc{G}_{(z,0)}$.
\begin{prop}
Let $\{\H,\bg,\bm\ell,\elltwo\}$ be metric hypersurface data and let $\Rcero_{ab}$ be the Ricci tensor of the metric connection $\nablacero$. Let $z\in\mc{F}^{\star}(\H)$ and $w=\log|z|$. Then,
\begin{equation}
	\label{Rcerotrans}
	\begin{aligned}
\mc{G}_{(z,0)}\Rcero_{ab} & = \Rcero_{ab} +\left(\tr_P\bU+\dfrac{1}{2}n(w)\right)\ell_{(a}\nablacero_{b)}w + \s_{(a}\left(\nablacero_{b)}w - n(w) \ell_{b)}\right) \\
&\quad\,+ \dfrac{1}{2}\ell_{b}\nablacero_{a} n(w)  - P^{cd}\U_{d(a}\ell_{b)}\nablacero_c w - \dfrac{1}{2}\nablacero_a\nablacero_b w + \dfrac{1}{2}n(w)\left(\F_{ab}+\elltwo \U_{ab}\right)\\
&\quad\, - \dfrac{1}{4}\nablacero_a w\nablacero_b w - \dfrac{1}{4}\big(n(w)\big)^2\ell_a\ell_b.
	\end{aligned}
\end{equation}
\begin{proof}
We use the well-known fact that the difference of the Ricci tensors of two connections $\nabla$ and $\nabla'$ is 
\begin{equation}
R'_{ab} - R_{ab} = \nabla_c S^c{}_{ab} - \nabla_b S^c{}_{ca} + S^c{}_{dc}S^{d}{}_{ba} - S^c{}_{db}S^d{}_{ca},
\end{equation}
where $S\d\nabla'-\nabla$. Particularizing this identity to $\nabla=\nablacero$ and $\nabla' = \mc{G}_{(z,0)}\nablacero = \nablacero +n\otimes \bm\ell \otimes_s dw$ (see \cite{Marc2}) gives (note that $S^c{}_{ab} = n^c\ell_{(a}\nablacero_{b)}w$)
\begin{align*}
\Rcero'_{ab} - \Rcero_{ab} = \nablacero_c\big(n^c\ell_{(a}\nablacero_{b)}w\big) - \nablacero_b\big(n^c\ell_{(c}\nablacero_{a)}w\big) + n^c n^d \ell_{(d}\nablacero_{c)}w\, \ell_{(b}\nablacero_{a)}w - n^c n^d \ell_{(d}\nablacero_{b)}w\, \ell_{(c}\nablacero_{a)}w.
\end{align*}
Using \eqref{derivadannull} and \eqref{nablaell} the first term is
\begin{align*}
\nablacero_c\big(n^c\ell_{(a}\nablacero_{b)}w\big) &= \big(\tr_P\bU\big)\ell_{(a}\nablacero_{b)}w +n^c\big(F_{c(a}-\elltwo\U_{c(a}\big)\nablacero_{b)}w +n^c \ell_{(a|}\nablacero_c\nablacero_{|b)}w\\
&= \big(\tr_P\bU\big)\ell_{(a}\nablacero_{b)}w + \s_{(a}\nablacero_{b)}w + n^c\ell_{(a}\nablacero_{b)}\nablacero_c w\\
&= \big(\tr_P\bU\big)\ell_{(a}\nablacero_{b)}w + \s_{(a}\nablacero_{b)}w + \ell_{(a}\nablacero_{b)}n(w) - \nablacero_c w\, \ell_{(a}\nablacero_{b)}n^c\\
&= \big(\tr_P\bU\big)\ell_{(a}\nablacero_{b)}w + \s_{(a}\nablacero_{b)}w + \ell_{(a}\nablacero_{b)}n(w) - n(w)\ell_{(a}\s_{b)} - P^{cd}\U_{d(b}\ell_{a)}\nablacero_c w,
\end{align*}
where in the second line we used that the Hessian is symmetric and in the last one we inserted \eqref{derivadannull}. Using again \eqref{nablaell} the second term is given by
\begin{align*}
\nablacero_b\big(n^c\ell_{(c}\nablacero_{a)}w\big) & = \dfrac{1}{2}\nablacero_b\big(\nablacero_{a}w + n(w)\ell_a\big)\\
&= \dfrac{1}{2}\nablacero_b\nablacero_a w +\dfrac{1}{2}\ell_{a}\nablacero_{b} n(w) + \dfrac{1}{2}n(w)\big(\F_{ba}-\elltwo\U_{ba}\big).
\end{align*}
The third term is simply given by $$n^c n^d \ell_{(d}\nablacero_{c)}w\, \ell_{(b}\nablacero_{a)}w = n(w) \ell_{(b}\nablacero_{a)}w.$$ Finally, the last term is 
\begin{align*}
n^c n^d \ell_{(d}\nablacero_{b)}w\, \ell_{(c}\nablacero_{a)}w &= \dfrac{1}{4}n^c n^d \big(\ell_d\nablacero_b w + \ell_b\nablacero_d w\big)\big(\ell_c\nablacero_a w + \ell_a\nablacero_c w\big)\\
&= \dfrac{1}{4}\nablacero_b w\nablacero_a w + \dfrac{1}{4}\big(n(w)\big)^2\ell_a\ell_b +\dfrac{1}{2}n(w)\ell_{(a}\nablacero_{b)}w.
\end{align*}
Combining everything \eqref{Rcerotrans} follows.
\end{proof}
\end{prop}

\begin{cor}
	\label{corRcero}
Let $\{\H,\bg,\bm\ell,\elltwo\}$ be metric hypersurface, $z\in\mc{F}^{\star}(\H)$ and $w=\log|z|$. Then, 
\begin{equation*}
	\Rcero'_{(ab)} + \s'_a\s'_b -\nablacero'_{(a}\s'_{b)} = \Rcero_{(ab)} + \s_a\s_b -\nablacero_{(a}\s_{b)} + \left(\tr_P\bU\right)\ell_{(a}\nablacero_{b)}w -P^{cd}\U_{d(a}\ell_{b)}\nablacero_c w+n(w)\elltwo \U_{ab}.
\end{equation*}
\begin{proof}
From $\bs' = \bs +\dfrac{1}{2}\left(n(w)\bm\ell - dw\right)$ (see \cite{Marc2}) it follows 
\begin{equation}
	\label{ss}
\s'_a \s'_b = \s_a\s_b +n(w)\s_{(a}\ell_{b)}-\s_{(a}\nablacero_{b)}w +\dfrac{1}{4}\big(n(w)\big)^2 \ell_a\ell_b -\dfrac{1}{2}n(w)\ell_{(a}\nablacero_{b)}w + \dfrac{1}{4}\nablacero_a w \nablacero_b w.
\end{equation} 
Using $\nablacero' = \nablacero +n\otimes \bm\ell \otimes_s dw$ gives 
\begin{align}
	\nablacero_{(a}'\s_{b)}' &= \nablacero_{(a}\s_{b)}' - \bs'(n) \ell_{(a}\nablacero_{b)}w\nonumber\\	
	&= \nablacero_{(a}\s_{b)} + \dfrac{1}{2}\nablacero_{(a}\big(n(w)\ell_{b)}-\nablacero_{b)}w\big) \nonumber\\
	&= \nablacero_{(a}\s_{b)} + \dfrac{1}{2}\ell_{(b}\nablacero_{a)}n(w) -\dfrac{1}{2}n(w)\elltwo\U_{ab} - \dfrac{1}{2}\nablacero_a\nablacero_b w,\label{nablas}
\end{align}
where in the second line we used $\bs'(n) = z\bs'(n')=0$ and in the third one we inserted \eqref{nablaell} and used $\F_{(ab)} = 0$. Combining \eqref{ss} and \eqref{nablas} with \eqref{Rcerotrans} the result follows.
\end{proof}
\end{cor}
	

%% file: reviewtransverse.tex
\section{Transverse expansion of the metric}
\label{section_higher}

In the first paper \cite{Mio3} of the series we computed the $m$-th transverse derivative of the ambient Ricci tensor on a general null hypersurface $\H$ in terms of transverse derivatives of the ambient metric $g$ at $\mc H$ up to order $m+2$ making explicit the dependence on the terms $\lie_{\xi}^{(m+2)}g$ and $\lie_{\xi}^{(m+1)}g$. In this section we review the main results of \cite{Mio3} that we shall need in the rest of the paper. In order to simplify the notation we introduced the tensors 
\begin{equation}
	\label{def_asymptoticexpansion}
	\mc{K}^{(m)}\d\lie_{\xi}^{(m)}g,\qquad \bY^{(m)}\d \dfrac{1}{2}\Phi^{\star}\mc{K}^{(m)}, \qquad \br^{(m)}\d \bY^{(m)}(n,\cdot), \qquad \kappa^{(m)}\d -\bY^{(m)}(n,n),
\end{equation} 
as well as 
\begin{equation}
	\label{def_asymptoticricci}
\hspace{-0.3cm}	\mc{R}^{(m)}\d \Phi^{\star}\big(\lie_{\xi}^{(m-1)}\ric\big),\quad \dot{\mc R}^{(m)}\d \Phi^{\star}\big(\lie_{\xi}^{(m-1)}\ric(\xi,\cdot)\big),\quad \ddot{\mc R}^{(m)}\d \Phi^{\star}\big(\lie_{\xi}^{(m-1)}\ric(\xi,\xi)\big).
\end{equation} 

In what follows we refer to the collection of tensors $\{\bY^{(k)}\}_{k\ge 1}$ as the \textit{transverse} or \textit{asymptotic expansion}. As proved in \cite{Mio3}, the remaining derivatives of the metric, namely the tensors $\mc{K}^{(m)}(\xi,\cdot)$ at $\mc H$, are given in terms of $a_{\xi}\d\nabla_{\xi}\xi$, metric hypersurface data and the tensors $\{\bY,\bY^{(2)},...,\bY^{(m-1)}\}$ due to the following proposition.
\begin{prop}
	\label{liegxi}
	Let $\mc H$ be a null hypersurface embedded in a semi-Riemannian manifold $(\mc M,g)$ with embedding $\Phi$ and rigging $\xi$ extended arbitrarily as a smooth vector field in a neighbourhood of $\Phi(\H)$ and define $a_{\xi}\d\nabla_{\xi}\xi$. Let $Z$ be any vector field tangent to $\Phi(\mc H)$. Then for every $m\in\mathbb{N}\cup\{0\}$,
	\begin{align}
		\mc{K}^{(m+1)}(\xi,\xi) &\st{\mc H}{=} 2\sum_{i=0}^m\binom{m}{i}\big(\mc{K}^{(m-i)}\big) \big(\lie_{\xi}^{(i)} a_{\xi},\xi\big),\label{liegxixi}\\
		\mc{K}^{(m+1)}(\xi,Z) & \st{\mc H}{=}  \dfrac{1}{2}Z\left(\mc{K}^{(m)}(\xi,\xi)\right) + \sum_{i=0}^m\binom{m}{i}\big(\mc{K}^{(m-i)}\big) \big(\lie_{\xi}^{(i)} a_{\xi},Z\big).\label{liegxiX}
	\end{align}
	In particular, when $a_{\xi}=0$ it follows $\mc{K}^{(1)}(\xi,Z) = \dfrac{1}{2}Z\big(\elltwo\big)$ and $\mc{K}^{(m+1)}(\xi,\xi)=\mc{K}^{(m+2)}(\xi,Z)=0$ for every $m\ge 0$.
\end{prop}

In the following lemma we recall a well-known identity for derivatives of products of any two objects $S$ and $T$. Note that when $S$ and $T$ are tensors, the expression also holds when contraction is allowed. 
\begin{lema}
	Let $S$ and $T$ be two objects, $S\circledast T$ any product of them and $\mc{D}$ any derivative operator. Then, 
	\begin{equation}
		\label{derivada}
		\mc{D}^{(m)}\big(S\circledast T\big) = \sum_{i=0}^m\binom{m}{i}\big( \mc{D}^{(i)} S\big)\circledast\big( \mc{D}^{(m-i)}T\big).
	\end{equation}
\end{lema}
In order to compute $\lie_{\xi}^{(m)}R_{\alpha\beta}$ up to order $m+1$ the strategy employed in \cite{Mio3} can be summarized as follows. Firstly, given any vector field $\xi$ one can introduce the tensor $\Sigma[\xi]\d \lie_{\xi}\nabla$, or in abstract index notation
\begin{equation}
	\label{Sigma}
	\Sigma[\xi]^{\alpha}{}_{\mu\nu} =\dfrac{1}{2}g^{\alpha\beta}\left(\nabla_{\mu}\mc{K}[\xi]_{\nu\beta} + \nabla_{\nu}\mc{K}[\xi]_{\mu\beta} - \nabla_{\beta}\mc{K}[\xi]_{\mu\nu}\right),
\end{equation} 
where $\mc{K}[\xi]$ is the so-called deformation tensor of $\xi$ defined by $\mc{K}[\xi]\d \lie_{\xi}g$. To compute $\lie_{\xi}^{(m)}R_{\alpha\beta}$ it suffices to apply the operator $\lie_{\xi}^{(m-1)}$ to the classical identity \cite{yano2020theory}
\begin{equation}
	\label{Yano}
	\lie_{\xi}R_{\alpha\beta} = \nabla_{\mu}\Sigma[\xi]^{\mu}{}_{\alpha\beta} - \nabla_{\beta}\Sigma[\xi]^{\mu}{}_{\alpha\mu}
\end{equation}
and express the result making $\lie^{(m+1)}_{\xi}g$ and $\lie^{(m+2)}_{\xi}g$ explicit. In order to do that one needs to commute $\lie_{\xi}^{(m-1)}$ and $\nabla$. This commutator is computed explicitly in the following proposition \cite{Mio3}. We introduce the notation $A^{(m)}\d \lie_{\xi}^{(m-1)}A$, $m\ge 1$, for any tensor field $A$.
\begin{prop}
	\label{propMarc}
	Let $\xi\in\X(\mc M)$ and $m\ge 1$ be a integer. Then, given any $(p,q)$ tensor $A^{\alpha_1\cdots\alpha_q}_{\beta_1\cdots\beta_p}$ the following identity holds
	\begin{equation*}
		\begin{aligned}
			\lie_{\xi}^{(m)}\nabla_{\gamma}A^{\alpha_1\cdots\alpha_q}_{\beta_1\cdots\beta_p} = \nabla_{\gamma} A^{(m+1)}{}^{\alpha_1\cdots\alpha_q}_{\beta_1\cdots\beta_p} + \sum_{k=0}^{m-1}\binom{m}{k+1}&\left(\sum_{j=1}^{q}A^{(m-k)}{}^{\alpha_1\cdots\alpha_{j-1}\sigma\alpha_{j+1}\cdots\alpha_q}_{\beta_1\cdots\beta_p}\Sigma^{(k+1)}{}^{\alpha_j}{}_{\sigma\gamma}\right. \\
			&\left.- \sum_{i=1}^{p}A^{(m-k)}{}^{\alpha_1\cdots\alpha_q}_{\beta_1\cdots\beta_{i-1}\sigma\beta_{i+1}\cdots\beta_p}\Sigma^{(k+1)}{}^{\sigma}{}_{\beta_i\gamma}\right).
		\end{aligned}
	\end{equation*}
\end{prop}
Applying $\lie_{\xi}^{(m-1)}$ to \eqref{Yano} one finds that $$\lie_{\xi}^{(m)} R_{\alpha\beta} = \nabla_{\mu}\Sigma[\xi]^{(m)}{}^{\mu}{}_{\alpha\beta} - \nabla_{\beta}\Sigma[\xi]^{(m)}{}^{\mu}{}_{\mu\alpha} + \text{lower order terms},\qquad m\ge 1,$$ 

where the ``lower order terms'' stand for derivatives of $g$ up to an including order $m$ along the transverse direction. Contracting this identity with two transverse vectors, one tangent and one transverse, and two tangent vectors, one arrives at the following result relating the tensors \eqref{def_asymptoticexpansion} and \eqref{def_asymptoticricci} \cite{Mio3}.
\begin{prop}
	\label{prop_derivadas}
	Let $\{\mc H,\bg,\bm\ell,\elltwo\}$ be null metric hypersurface data $(\Phi,\xi)$-embedded in $(\mc M,g)$ (Def. \ref{defi_embedded}) and extend $\xi$ arbitrarily off $\Phi(\mc H)$. Let $m\ge 1$ be an integer. Then,
	\begin{multicols}{2}
		\noindent
		\begin{equation}
			\label{ddotR}
			\ddot{\mc R}^{(m)} = - \tr_P\bY^{(m+1)} + \mc{O}^{(m)}(\bY^{\le m}),
		\end{equation}
		\begin{equation}
			\label{dotR}
			\dot{\mc R}^{(m)}_a = \r^{(m+1)}_a + \mc{O}^{(m)}_a(\bY^{\le m}),
		\end{equation}
	\end{multicols}
	\vskip -0.4cm
	\begin{equation}
		\label{R}
		\begin{aligned}
			\mc{R}^{(m+1)}_{ab} &= -2\lie_n\Y^{(m+1)}_{ab} - \left(2(m+1)\kappa_n + \tr_P\bU\right)\Y^{(m+1)}_{ab}- (\tr_P\bY^{(m+1)})\U_{ab}\\
			&\quad\,  + 4P^{cd}\U_{c(a}\Y^{(m+1)}_{b)d}  +4(\s-\r)_{(a} \r^{(m+1)}_{b)}+ 2\nablacero_{(a}\r^{(m+1)}_{b)}\\
			&\quad\, -2\kappa^{(m+1)}\Y_{ab} + \mc{O}^{(m)}_{ab}(\bY^{\le m}),
		\end{aligned}
	\end{equation}
	where $\mc{O}^{(m)}$, $\mc{O}^{(m)}_a$ and $\mc{O}^{(m)}_{ab}$ are, respectively, a scalar, a one-form and a $(0,2)$ symmetric tensor with the property that when $\nabla_{\xi}\xi=0$ they only depend on metric data $\{\bg,\bm\ell,\elltwo\}$ and $\{\bY,...,\bY^{(m)}\}$. Moreover,
		\begin{align}
		\hskip-0.3cm	\mc{R}^{(m+1)}_{ab} n^b &= -\lie_n\r^{(m+1)}_a - \big(2m\kappa_n + \tr_P\bU\big)\r^{(m+1)}_a-\nablacero_a\kappa^{(m+1)} + \mc{O}^{(m)}_{ab}n^b,\label{elierictangn}\\
		\hskip-0.3cm		P^{ab}\mc{R}^{(m+1)}_{ab} &= -2\lie_n\big(\tr_P\bY^{(m+1)}\big) - 2\left((m+1)\kappa_n + \tr_P\bU\right)\tr_P\bY^{(m+1)} \nonumber\\
				&\quad\, + 2\kappa^{(m+1)}\big(n(\elltwo)-\tr_P\bY\big)-4P\big(\br+\bs,\br^{(m+1)}\big) + 2\div_P\br^{(m+1)} + P^{ab}\mc{O}^{(m)}_{ab}.	\label{ePcontractioneq}
			\end{align}
\end{prop}

%
%
%

As one can see, the scalar $\tr_P\bY^{(m)}$ and the one-form $\br^{(m)}$ appear both in identities \eqref{ddotR}, \eqref{dotR} and in \eqref{elierictangn}, \eqref{ePcontractioneq}. It is straightforward to combine them and obtain
\begin{equation}
	\label{B1}
	\mc{R}_{ab}^{(m+1)}n^b + \lie_n \dot{\mc R}^{(m)}_a + \big(2m\kappa_n + \tr_P\bU\big) \dot{\mc R}^{(m)}_a - \nablacero_a\big(\dot{\mc R}^{(m)}_b n^b\big) = \T^{(m)}_a,
\end{equation}
and
\begin{equation}
	\label{B2}
	\begin{aligned}
		P^{ab}\mc{R}^{(m+1)}_{ab} - 2\lie_n \ddot{\mc R}^{(m)} - 2\big((m+1)\kappa_n + \tr_P\bU\big)\ddot{\mc R}^{(m)}&\\
		+ 2 \big(n(\elltwo)-\tr_P\bY\big) \dot{\mc R}^{(m)}_a n^a + 4 P\big(\br+\bs, \dot{\mc R}^{(m)}\big) -2\div_P\dot{\mc R}^{(m)} &= \T^{(m)},
	\end{aligned}
\end{equation}
where $\T_a^{(m)}$ and $\T^{(m)}$ are, respectively, a one-form and a scalar function that depend only on metric data and $\{\bY,...,\bY^{(m)}\}$ as well as in the way the rigging has been extended (this dependence drops out completely when $\xi$ is extended geodesically). Identities \eqref{B1}-\eqref{B2} are a manifestation of the (ambient) second contracted Bianchi identity order by order on $\Phi(\H)$. The dependence of $\T_a^{(m)}$ and $\T^{(m)}$ in terms of the tensors $\mc{O}^{(m)}$, $\mc{O}^{(m)}_a$ and $\mc{O}^{(m)}_{ab}$ can be easily read from equations \eqref{B1}-\eqref{B2}. We do not write its explicit form since it will not be needed.\\

For later use we need to compute the ``gauge'' transformations of the tensors $\mc{O}^{(m)}(\bY^{\le m})$, $\mc{O}^{(m)}_a(\bY^{\le m})$ and $\mc{O}^{(m)}_{ab}(\bY^{\le m})$ under the change $\xi'=z\xi$ with $\xi(z)=0$. In order not to overload the notation we drop the arguments ``$\bY^{\le m}$'' and ``$\bY'{}^{\le m}$''. When a prime is placed on the quantities $\mc{O}^{(m)}{}'$, $\mc{O}_a^{(m)}{}'$ and $\mc{O}_{ab}^{(m)}{}'$ it means that the object has been constructed with $\xi'$ according to \eqref{ddotR}-\eqref{R}.

\begin{prop}
	\label{propgaugeO}
Let $\{\H,\bg,\bm\ell,\elltwo\}$ be null metric hypersurface data $(\Phi,\xi)$-embedded in $(\mc M,g)$ and extend $\xi$ off $\Phi(\H)$ by $\nabla_{\xi}\xi = 0$. Assume $\elltwo=0$ and let $z\in\mc{F}^{\star}(\mc M)$ satisfying $\xi(z)=0$ and define $\xi' \d z\xi$. Then, for every $m\ge 1$,
\begin{align}
\mc{O}^{(m)}{}' & = z^{m+1}\mc{O}^{(m)},\label{transO}\\
\mc{O}_a^{(m)}{}' & = z^{m}\mc{O}_a^{(m)} - (m-1)z^{m-1}\left(\tr_P\bY^{(m)}-\mc{O}^{(m-1)}\right)\nablacero_a z,\label{transOa}\\
\mc{O}_{ab}^{(m)}{}' & = z^{m}\mc{O}_{ab}^{(m)} + 2m z^{m-1} \mc{O}^{(m)}_{(a}\nablacero_{b)} z - m(m-1) z^{m-2}\left(\tr_P\bY^{(m)}-\mc{O}^{(m-1)}\right)\nablacero_az \nablacero_b z.\label{transOab}
\end{align}
\begin{proof}
The idea of the proof is to compute the transformations of the expansion $\{\bY^{(m)}\}$ and of the tensors $\ddot{\mc R}^{(m)}$, $\dot{\mc R}^{(m)}_a$ and $\mc{R}_{ab}^{(m+1)}$ and insert them into \eqref{ddotR}-\eqref{R} to obtain the expressions for $\mc{O}^{(m)}{}'$, $\mc{O}_a^{(m)}{}'$ and $\mc{O}_{ab}^{(m)}{}'$. Under the rescaling $\xi' = z\xi$ the tensors $n$, $P$, $\bU$, $\bs-\br$ and the metric connection $\nablacero$ transform by \cite{Marc2,Mio2}
\begin{equation}
	\label{transUP}
n' = z^{-1}n,\qquad \bU ' = z^{-1}\bU,\qquad P' = P,\qquad \bs'-\br' = \bs - \br - z^{-1} d z,
\end{equation}
\begin{equation}
	\label{transnablcaero}
\nablacero' = \nablacero + z^{-1} n\otimes \bm\ell \otimes_s dz.
\end{equation}
In order to compute the transformations of $\{\bY^{(m)}\}$ and of $\{\ddot{\mc R}^{(m)},\dot{\mc R}^{(m)}_a,\mc{R}_{ab}^{(m+1)}\}$ we first prove the following identity valid for every symmetric $(0,2)$ tensor field $T$,
\begin{equation}
	\label{indu}
	\begin{aligned}
		\lie_{z\xi}^{(m)}T_{\alpha\beta} &= z^m \lie_{\xi}^{(m)}T_{\alpha\beta} + 2m z^{m-1} \xi^{\mu}\lie_{\xi}^{(m-1)}T_{\mu(\alpha}\nabla_{\beta)}z \\
		&\quad\, + m(m-1) z^{m-2} \xi^{\mu}\xi^{\nu}\lie^{(m-2)}_{\xi}T_{\mu\nu} \, \nabla_{\alpha}z\nabla_{\beta} z.
	\end{aligned}
\end{equation}
For $m=1$ it is clearly true because it reduces to the standard identity $$\lie_{z\xi} T_{\alpha\beta} = z\lie_{\xi}T_{\alpha\beta} + 2\xi^{\mu}T_{\mu(\alpha}\nabla_{\beta)}z.$$

Let us assume it holds for $m\ge 2$. Then, applying $\lie_{z\xi}$ to \eqref{indu} and using $\xi(z)=0$ gives 
\begin{align*}
	\lie_{z\xi}^{(m+1)}T_{\alpha\beta} & = z\lie_{\xi} \lie_{z\xi}^{(m)}T_{\alpha\beta} + 2\xi^{\mu}\lie_{z\xi}^{(m)}T_{\mu(\alpha}\nabla_{\beta)} z\\	
	&= z^{m+1}\lie_{\xi}^{(m+1)}T_{\alpha\beta}  + 2m z^{m} \xi^{\mu}\lie_{\xi}^{(m)}T_{\mu(\alpha}\nabla_{\beta)}z + m(m-1) z^{m-1} \xi^{\mu}\xi^{\nu}\lie^{(m-1)}_{\xi}T_{\mu\nu} \, \nabla_{\alpha}z\nabla_{\beta} z\\
	&\quad\, + 2z^m \xi^{\mu}\lie_{\xi}^{(m)}T_{\mu(\alpha}\nabla_{\beta)}z+ 2mz^{m-1} \xi^{\mu}\xi^{\nu}\lie^{(m-1)}_{\xi}T_{\mu\nu} \, \nabla_{\alpha}z\nabla_{\beta} z \\
	& = z^{m+1}\lie_{\xi}^{(m+1)}T_{\alpha\beta} + 2(m+1) z^{m} \xi^{\mu}\lie_{\xi}^{(m)}T_{\mu(\alpha}\nabla_{\beta)}z +(m+1)m z^{m-1} \xi^{\mu}\xi^{\nu}\lie^{(m-1)}_{\xi}T_{\mu\nu} \, \nabla_{\alpha}z\nabla_{\beta} z.
\end{align*}
Hence \eqref{indu} follows by induction. Applying \eqref{indu} to $T=g$ and using Proposition \ref{liegxi} with $a_{\xi}=0$ (so that $\mc{K}^{(m)}(\xi,\cdot)=0$ for $m\geq 1$) and $\elltwo=0$ gives
\begin{equation*}
	\mc{K}' = z\mc{K} +2dz\otimes_s g(\xi,\cdot),\qquad \mc{K}^{(k)}{}' = z^k \mc{K}^{(k)} \quad (k\ge 2),
\end{equation*}
which becomes
\begin{equation}
	\label{trans}
	\bY' = z\bY + d z \otimes_s \bm\ell,\qquad  \bY^{(k)}{}'= z^{k} \bY^{(k)} \quad (k\ge 2)
\end{equation}
after pulling it back to $\H$. Consequently,
\begin{multicols}{2}
	\noindent
\begin{align*}
\br' &= \br + \dfrac{1}{2z}\left(dz + n(z)\bm\ell\right),\\
\kappa'_n &= z^{-1}\kappa_n - z^{-2} n(z),
\end{align*}
\begin{align}
\qquad \br^{(k)}{}' = z^{k-1}\br^{(k)},\\
\qquad \kappa^{(k)}{}' = z^{k-2}\kappa^{(k)}.\label{transkappam}
\end{align}
\end{multicols}
To calculate the transformation of $\{\ddot{\mc R}^{(m)},\dot{\mc R}^{(m)}_a,\mc{R}_{ab}^{(m+1)}\}$ we particularize \eqref{indu} to $T_{\alpha\beta}=R_{\alpha\beta}$ and contract it with $\xi'{}^{\alpha}\xi'{}^{\beta}$, $e_a^{\alpha}\xi'{}^{\beta}$ and $e_a^{\alpha}e_b^{\beta}$, which gives
\begin{align}
	\ddot{\mc R}^{(m)}{}' & = z^{m+1}\ddot{\mc R}^{(m)},\label{transR}\\
	\dot{\mc R}^{(m)}_a{}' & = z^{m} \dot{\mc R}^{(m)}_a + (m-1)z^{m-1}\ddot{\mc R}^{(m-1)} \nablacero_a z,\\
	\mc{R}_{ab}^{(m+1)}{}' & = z^m \mc{R}_{ab}^{(m+1)} + 2m z^{m-1}\dot{\mc R}^{(m)}_{(a}\nablacero_{b)}z +m(m-1) z^{m-2}\ddot{\mc R}^{(m-1)}\nablacero_a z \nablacero_b z.\label{transRab}
\end{align}
In order to compute the transformed tensors $\mc{O}^{(m)}{}'$, $\mc{O}_a^{(m)}{}'$ and $\mc{O}_{ab}^{(m)}{}'$ one possibility is to insert the transformations we have just computed into \eqref{ddotR}-\eqref{R}, which gives \eqref{transO}-\eqref{transOab} after a somewhat long computation. Fortunately there is a quicker way to deduce \eqref{transO}-\eqref{transOab} without computing them explicitly. Let us rewrite \eqref{ddotR}-\eqref{R} formally as 
\begin{equation}
	\label{bes}
	\ddot{\mc R}^{(m)} = \mc{B}^{(m+1)} + \mc{O}^{(m)},\qquad \dot{\mc R}_a^{(m)} = \mc{B}_a^{(m+1)} + \mc{O}^{(m)}_a,\qquad \mc{R}_{ab}^{(m+1)} = \mc{B}^{(m+1)}_{ab} + \mc{O}^{(m)}_{ab},
\end{equation}
where $\mc{B}^{(m+1)}$, $\mc{B}_a^{(m+1)}$ and $\mc{B}^{(m+1)}_{ab}$ depend on $\bY^{(m+1)}$. Applying the rescaling $\xi' = z\xi$ gives $$\ddot{\mc R}^{(m)}{}' = \mc{B}^{(m+1)}{}' + \mc{O}^{(m)}{}',\qquad \dot{\mc R}_a^{(m)}{}' = \mc{B}_a^{(m+1)}{}' + \mc{O}_a^{(m)}{}',\qquad \mc{R}_{ab}^{(m+1)}{}' = \mc{B}_{ab}^{(m+1)}{}' + \mc{O}_{ab}^{(m)}{}'.$$ 

Inserting \eqref{transR}-\eqref{transRab} in the left hand sides and replacing \eqref{bes} we get identities that must be true for every $z$ and every collection $\{\bY^{(m)}\}$. In particular, terms involving \textit{only} $\bY^{(m+1)}$ must cancel each other. This also means that the terms that \textit{do not} depend on $\bY^{(m+1)}$ must also cancel each other. The quantities $\mc{B}^{(m+1)}$, $\mc{B}^{(m+1)}_a$ and $\mc{B}_{ab}^{(m+1)}$ and their primed counterparts are homogeneous in $\bY^{(m+1)}$ (i.e. they vanish identically when $\bY^{(m+1)}$ vanishes identically). Thus, the tensors $\mc{O}^{(m)}{}'$, $\mc{O}_a^{(m)}{}'$ and $\mc{O}_{ab}^{(m)}{}'$ must agree with those terms in \eqref{transR}-\eqref{transRab} involving only $\{\bY^{k\leq m}\}$. Using \eqref{ddotR} and \eqref{dotR} gives
\begin{align*}
	\mc{O}^{(m)}{}' & = z^{m+1}\mc{O}^{(m)},\\
	\mc{O}_a^{(m)}{}' & = z^{m}\mc{O}_a^{(m)} - (m-1)z^{m-1}\left(\tr_P\bY^{(m)}-\mc{O}^{(m-1)}\right)\nablacero_a z,\\
	\mc{O}_{ab}^{(m)}{}' & = z^{m}\mc{O}_{ab}^{(m)} + 2m z^{m-1} \mc{O}^{(m)}_{(a}\nablacero_{b)} z - m(m-1) z^{m-2}\left(\tr_P\bY^{(m)}-\mc{O}^{(m-1)}\right)\nablacero_az \nablacero_b z,
\end{align*}
which proves \eqref{transO}-\eqref{transOab}.
\end{proof}
\end{prop}

\begin{rmk}
	\label{rmk_imp}
As shown in \cite{Mio3} the objects $\mc{O}^{(m)}$, $\mc{O}_a^{(m)}$ and $\mc{O}_{ab}^{(m)}$ are universal, which means that the quantities $\mc{O}^{(m)}{}'$, $\mc{O}_a^{(m)}{}'$ and $\mc{O}_{ab}^{(m)}{}'$ constructed with $\xi'=z\xi$ (extended geodesically) according to \eqref{ddotR}-\eqref{R} are exactly
\begin{gather*}
\mc{O}^{(m)}{}' = \mc{O}^{(m)}\big(\bg',\bm\ell',\elltwo{}',\bY'^{(\le m)}\big), \qquad \mc{O}_a^{(m)}{}' = \mc{O}_a^{(m)}\big(\bg',\bm\ell',\elltwo{}',\bY'^{(\le m)}\big),\\
\mc{O}_{ab}^{(m)}{}' = \mc{O}_{ab}^{(m)}\big(\bg',\bm\ell',\elltwo{}',\bY'^{(\le m)}\big),
\end{gather*} 	
where in the universal functions on the right hand sides all the quantities associated to metric data (such as $\nablacero$, $\bU$, etc.) and the expansion ($\bY$, $\bY^{(2)}$, etc.) are replaced by their primed counterparts according to (cf. \eqref{transgamma}-\eqref{transell2} and \eqref{trans}) $$\bg'=\bg,\qquad \bm\ell' = z\bm\ell,\qquad \elltwo{}' = z^2\elltwo, \qquad \bY' = z\bY + \bm\ell\otimes_s dz, \qquad \bY'^{(k)}=z^k\bY^{(k)} \quad (k\ge 2).$$
\end{rmk}


In some situations one has a privileged vector field $\eta$ on $(\mc M,g)$ whose deformation tensor $\mc{K}[\eta]\d\lie_{\eta} g$ is known. This is in general a very valuable information that one may want to incorporate into our identities. Let us assume that $\eta|_{\Phi(\mc H)}$ is null and tangent to $\Phi(\mc H)$, so there exists a function $\alpha\in\mc{F}(\mc H)$ such that $\eta|_{\Phi(\mc H)} = \alpha\nu$. Denoting by $\bar\eta$ the vector field on $\mc H$ such that $\Phi_{\star}\bar\eta = \eta|_{\Phi(\mc H)}$, it follows $\bar\eta = \alpha n$. The gauge transformation of $\alpha$ is a consequence of \eqref{transn} and reads $\alpha'=z\alpha$. Following \cite{tesismiguel}, one can introduce the gauge invariant scalar function $\kappa$ by
\begin{equation}
	\label{kappasgeneral}
	\kappa \d n(\alpha)+\alpha\kappa_n.
\end{equation}
This quantity extends everywhere on $\H$ the standard notion of surface gravity $\wt{\kappa}$ defined only at points where the vector field $\eta$ does not vanish by means of
\begin{equation}
	\label{nablaetageneral}
	\nabla_{\eta}\eta \st{\mc H}{=} \wt{\kappa}\eta.
\end{equation}
Indeed, inserting $\eta |_{\mc H} = \alpha\nu$ into \eqref{nablaetageneral} and using \eqref{connections} it follows that $\wt{\kappa} = \kappa$ on the subset of $\mc H$ where $\eta\neq 0$. We emphasize that $\kappa$ is well defined and smooth everywhere on $\mc H$.\\

Using information on the deformation tensor of $\eta$ the constraint tensor \eqref{constraint} can be rewritten so that the dependence on the tensor $\bY$ is algebraic instead of via a transport equation. The strategy put forward in \cite{tesismiguel,manzano2024embedded} was to employ the identity $\lie_{\eta}\lie_{\xi}g = \lie_{\xi}\lie_{\eta}g + \lie_{[\eta,\xi]}g$ together with
\begin{equation}
	\label{liexieta}
\lie_{\xi}\eta \st{\mc H}{=} -A_{\eta}\xi -\Phi_{\star} X_{\eta},
\end{equation}
where $$A_{\eta}\d -\mc{K}[\eta](\xi,\nu) + n(\alpha),\qquad X_{\eta}^a \d \dfrac{1}{2}\mc{K}[\eta]\big(\xi,n^a\xi-2\theta^a\big) + \dfrac{1}{2}\alpha n(\elltwo) n^a + 2\alpha P^{ab}\s_b + P^{ab}\nablacero_b\alpha,$$ to get 
\begin{equation}
\label{alfalieY}
\alpha	\lie_{n}\bY = A_{\eta}\bY -2d\alpha\otimes_s\br +\bm\ell\otimes_s dA_{\eta} + \dfrac{1}{2}\lie_{X_{\eta}}\bg + \dfrac{1}{2}\Phi^{\star}\big(\lie_{\xi}\mc{K}[\eta]\big),
\end{equation}
which inserted into \eqref{constraint} gives the so-called \textit{generalized master equation}
\begin{equation}
	\label{constraint22}
	\begin{aligned}
		\alpha \mc R_{ab}& = -\big(2\kappa+\alpha\tr_P\bU-2\mf{w}\big)\Y_{ab} + 2\alpha P^{cd}\U_{d(a}\big(2\Y_{b)c}+\F_{b)c}\big) -2\mf{I}_{ab}\\
		&\quad\, +\big(\mf{p} - \alpha(\tr_P\bY)-\alpha n(\elltwo)\big)\U_{ab}-2\alpha\nablacero_{(a}(\s-\r)_{b)} -4(\s-\r)_{(a}\nablacero_{b)}\alpha      \\
		&\quad\, -2\alpha(\s-\r)_a(\s-\r)_b - 2\nablacero_a\nablacero_b\alpha - \alpha\nablacero_{(a}\s_{b)}+\alpha\s_a\s_b  +2\nablacero_{(a}\mf{q}_{b)} +\alpha \Rcero_{(ab)},
	\end{aligned}
\end{equation}
where the tensorial quantities $\mf{w}, \mf{p}, \mf{q}$ and $\mf{I}$ codify information on the deformation of $\eta$ and are defined by 
\begin{equation}
	\label{hebrew}
	\mf{w}\d\mc{K}[\eta](\xi,\nu),\qquad \mf{p} \d \mc{K}[\eta](\xi,\xi),\qquad \mf{q} \d \Phi^{\star}\big(\mc{K}[\eta](\xi,\cdot)\big), \qquad \mf{I} \d \dfrac{1}{2}\Phi^{\star}\big(\lie_{\xi}\mc{K}[\eta]\big).
\end{equation} 
The contraction of \eqref{constraint22} with $n$ is \cite{manzano2024embedded}
\begin{equation}
	\label{constraintn2}
	\begin{aligned}
		\alpha\mc R_{ab}n^b &= \big(\mf{w}-\alpha\tr_P\bU\big)\r_a -\nablacero_a\kappa -\mf{I}_{ab}n^b + P^{cd}\nablacero_c\big(\alpha\U_{ad} \big) + \alpha(\tr_P\bU)\s_a\\
		&\quad\,- \alpha\nablacero_a\tr_P\bU + \dfrac{1}{2}\left(\nablacero_n\mf{q}_a + \nablacero_a\mf{w} - \mf{w}\s_a - P^{bc}\U_{ca}\mf{q}_b\right).
	\end{aligned}
\end{equation}
In \cite{Mio3} we extended the same idea to the higher order derivatives of the Ricci tensor. Introducing the tensors 
\begin{equation}
	\label{hebreasm}
	\hspace{-0.2cm} \mf{w}^{(m-1)}\d\mc{K}[\eta]^{(m)}(\xi,\nu),\quad \mf{p}^{(m-1)} \d \mc{K}[\eta]^{(m)}(\xi,\xi),\quad \mf{q}^{(m-1)} \d \Phi^{\star}\big(\mc{K}[\eta]^{(m)}(\xi,\cdot)\big),
\end{equation}
\begin{equation}
	\label{hebreasm2}
	\mf{I}^{(m)} \d \dfrac{1}{2}\Phi^{\star}\big(\mc{K}[\eta]^{(m+1)}\big),
\end{equation}
we proved that the tensor $\mc{R}^{(m+1)}_{ab}$, $m\ge 1$, can be written algebraically in terms of the expansion as 
\begin{equation}
	\label{lierictang3}
	\begin{aligned}
		\alpha \mc{R}^{(m+1)}_{ab} &= -\left(2(m+1)(\kappa-\mf{w})+\alpha\tr_P\bU\right)\Y^{(m+1)}_{ab}  - 2\mf{I}^{(m+1)}_{ab}\\
		&\quad\, -\alpha\big(\tr_P\bY^{(m+1)}\big)\U_{ab} + 4\alpha P^{cd}\U_{c(a}\Y^{(m+1)}_{b)d}  +4\alpha (\s-\r)_{(a} \r^{(m+1)}_{b)} \\
		&\quad\, +4\r^{(m+1)}_{(a}\nablacero_{b)}\alpha + 2\alpha\nablacero_{(a}\r^{(m+1)}_{b)}-2\alpha\kappa^{(m+1)}\Y_{ab} + \alpha\mc{O}^{(m)}_{ab} + \mc{P}^{(m)},
	\end{aligned}
\end{equation}
where recall that when $\nabla_{\xi}\xi=0$ the tensor $\mc{O}^{(m)}$ depends \textit{only} on metric hypersurface data and $\{\bY,...,\bY^{(m)}\}$. The tensor $\mc{P}^{(m)}$ depends on the same data, and in addition on the vectors $\lie_{\xi}\eta,...,\lie_{\xi}^{(m+1)}\eta$ on $\Phi(\mc H)$. As proven in \cite{Mio3}, the main property of $\{\mc{P}^{(m)}\}$ is that they vanish identically whenever $X_{\eta}=0$ (cf. \eqref{liexieta}) and $\lie_{\xi}^{(i)}\eta \st{\H}{=} 0$ for every $i\ge 2$.

%% file: Existence.tex
\section{Existence of an ambient space}
\label{sec_existence}

In this section we show that by prescribing null metric hypersurface data $\{\mc H,\bg,\bm\ell,\elltwo\}$ together with the full transverse expansion it is possible to construct a smooth semi-Riemannian manifold $(\mc M,g)$ where the data is embedded in the following sense: (i) the metric hypersurface data is $(\Phi,\xi)$-embedded and (ii) the given expansion agrees with the pullback of $\mc{K}^{(k)}$ for all $k\ge 1$. At this point it is crucial to distinguish the notion of expansion as abstract data from the concept of expansion as the pullback of $\frac{1}{2}\lie_{\xi}^{(k)}g$, $k\ge 1$. For the former we will employ the notation $\bcY^{(k)}$, and for the latter we continue using $\bY^{(k)}$. As usual, we introduce the tensors $\bcr^{(k)}\d\bcY^{(k)}(n,\cdot)$ and $\bck^{(k)}\d -\bcr^{(k)}(n)$.\\

The construction of $(\mc M,g)$ that we present does not require any field equations. If one is interested in $(\mc M,g)$ to solve field equations such as vacuum, then the prescribed transverse expansion needs to satisfy suitable restrictions. In Subsection \ref{subsection} we study the $\Lambda$-vacuum case and find necessary \textit{and sufficient} conditions on the transverse expansion that guarantees that the $(\mc M,g)$ constructed in Theorem \ref{borel} below satisfies the equations at $\Phi(\mc H)$ to infinite order. Before proving Theorem \ref{borel} we review a result due to Borel that we shall need, see \cite[Lemma 2.5]{golubitsky2012stable}.

\begin{lema}[Borel]
	\label{Borellemma}
	Let $\{F_n(x)\}_{n\ge 0}$ be a sequence of smooth functions defined on a given neighbourhood of $0$ in $\real^n$. Then there exists a smooth function $F(r,x)$ defined on a neighbourhood of $0$ in $\real\times\real^n$ such that 
	\begin{equation}
		\label{auxborel}
		\left.\dfrac{\partial^k F}{\partial r^k}\right|_{(0,x)} = F_k(x) \qquad \text{for all } k\ge 0.
	\end{equation}
	\begin{proof}
		We sketch the argument provided in \cite{golubitsky2012stable} since it will help us in the proof of Theorem \ref{borel}. Choose any smooth function $\rho:\real\to\real$ satisfying $$\rho(t) = \left\lbrace\begin{array}{c} 1 \qquad |r|\le \frac{1}{2} \\ 0 \qquad |r|\ge 1\end{array}\right.$$ and set 
		\begin{equation}
			\label{borelsum}
			F(r,x) = \sum_{k=0}^{\infty} \dfrac{r^k}{k!}\rho(\mu_k r) F_k(x),
		\end{equation} 
		where $\{\mu_k\}_{k\ge 0}$ is an increasing and unbounded sequence of real values suitably chosen to make $F(r,x)$ smooth. By construction $F(r,x)$ satisfies \eqref{auxborel}.
	\end{proof}
\end{lema}

\begin{teo}
	\label{borel}
	Let $\{\mc H,\bg,\bm\ell,\elltwo\}$ be null metric hypersurface data (Def. \ref{def_hypersurfacedata} with $\ntwo=0$) and $\{\bcY^{(k)}\}_{k\ge 1}$ a sequence of $(0,2)$ symmetric tensor fields on $\mc H$. Then there exists a semi-Riemannian manifold $(\mc M,g)$, an embedding $\Phi:\mc H\hookrightarrow\mc M$ and a rigging vector $\xi$ satisfying $a_{\xi}=\nabla_{\xi}\xi = 0$ on $\mc M$ such that (i) $\{\mc H,\bg,\bm\ell,\elltwo\}$ is null metric hypersurface data $(\Phi,\xi)$-embedded in $(\mc M,g)$ (Def. \ref{defi_embedded}) and (ii) $\{\bcY^{(k)}\}_{k\ge 1}$ is the transverse expansion of $g$ at $\Phi(\mc H)$, i.e. $\bcY^{(k)} = \bY^{(k)}\d \frac{1}{2}\Phi^{\star}\big(\lie^{(k)}_{\xi}g\big)$ for every $k\ge 1$.
	\begin{proof}
Define $\wt{\mc M}\d (-\eps,\eps)\times \mc H$ and let us construct a local coordinate system on $\wt{\mc M}$ around any point of $\{0\}\times\mc H$ as follows. Let $r$ be a coordinate in the first factor of $\wt{\mc M}$. Pick any $p\in\mc H$ and choose any local section $\mc S$ of $\mc H$ containing $p$ with (local) coordinates $\{x^A\}$. Let $u$ be a local function on $\mc H$ satisfying $u|_{\mc S}=0$ and $n(u)=1$ defined in a neighbourhood $\mc V\subset \mc H$ and extend $\{x^A\}$ to $\mc V$ by means of $n(x^A)=0$. Then $\{u,x^A\}$ are coordinates on $\mc V$ (note that $n|_{\mc V}=\partial_u$, so in these coordinates $\ell_u = \bm\ell(\partial_u) = \bm\ell(n)=1$). Finally, we extend trivially $\{u,x^A\}$ to $\wt{\mc U}\d (-\eps,\eps)\times\mc V$ so that $\{r,u,x^A\}$ are coordinates on $\wt{\mc U}$. In this coordinate system the natural embedding $\Phi_{\wt{\mc U}}:\mc V\hookrightarrow\wt{\mc U}$ is given by $\Phi_{\wt{\mc U}}(u,x^A) = (0,u,x^A)$.\\
		
The idea now is to construct smooth functions $f$, $f_A$, $h$, $h_A$ and $H_{AB}$ on a neighbourhood $\mc U\subset \wt{\mc U}$ of $p$ in terms of metric data and $\{\bcY^{(k)}\}_{k\ge 1}$ using \eqref{borelsum} so that items (i) and (ii) of the theorem follow. In order to do so we just fix the function $\rho$ and the sequence $\{\mu_k\}$ of the proof of Lemma \ref{Borellemma} and define $f$ on $\mc U$ from the sequence $\{0,-2\bck^{(i)}\}$, $f_A$ from $\{\ell_A,0,0,...\}$, $h$ from $\{\elltwo,0,0,...\}$, $h_A$ from $\{0,2\bcr^{(i)}_A\}$ and $H_{AB}$ from $\{\gamma_{AB},2\bcY^{(i)}_{AB}\}$. This guarantees that 
		\begin{equation}
			\label{order0}
			\hspace{-0.0cm} f|_{r=0} = 0, \qquad f_A|_{r=0} = \ell_A,\qquad h|_{r=0}=\elltwo,\qquad h_A|_{r=0} = 0,\qquad H_{AB}|_{r=0} = \gamma_{AB},
		\end{equation}
		as well as 
		\begin{equation}
			\label{orderk}
			\begin{gathered}
				\partial_r^{(k)}f|_{r=0} = -2\bck^{(k)},\qquad \partial_r^{(k)} f_A|_{r=0} = 0,\qquad \partial_r^{(k)}h|_{r=0} = 0,\qquad \partial_r^{(k)} h_A|_{r=0} = 2\bcr^{(k)}_A,\\
				\partial_r^{(k)} H_{AB}|_{r=0} = 2\bcY^{(k)}_{AB},
			\end{gathered}
		\end{equation}
for every $k\ge 1$. Now we define the following smooth tensor on $\mc U$
		\begin{equation}
			\label{metric0}
			g_{\mc U} = 2 dr du +h dr^2 +2f_A dr dx^A + 2 h_A du dx^A +f du^2 + H_{AB}dx^A dx^B.
		\end{equation} 
		By \eqref{order0} it follows $\Phi^{\star}_{\mc U} g_{\mc U} = \gamma_{AB}dx^A dx^B$, $\Phi_{\mc U}^{\star}\big(g_{\mc U}(\partial_r,\cdot)\big) = \bm\ell$ and $\Phi^{\star}_{\mc U}\big(g_{\mc U}(\partial_r,\partial_r)\big)=\elltwo$, so from Definition \ref{def_hypersurfacedata} the determinant of $g$ at $\Phi(\mc H)$ coincides with the determinant of the tensor $\mc A$, and hence is non-vanishing. Since this is an open condition, by shrinking $\mc U$ if necessary $g_{\mc U}$ is a smooth semi-Riemannian metric on $\mc U$. Moreover, $\{\mc V,\bg,\bm\ell,\elltwo\}$ is $(\Phi_{\mc U},\partial_r)$-embedded null metric hypersurface data in $(\mc U,g_{\mc U})$ and from \eqref{orderk} it is clear that the (embedded) transverse expansion of $g_{\mc U}$ at $\Phi(\mc V)$ agrees with $\{\bcY^{(k)}\}_{k\ge 1}$, i.e. $\bcY^{(k)} = \bY^{(k)}$ for all $k\ge 1$. \\
		
		Define $\mc M$ to be the union of all the $\mc U$'s. In order to finish the proof we need to check that two different metrics $g_{\mc U}$ and $\wh{g}_{\wh{\mc U}}$ constructed in this way agree on $\mc U\cap \wh{\mc U}$ (i.e. that $g_{\mc U\cap \wh{\mc U}}$ and $\wh g_{\mc U\cap \wh{\mc U}}$ are related by a change of coordinates), because this will define a metric on $\mc M$. So let us consider two coordinate patches $\mc V$ and $\wh{\mc V}$ on $\mc H$ with respective coordinates $\{u,x^A\}$ and $\{\wh u,\wh{x}^A\}$ such that $\mc V\cap \wh{\mc V}\neq \emptyset$. Since in the intersection $n = \partial_{u} =\partial_{\wh u}$ it follows that the most general transformation $\{u,x^A\}\mapsto \{\wh u,\wh{x}^A\}$ must be of the form $$\left\{
		\begin{aligned}
			\wh u &= u + \phi(x^A), \\ 
			\wh{x}^A &= \wh{x}^A(x^B),
		\end{aligned}
		\right.$$ 
		
		where $\phi(x^A)$ is arbitrary and $\wh{x}^A(x^B)$ must be invertible (to define a change of coordinates). Under this transformation, $\elltwo$, $\ell_A$ and $\{\bck^{(i)},\bcr^{(i)}_A,\bcY_{AB}^{(i)}\}_{i\ge 1}$ change by
		\begin{equation}
			\begin{gathered}
				\label{transTs}
				\elltwo = \wh{\ell}^{(2)},\qquad \ell_A = \dfrac{\partial\wh{x}^B}{\partial x^A} \wh{\ell}_B + \dfrac{\partial\phi}{\partial x^A},\qquad \bck^{(i)} = \wh{\bck}^{(i)},\qquad \bcr_A^{(i)} = \dfrac{\partial\wh{x}^B}{\partial x^A} \wh{\bcr}_B^{(i)} - \dfrac{\partial\phi}{\partial x^A} \wh{\bck}^{(i)},\\
				\bcY_{AB}^{(i)} = \dfrac{\partial\wh{x}^C}{\partial x^A}\dfrac{\partial\wh{x}^D}{\partial x^B}\wh{\bcY}_{CD}^{(i)} +  \dfrac{\partial\phi}{\partial x^A} \dfrac{\partial\wh{x}^C}{\partial x^B} \wh{\bcr}^{(i)}_C +  \dfrac{\partial\wh{x}^C}{\partial x^A}\dfrac{\partial\phi}{\partial x^B}\wh{\bcr}^{(i)}_C - \dfrac{\partial\phi}{\partial x^A}\dfrac{\partial\phi}{\partial x^B}\wh{\bck}^{(i)}.
			\end{gathered}
		\end{equation}
		Inserting these expressions into \eqref{borelsum} and taking into account that we used the same $\rho$ to construct all the functions, it follows
		\begin{equation}
			\begin{gathered}
				\label{transg}
				h = \wh{h},\qquad f_A =\dfrac{\partial\wh{x}^B}{\partial x^A} \wh{f}_B + \dfrac{\partial\phi}{\partial x^A},\qquad  f = \wh{f}, \qquad h_A = \dfrac{\partial\wh{x}^B}{\partial x^A} \wh{h}_A+ \dfrac{\partial\phi}{\partial x^A} f,\\
				H_{AB} = \dfrac{\partial\wh{x}^C}{\partial x^A}\dfrac{\partial\wh{x}^D}{\partial x^B}\wh{H}_{CD} +  \dfrac{\partial\phi}{\partial x^A} \dfrac{\partial\wh{x}^C}{\partial x^B} \wh{f}_C +  \dfrac{\partial\wh{x}^C}{\partial x^A}\dfrac{\partial\phi}{\partial x^B}\wh{f}_C + \dfrac{\partial\phi}{\partial x^A}\dfrac{\partial\phi}{\partial x^B}f.
			\end{gathered}
		\end{equation}
		Using these expressions, the coordinate change $\{r,u,x^A\}\mapsto \{r,\wh u,\wh{x}^A\}$ applied to the metric $g_{\mc U\cap \wh{\mc U}}$ gives precisely the metric $\wh g_{\mc U\cap \wh{\mc U}}$. So, the two tensorial objects are just different coordinate expressions of a single tensor field. This proves the existence of a metric $g$ on $\mc M$.
	\end{proof}
\end{teo}

\begin{rmk}
	\label{rmkunique}
	It is worth mentioning that Borel's theorem does not provide unique functions $f$, $f_A$, $h$, $h_A$ and $H_{AB}$, and thus the manifold $(\mc M,g)$ in the previous proposition is not unique. This is due to the fact that we can choose any $\rho$ and any sequence $\{\mu_k\}$ fulfilling the conditions required in the proof of Lemma \ref{Borellemma}.
\end{rmk}

\subsection{$\Lambda$-vacuum case}
\label{subsection}

The identities of Section \ref{section_higher} allow us to write down the necessary and sufficient conditions (i.e. the ``constraint'' equations) on the sequence $\{\bcY^{(k)}\}_{k\ge 1}$ in order for the ambient space to be $\Lambda$-vacuum to infinite order on $\mc H$. A similar analysis could be done for other field equations.
\begin{teo}
	\label{existencia}
	Let $\{\mc H,\bg,\bm\ell,\elltwo\}$ be null metric hypersurface data (Def. \ref{def_hypersurfacedata} with $\ntwo=0$) and $\{\bcY^{(k)}\}_{k\ge 1}$ a sequence of $(0,2)$ symmetric tensors satisfying
	\begin{equation}
		\label{xconstraint}
		\begin{aligned}
			\lambda\gamma_{ab}& = \accentset{\circ}{R}_{(ab)} -2\lie_n \bcY_{ab} - (2\bck_n+\tr_P\bU)\bcY_{ab} + \nablacero_{(a}\left(\s_{b)}+2\bcr_{b)}\right)\\
			&\quad -2\bcr_a\bcr_b + 4\bcr_{(a}\s_{b)} - \s_a\s_b - (\tr_P\bcY)\U_{ab} + 2P^{cd}\U_{d(a}\left(2\bcY_{b)c}+\F_{b)c}\right),
		\end{aligned}
	\end{equation}
	\begin{multicols}{2}
		\noindent
		\begin{equation}
			\label{xxixi}
			\lambda\elltwo \delta^m_1 = - \tr_P\bcY^{(m+1)} + \mc{O}^{(m)}(\bcY^{\le m}),
		\end{equation}
		\begin{equation}
			\label{xxix}
			\lambda \ell_a \delta_1^m = \bcr_a^{(m+1)} + \mc{O}^{(m)}(\bcY^{\le m})_a,
		\end{equation}
	\end{multicols}
	\vspace{-0.4cm}
	\begin{equation}
		\label{xtt}
		\begin{aligned}
			\hskip -0.5cm 2\lambda\bcY^{(m)}_{ab} &= -2\lie_n\bcY^{(m+1)}_{ab} - \left(2(m+1)\bck_n + \tr_P\bU\right)\bcY^{(m+1)}_{ab}- (\tr_P\bcY^{(m+1)})\U_{ab}\\
			&\quad\,  + 4P^{cd}\U_{c(a}\bcY^{(m+1)}_{b)d}  +4(\s-\bcr)_{(a} \bcr^{(m+1)}_{b)}+ 2\nablacero_{(a}\bcr^{(m+1)}_{b)}\\
			&\quad\, -2\bck^{(m+1)}\bcY_{ab} + \mc{O}^{(m)}_{ab}(\bcY^{\le m}),
		\end{aligned}
	\end{equation}
	for every $m\ge 1$, where the tensors $\mc{O}^{(m)}$, $\mc{O}^{(m)}_a$ and $\mc{O}^{(m)}_{ab}$ are the same as in Proposition \ref{prop_derivadas} with the $\bY^{(k)}$'s replaced by the $\bcY^{(k)}$'s, $\bcr^{(k)}$ and $\kappa^{(k)}$ defined from $\bcY^{(k)}$ by $\bcr^{(k)}\d \bcY^{(k)}(n,\cdot)$ and $\bck^{(k)}\d-\bcr^{(k)}(n)$, and where $n$, $P$ and $\bU$ are defined in \eqref{gamman}-\eqref{Pgamma} and \eqref{defK}, respectively. Then, the semi-Riemannian manifold $(\mc M,g)$ constructed from $\{\bcY^{(k)}\}$ as in Theorem \ref{borel} solves the $\Lambda$-vacuum equations to infinite order on $\Phi(\mc H)$, i.e. $R_{\mu\nu}^{(i+1)}\d \lie_{\xi}^{(i)}R_{\mu\nu} \st{\H}{=} \lambda\mc{K}_{\mu\nu}^{(i)}=\lambda \lie_{\xi}^{(i)}g_{\mu\nu}$ for every $i\ge 0$.
	\begin{proof}
		By equations \eqref{constraint}, \eqref{ddotR} and \eqref{dotR} as well as \eqref{xconstraint}, \eqref{xxixi} and \eqref{xxix} for $m=1$ together with the fact that the hypersurface data is $(\Phi,\xi)$-embedded (see Def. \ref{defi_embedded}) it follows that $R_{\mu\nu} \st{\mc H}{=} \lambda g_{\mu\nu}$. Moreover, by Proposition \ref{liegxi}, $\mc{K}^{(i-1)}_{\alpha\beta}\xi^{\alpha} = 0$ for every $i\ge 2$. Now, \eqref{xxixi} and \eqref{xxix} combined with \eqref{ddotR} and \eqref{dotR} gives $R^{(i)}_{\alpha\beta}\xi^{\alpha} = 0$ for $i\ge 2$ after taking into account $\bcY^{(k)} = \bY^{(k)}$ for every $k\ge 1$. Thus, $R^{(i)}_{\alpha\beta}\xi^{\alpha} = \lambda \mc{K}^{(i-1)}_{\alpha\beta}\xi^{\alpha}$ for every $i\ge 1$. Finally, combining equations \eqref{xtt} and \eqref{R} and using again $\bcY^{(k)} = \bY^{(k)}$ for $k\ge 1$ gives $\mc{R}^{(i)}_{ab} = 2\lambda\Y^{(i-1)}_{ab} = \lambda \mc{K}_{ab}^{(i-1)}$ for every $i\ge 1$, which finishes the proof. 
	\end{proof}
\end{teo}
It is worth emphasizing that in Theorem \ref{existencia} the topology of the null hypersurface does not need to be a product. This is particularly relevant because there are examples of null hypersurfaces whose topology is not a product, such as certain compact Cauchy horizons (see e.g. \cite{friedrich1999rigidity}). In these cases the constraint equations of Theorem \ref{existencia} may or mat not have solutions. However, when $\mc H$ admits a cross-section (and hence its topology is $\real\times \mc S$), these constraint equations can always be integrated given initial conditions at a section. Note however that due to \eqref{elierictangn}-\eqref{ePcontractioneq} there are more constraints than independent components of $\{\bcY^{(m)}\}$ to integrate. In the following proposition we show that some of the constraints are redundant.

\begin{prop}
	\label{teo_bianchi}
	Let $(\mc M,g)$ be an $\mf n$-dimensional semi-Riemannian manifold and $\Phi:\mc H\hookrightarrow\mc M$ an embedded null hypersurface with rigging $\xi$ extended off $\Phi(\mc H)$ arbitrarily. Fix a natural number $m\ge 1$ and assume the Ricci tensor of $g$ satisfies ${R}^{(i)}_{\mu\nu} \st{\mc H}{=} \lambda \mc{K}^{(i-1)}_{\mu\nu}\st{\mc H}{=} \lambda\lie^{(i-1)}_{\xi}g_{\mu\nu}$ for every $i=1,...,m$. Then,
	\begin{equation}
		\label{equationsbianchi}
		\mc{R}^{(m+1)}_{ab}n^a \st{\mc H}{=} \lambda \mc{K}^{(m)}_{ab} n^a,\qquad P^{ab}\mc{R}^{(m+1)}_{ab} \st{\mc H}{=} \lambda P^{ab}\mc{K}^{(m)}_{ab},
	\end{equation}
where $\mc{R}^{(m+1)}_{ab}$ is defined in \eqref{def_asymptoticricci}.
	\begin{proof}
		The idea is to prove the following identity 
		\begin{equation}
			\label{bianchi0}
			\nu^{\alpha}R^{(m+1)}_{\alpha\beta} - \dfrac{1}{2}g^{\mu\alpha}R^{(m+1)}_{\mu\alpha} \nu_{\beta} \st{\mc H}{=} \lambda\left(\mc{K}^{(m)}_{\alpha\beta}\nu^{\alpha} - \dfrac{1}{2}g^{\mu\alpha}\mc{K}^{(m)}_{\mu\alpha}\nu_{\beta}\right),
		\end{equation}		
		since then its contraction with a tangent vector $\wh{e}^{\beta}_b$ is (recall $\nu_{\beta}$ is normal to $\mc H$) $$\wh{e}^{\beta}_b\nu^{\alpha}R^{(m+1)}_{\alpha\beta} \st{\mc H}{=} \lambda \mc{K}^{(m)}_{\alpha\beta}\nu^{\alpha}\wh{e}^{\beta}_b,$$
		
		 which proves the first equation in \eqref{equationsbianchi}, and its contraction with $\xi^{\beta}$ gives, after inserting \eqref{inversemetric}, $$\xi^{\beta}\nu^{\alpha}R^{(m+1)}_{\alpha\beta} - \xi^{\mu}\nu^{\alpha}R^{(m+1)}_{\mu\alpha} - \dfrac{1}{2} P^{ab}\wh{e}^{\mu}_a\wh e^{\alpha}_b R^{(m+1)}_{\mu\alpha} \st{\mc H}{=} \lambda\left( \xi^{\beta} \nu^{\alpha}\mc{K}^{(m)}_{\alpha\beta} -  \xi^{\mu} \nu^{\alpha}\mc{K}^{(m)}_{\mu\alpha} - \dfrac{1}{2} P^{ab}\wh{e}^{\mu}_a\wh e^{\alpha}_b\mc{K}^{(m)}_{\mu\alpha}\right),$$ 
		 
		 so $P^{ab}\wh{e}^{\mu}_a\wh e^{\alpha}_b R^{(m+1)}_{\mu\alpha} \st{\mc H}{=} \lambda P^{ab}\wh{e}^{\mu}_a\wh e^{\alpha}_b\mc{K}^{(m)}_{\mu\alpha}$ and hence the second equation in \eqref{equationsbianchi} is also established. So let us prove \eqref{bianchi0}. By hypothesis $\lie^{(i-1)}_{\xi}{R}_{\mu\nu} \st{\mc H}{=}  \lambda\lie_{\xi}^{(i-1)}g_{\mu\nu}$ for every $i=1,...,m$, so for any $j\in\{0,...,m-1\}$ it follows (cf. \eqref{derivada}) $$\lie^{(j)}_{\xi}(R^{\alpha}{}_{\beta}) = \lie^{(j)}_{\xi}(R_{\mu\beta}g^{\mu\alpha}) = \sum_{k=0}^{j}\binom{j}{k} \lie_{\xi}^{(k)}R_{\mu\beta} \lie_{\xi}^{(j-k)}g^{\mu\alpha} \st{\mc H}{=} \lambda \sum_{k=0}^{j}\binom{j}{k} \lie_{\xi}^{(k)}g_{\mu\beta} \lie_{\xi}^{(j-k)}g^{\mu\alpha} = \lambda \lie_{\xi}^{(j)}\delta_{\beta}^{\alpha},$$ 
		 
		 and thus
		\begin{equation}
			\label{derivadasR}
			\lie^{(j)}_{\xi}(R^{\alpha}{}_{\beta}) \st{\mc H}{=} \left\{\begin{aligned}
				&\lambda \delta_{\beta}^{\alpha} &  &j=0,\\
				&0 & &1\le j \le m-1,
			\end{aligned}\right. \qquad \lie_{\xi}^{(j)}R \st{\mc H}{=} \left\{\begin{aligned}
				&\lambda \mf n & &j=0,\\
				&0 & &1\le j \le m-1.
			\end{aligned}\right.
		\end{equation}
		Therefore the derivatives of the Einstein tensor $G_{\mu\nu} = R_{\mu\nu} - \frac{1}{2}R g_{\mu\nu}$ satisfy 
		\begin{equation}
			\label{lieG}
			\lie_{\xi}^{(j)} (G^{\alpha}{}_{\beta}) \st{\mc H}{=} \left\{\begin{aligned}
				&\dfrac{2-\mf n}{2}\lambda \delta_{\beta}^{\alpha} &  &j=0,\\
				&0 & &1\le j \le m-1.
			\end{aligned}\right.
		\end{equation}
		Applying $\lie_{\xi}^{(m-1)}$ to the contracted Bianchi identity $\nabla_{\alpha}G^{\alpha}{}_{\beta} = 0$ and using Proposition \ref{propMarc} gives
		\begin{align*}
			0 & = \lie_{\xi}^{(m-1)} \nabla_{\alpha}G^{\alpha}{}_{\beta}\\
			& = \nabla_{\alpha}  (\lie_{\xi}^{(m-1)}G^{\alpha}{}_{\beta}) + \sum_{k=0}^{m-2}(\lie_{\xi}^{(m-2-k)}G^{\sigma}{}_{\beta})\Sigma^{(k+1)}{}^{\alpha}{}_{\alpha\sigma} - \sum_{k=0}^{m-2}(\lie_{\xi}^{(m-2-k)}G^{\alpha}{}_{\sigma})\Sigma^{(k+1)}{}^{\sigma}{}_{\alpha\beta}\\
			&\st{\mc H}{=} \nabla_{\alpha}  (\lie_{\xi}^{(m-1)}G^{\alpha}{}_{\beta}) + \dfrac{2-\mf n}{2}\lambda \delta_{\beta}^{\sigma}\Sigma^{(m-1)}{}^{\alpha}{}_{\alpha\sigma} - \dfrac{2-\mf n}{2}\lambda \delta_{\sigma}^{\alpha}\Sigma^{(m-1)}{}^{\sigma}{}_{\alpha\beta}\\
			&\st{\mc H}{=}\nabla_{\alpha}  (\lie_{\xi}^{(m-1)}G^{\alpha}{}_{\beta}),
		\end{align*}
		where in the third equality we used \eqref{lieG}. We now compute the trace in this equation by inserting $\delta^{\alpha}_{\mu} = \wh{e}^{\alpha}_a\theta^a_{\mu} + \xi^{\alpha}\nu_{\mu}$ and get $$\nabla_{\alpha}  (\lie_{\xi}^{(m-1)}G^{\alpha}{}_{\beta})\st{\H}{=} \left(\wh{e}^{\alpha}_a\theta^a_{\mu} + \xi^{\alpha}\nu_{\mu}\right)\nabla_{\alpha}(\lie^{(m-1)}_{\xi}G^{\mu}{}_{\beta}) \st{\mc H}{=} 0.$$  
		
		The first term vanishes because we can take tangential derivatives of \eqref{lieG}, and therefore $\xi^{\alpha}\nu_{\mu}\nabla_{\alpha}(\lie^{(m-1)}_{\xi}G^{\mu}{}_{\beta}) \st{\mc H}{=} 0$. Using again \eqref{lieG} together with the well-known identity $$\nabla_{\xi}T_{\lambda\gamma} = \lie_{\xi}T_{\lambda\gamma} - 2 T_{\mu(\lambda}\nabla_{\gamma)}\xi^{\mu},$$ 
		
		valid for every $(0,2)$ symmetric tensor $T$, gives 
		\begin{equation}
			\label{lieG2}
			\nu_{\mu}(\lie^{(m)}_{\xi}G^{\mu}{}_{\beta} )\st{\mc H}{=} 0.
		\end{equation}
		Moreover, equations \eqref{derivada} and \eqref{lieG} yield $$ g_{\mu\alpha}(\lie^{(m)}_{\xi}G^{\mu}{}_{\beta}) = \lie_{\xi}^{(m)}G_{\alpha\beta} - \sum_{j=0}^{m-1}\binom{m}{j} (\lie_{\xi}^{(j)}G^{\mu}{}_{\beta} )\mc{K}^{(m-j)}_{\mu\alpha} \st{\mc H}{=} \lie_{\xi}^{(m)}G_{\alpha\beta} - \dfrac{2-\mf n}{2}\lambda\mc{K}^{(m)}_{\alpha\beta}.$$ 
		
		Contracting with $\nu^{\alpha}$ and using \eqref{lieG2} gives $\nu^{\alpha}\lie_{\xi}^{(m)}G_{\alpha\beta} \st{\mc H}{=} \dfrac{2-\mf n}{2}\lambda\nu^{\alpha}\mc{K}^{(m)}_{\alpha\beta}$. Then, inserting $G_{\alpha\beta} = R_{\alpha\beta} - \dfrac{1}{2}R g_{\alpha\beta}$ and using the second expression in \eqref{derivadasR} we conclude
		\begin{equation}
			\label{bianchicasi}
			\nu^{\alpha} \lie_{\xi}^{(m)}R_{\alpha\beta} - \dfrac{1}{2}g_{\alpha\beta}\nu^{\alpha}\lie^{(m)}_{\xi} R - \dfrac{\mf n}{2}\lambda \mc{K}^{(m)}_{\alpha\beta}\nu^{\alpha} \st{\mc H}{=} \dfrac{2-\mf n}{2}\lambda\nu^{\alpha}\mc{K}^{(m)}_{\alpha\beta}.
		\end{equation} 
		The last step is to relate $\lie_{\xi}^{(m)}R$ to the term of $\lie_{\xi}^{(m)}R_{\alpha\beta}$. In order to do so we use again \eqref{derivada} and \eqref{derivadasR} to write $$g^{\mu\alpha}\lie^{(m)}_{\xi}R_{\mu\alpha} = g^{\mu\alpha}\lie^{(m)}_{\xi}\big(R^{\nu}{}_{\alpha} g_{\nu\mu}\big) = g^{\mu\alpha}g_{\nu\mu}\lie_{\xi}^{(m)}(R^{\nu}{}_{\alpha}) + \lambda g^{\mu\alpha}\delta_{\alpha}^{\nu}\mc{K}^{(m)}_{\nu\mu} = \lie_{\xi}^{(m)}R + \lambda g^{\mu\nu}\mc{K}^{(m)}_{\mu\nu}.$$ 
		
		It is immediate to combine this and \eqref{bianchicasi} to get \eqref{bianchi0}, which proves the proposition.		
	\end{proof}
\end{prop}

\begin{rmk}
	Some of the (``constraint'') equations that $\{\bcY^{(k)}\}$ need to satisfy in Theorem \ref{existencia} are redundant. By Proposition \ref{teo_bianchi} the components $n^b\mc{R}^{(i)}_{ab} = 2\lambda n^b\Y^{(i-1)}_{ab}$ and $P^{ab}\mc{R}^{(i)}_{ab} = 2\lambda P^{ab}\Y^{(i-1)}_{ab}$ of the equation $\mc{R}^{(i)}_{ab} = 2\lambda \Y^{(i-1)}_{ab}$ for $i\ge 2$ are automatically satisfied. Hence, one only needs to impose as constraint equations \eqref{xxixi}, \eqref{xxix} and the part of \eqref{xtt} that lies in the kernel of the so-called energy-momentum map introduced in \cite{MarcAbstract}. We will however not make use of this last fact in the following.
\end{rmk}

Now we have all the ingredients to restate Theorem \ref{existencia} when $\H$ admits cross sections.

\begin{teo}
	\label{teoexistencia}
	Let $\{\mc H,\bg,\bm\ell,\elltwo\}$ be null metric hypersurface data (Def. \ref{def_hypersurfacedata} with $\ntwo=0$) admitting a section $\iota:\mc S\hookrightarrow\mc H$ with metric $h\d\iota^{\star}\bg$ and assume there exists a smooth function $f$ on $\mc H$ satisfying 
	\begin{equation}
		\label{equationf}
		-n(\tr_P\bU) + (\tr_P\bU)f - P^{ab}P^{cd}\U_{ac}\U_{bd}=0,
	\end{equation}
where $n$, $P$ and $\bU$ are defined in \eqref{gamman}-\eqref{Pgamma} and \eqref{defK}, respectively. Let $\chi$ and $\bm\beta$ be a function and a one-form on $\mc S$, respectively, and $\{\S^{(k)}_{AB}\}_{k\ge 1}$ a sequence of symmetric tensors on $\mc S$ traceless w.r.t. $h$. Then, there exists a semi-Riemannian manifold $(\mc M,g)$, an embedding $\Phi:\mc H\hookrightarrow\mc M$ and a rigging $\xi$ such that (i) $\{\mc H,\bg,\bm\ell,\elltwo\}$ is null metric hypersurface data $(\Phi,\xi)$-embedded in $(\mc M,g)$ (Def. \ref{defi_embedded}), (ii) $(\mc M,g)$ solves the $\Lambda$-vacuum equations to infinite order on $\Phi(\mc H)$ and (iii) its asymptotic expansion $\{\bY^{(k)}\}_{k\ge 1}$ (see \eqref{def_asymptoticexpansion}) satisfies $\tr_P\bY|_{\mc S} = \chi$, $\iota^{\star}\br = \bm\beta$, $\kappa_n = f$ and $\mbox{tf }\big(\iota^{\star}\bY^{(k)}\big) = \bS^{(k)}$ for every $k\ge 1$, where $\mbox{tf}$ stands for the trace-free part of a tensor w.r.t. $h$.
	\begin{proof}
		The strategy is to construct explicitly the expansion $\{\bcY^{(k)}\}_{k\ge 1}$ in order to satisfy the hypothesis of Theorem \ref{existencia} and such that the initial conditions on $\mc S$ satisfy (iii). Let us start by constructing the tensor $\bcY$. Define $\bm\ell_{\para}\d\iota^{\star}\bm\ell$, $\ell^{\sharp}\d h^{\sharp}(\bm\ell_{\para},\cdot)$ and $\elltwo_{\sharp}\d h^{AB}\ell_A\ell_B$. The tensor $P$ can be decomposed as \cite{Mio1} 
		\begin{equation}
			\label{Pdecomp}
			P = h^{AB} e_A\otimes e_B - 2\ell^A n\otimes_s e_A - \big(\elltwo-\elltwo_{\sharp}\big)n\otimes n,
		\end{equation}
		where $\{e_A\}$ is a (local) basis on $\mc S$, and hence given a $(0,2)$ tensor $\T_{ab}$ on $\mc{H}$,
		\begin{equation}
			\label{trazas}
			\tr_P\bT = h^{AB}\T_{AB} - \bT(n,\iota_{\star}\ell^{\sharp}) - \bT(\iota_{\star}\ell^{\sharp},n) - \big(\elltwo-\elltwo_{\sharp}\big)\bT(n,n).
		\end{equation}
		We start with the first order. The idea is to integrate equation \eqref{xconstraint} to build the tensor $\bcY$. The difficulty is that this equation is a PDE due to the presence of $\nablacero$-derivatives of $\bcr$, which recall is the contraction of $\bcY$ with $n$. However, the equation has an interesting decoupling property that we shall exploit. Before integrating \eqref{xconstraint} we shall integrate the equations obtained by contracting it with $n^a$ and $P^{ab}$. A priori the quantities we obtain in this way are not related to $\bcY$, so we give them completely different names. A posteriori we shall prove that this intermediate quantities are in fact the correct contractions.\\
		
		Let us start with the contraction of \eqref{xconstraint} with $n$. That is, we solve the following equation for a one-form $\bm c$
		\begin{equation}
			\label{xconstraintn}
			\begin{aligned}
				0 = -\lie_n(c_b-\s_b) - \nablacero_b f - (\tr_P\bU) (c_b-\s_b)- \nablacero_b (\tr_P\bU) + P^{cd}\nablacero_c\U_{bd} - 2P^{cd}\U_{bd}s_c,
			\end{aligned}
		\end{equation}
		with initial conditions $\iota^{\star}\bm c = \bm\beta$ and $\mf{c}\d -\bm c(n) = f$ on $\mc S$. Let us first prove that in fact $\mf{c}=f$ everywhere. To that aim, the contraction of \eqref{xconstraintn} with $n^b$ gives $$n(\mf{c})-n(f) +(\tr_P\bU)\mf{c}-n(\tr_P\bU)  - P^{ab}P^{cd}\U_{ac}\U_{bd} = 0,$$ and subtracting \eqref{equationf} it follows $$n(\mf{c} - f) +(\tr_P\bU)(\mf{c}-f) = 0.$$ 
		
Since $\mf{c}=f$ initially, then $\mf{c} = f$ everywhere on $\mc H$. The second quantity we shall need is the would-be trace of $\bcY$ w.r.t. $P$. We use $t$ for its replacement and solve the equation that comes from taking the trace of \eqref{xconstraint} with $P^{ab}$
		\begin{equation}
			\label{xtrPR}
			\begin{aligned}
				\lambda (d-1) &= \tr_P \Rcero -2n(t)  - 2\big(f+\tr_P\bU\big)t+ \div_P(\bs + 2\bm c) -2P(\bm c,\bm c)\\
				&\quad -4P(\bm c,\bs)-P(\bs,\bs)+2f n(\elltwo),
			\end{aligned}
		\end{equation}
		where $d=\mf n-1$ is the dimension of $\mc H$, with initial condition $t = \chi$ on $\mc S$. Having replaced $\bcr$ for $\bm c$ and
		$\tr_P \bcY$ for $t$, we may integrate the following equation for $\bcY$, obtained directly from \eqref{xconstraint} with the replacements above,
		\begin{equation}
			\label{xconstraint2}
			\begin{aligned}
				\lambda\gamma_{ab}& = \accentset{\circ}{R}_{(ab)} -2\lie_n \bcY_{ab} - (2f+\tr_P\bU)\bcY_{ab} + \nablacero_{(a}\left(\s_{b)}+2c_{b)}\right)-2c_a c_b \\
				&\quad + 4c_{(a}\s_{b)} - \s_a\s_b - t\U_{ab} + 2P^{cd}\U_{d(a}\left(2\bcY_{b)c}+\F_{b)c}\right).
			\end{aligned}
		\end{equation}
		This equation is no longer a PDE, but an ODE along each generator, so it admits a unique solution $\bcY$ with initial conditions $\iota^{\star}\big(\bcY(n,\cdot)\big) = \iota^{\star}\big(\bcY(\cdot,n)\big)=\bm\beta$, $\bcY(n,n) = -f$ on $\mc S$ and (see \eqref{trazas}) 
\begin{equation}
	\label{YonS}
	\iota^{\star}\bcY_{AB} = \S^{(1)}_{AB} + \dfrac{1}{d-1}\left(t+2\bm c(\iota_{\star}\ell^{\sharp})-(\elltwo-\elltwo_{\sharp})f\right)h_{AB}.
\end{equation}		
Having obtained a unique $\bcY_{ab}$ on $\mc H$ let us now prove that $t=\tr_P\bcY$ and $\bm c =\bcr$ on $\mc H$. Contracting \eqref{xconstraint2} with $n^a$ and $P^{ab}$ respectively, using \eqref{nnablacerotheta}, \eqref{Rceron}, \eqref{lienP} and $\mf c=f$ it follows that $\bcr$ and $\tr_P\bcY$ satisfy the equations
		\begin{align}
			\hspace{-0.5cm} 0 &= \lie_n(\s_b+c_b-2\bcr_b) - \nablacero_b f - (\tr_P\bU) (\bcr_b-\s_b)- \nablacero_b (\tr_P\bU) + P^{cd}\nablacero_c\U_{bd}\nonumber\\
			&\quad\,  - 2P^{cd}\U_{bc}(\s_d+c_d-\bcr_d) + 2f(c_b-\bcr_b),\label{r1}\\
			\hspace{-0.5cm}  \lambda (d-1) &= \tr_P \Rcero -2\lie_n(\tr_P\bcY)  - \big(2f+\tr_P\bU\big)\tr_P\bcY+ \div_P(\bs + 2\bm c) -2P(\bm c,\bm c)\nonumber\\
			&\quad +4P(\bs,\bm c)-8P(\bs,\bcr)-P(\bs,\bs)-t\tr_P\bU+2\bck_n n(\elltwo).\label{tr}
		\end{align}
		Subtracting \eqref{r1} from \eqref{xconstraintn} gives $$2\lie_n (c_b-\bcr) + (\tr_P\bU)(c_b-\bcr_b) - 2P^{cd}\U_{bc}(c_d-\bcr_d) + 2f(c_b-\bcr_b) = 0,$$ 
		
		which is a homogeneous transport equation for the one-form $\bm c-\bcr$, and since the initial condition for $\bm c$ and $\bcr$ are the same, we conclude that $\bm c = \bcr$ on all $\mc H$. Subtracting \eqref{tr} from \eqref{xtrPR} and using $\bcr=\bm c$ and $\bck_n=f$ yields $$2n\big(t-\tr_P\bcY\big) + 2f\big(t-\tr_P\bcY\big) + (\tr_P\bU) \big(t-\tr_P\bcY\big) =0,$$ 
		
		so a similar argument proves $\tr_P\bcY = t$ on $\mc H$. Hence we have constructed a tensor field $\bcY_{ab}$ that satisfies equation \eqref{xconstraint}.\\
		
		For the higher orders we follow a similar strategy, but with some important differences, particularly on how $\bm{c}^{(m)}$ and $t^{(m)}$ are constructed. Let us start by discussing the second order. Now, we define a scalar $t^{(2)}$ and a one-form $\bm{c}^{(2)}$ by means of (cf. \eqref{xxixi}-\eqref{xxix})
		\begin{multicols}{2}
			\noindent
			\begin{equation}
				\label{xxixi2}
				\lambda\elltwo = - t^{(2)} + \mc{O}^{(1)}(\bcY),
			\end{equation}
			\begin{equation}
				\label{xxix2}
				\lambda \ell_a  = c_a^{(2)} + \mc{O}^{(1)}(\bcY)_a,
			\end{equation}
		\end{multicols}
		and the tensor field $\bcY^{(2)}$ by integrating (cf. \eqref{xtt})
		\begin{equation}
			\label{xtt2}
			\begin{aligned}
				\hskip -0.5cm 2\lambda\bcY_{ab} &= -2\lie_n\bcY^{(2)}_{ab} - \left(4\bck_n + \tr_P\bU\right)\bcY^{(2)}_{ab} - t^{(2)}\U_{ab} + 4P^{cd}\U_{c(a}\bcY^{(2)}_{b)d}  \\
				&\quad\, +4(\s-\bcr)_{(a} c^{(2)}_{b)}+ 2\nablacero_{(a}c^{(2)}_{b)}-2c^{(2)}\bcY_{ab} + \mc{O}^{(1)}_{ab}(\bcY),
			\end{aligned}
		\end{equation}
		where $\mf{c}^{(2)}\d -\bm{c}^{(2)}(n)$, with initial conditions $\iota^{\star}\big(\bcY^{(2)}(n,\cdot)\big) = \iota^{\star}\big(\bcY^{(2)}(\cdot,n)\big) = \iota^{\star}\big(\bm c^{(2)}\big)$, $\bcY^{(2)}(n,n)=-\mf{c}^{(2)}$ on $\mc S$ and 
		\begin{equation}
			\label{incon}
			\iota^{\star}\bcY^{(2)}_{AB} = \S^{(2)}_{AB} + \dfrac{1}{d-1}\big(t^{(2)}+2\bm c^{(2)}(\iota_{\star}\ell^{\sharp}) -\big(\elltwo-\elltwo_{\sharp}\big)\mf{c}^{(2)} \big) h_{AB}.
		\end{equation} 
		The equations that $\bcr^{(2)}\d\bcY^{(2)}(n,\cdot)$ and $\tr_P\bcY^{(2)}$ satisfy are obtained, respectively, after contracting \eqref{xtt2} with $n^b$ and $P^{ab}$ and using \eqref{Rceron}, \eqref{nnablacerotheta}, \eqref{derivadannull} and \eqref{lienP},
		\begin{align}
			2\lambda\bcr_a &= -2\lie_n\bcr^{(2)}_a - \big(4\bck_n + \tr_P\bU\big)\bcr^{(2)}_a + 2P^{cd}\U_{ca}\bcr^{(2)}_d + 2\bck_n c_a^{(2)} + \lie_n c_a^{(2)} -\nablacero_a \mf{c}^{(2)} \nonumber\\
			&\quad\, -2P^{cd}\U_{ca}c^{(2)}_d+\mc O^{(1)}_{ab}(\bcY) n^b  ,\label{cn}\\
			2\lambda\tr_P\bcY & = -2\lie_n\big(\tr_P\bcY^{(2)}\big) - 8P(\bs,\bcr^{(2)}) +2\bck^{(2)} n(\elltwo) - \left(4\bck_n + \tr_P\bU\right)\tr_P\bcY^{(2)}\nonumber\\
			&\quad\,  - t^{(2)}\tr_P\bU  + 4P(\bs-\bcr,\bm{c}^{(2)}) + 2\div_P\bm c^{(2)}- 2\mf{c}^{(2)}\tr_P\bcY + \tr_P\mc{O}^{(1)}(\bcY).\label{trc} 
		\end{align}
		Another contraction of \eqref{cn} with $n^a$ yields 
		\begin{equation}
			\label{cnn}
			-2\lambda\bck_n  = 2n\big(\bck^{(2)}-\mf{c}^{(2)}\big)+ \big(4\bck_n + \tr_P\bU\big)\bck^{(2)} - 2\bck_n \mf{c}^{(2)} +\mc O^{(1)}_{ab}(\bcY) n^a n^b.
		\end{equation}
		Let us now check that $c^{(2)}_a =  \bcr^{(2)}_a$ and $t^{(2)} = {\tr_P\bcY}^{(2)}$ everywhere. Define the tensor field $$\T^{(2)}_{ab} \d \dfrac{t^{(2)}-\mf{c}^{(2)}\elltwo}{d-1}\gamma_{ab} + 2c^{(2)}_{(a}\ell_{b)} +\mf{c}^{(2)}\ell_a\ell_b.$$ 
		
		It is a matter of simple computation to check that
		\begin{equation}
			\label{equationsT2}
			P^{ab}\T^{(2)}_{ab} = t^{(2)},\qquad \T^{(2)}_{ab}n^a = \T^{(2)}_{ba}n^a = c^{(2)}_b.
		\end{equation}
		We now construct an auxiliary semi-Riemannian manifold $(\wt{\mc M},\wt g)$ using Theorem \ref{borel} with the sequence $\{\bcY,\bT^{(2)},0,0,...\}$ and denote its transverse expansion by $\wt{\bY}{}^{(k)} \d \frac{1}{2}\wt{\Phi}^{\star}\big(\lie_{\xi}^{(k)}\wt g\big)$. We shall use this space to relate the contractions $\tr_P\bcY^{(2)}$ and $\bcr^{(2)}$ with $t^{(2)}$ and $\bm c^{(2)}$. It is important to note that these relations hold true independently of the auxiliary space we construct.\\
		
		Since $\wt\bY=\bcY$ and $\wt\bY{}^{(2)}=\bT^{(2)}$, it is clear from Def. \ref{defi_embedded} and equations \eqref{constraint}, \eqref{ddotR}, \eqref{dotR} for $m=1$ as well as \eqref{xconstraint}, \eqref{xxixi2}, \eqref{xxix2} that the Ricci tensor of $\wt g$ satisfies $\wt R_{\mu\nu} = \lambda \wt g_{\mu\nu}$ on $\Phi(\mc H)$. Hence, by Proposition \ref{teo_bianchi} it then follows $n^b \wt{\mc{R}}^{(2)}_{ab} = \lambda n^b {\mc{K}}_{ab}$ as well as $P^{ab}\wt{\mc{R}}^{(2)}_{ab} = \lambda P^{ab}{\mc{K}}_{ab}$. Using the fact that $\mc K_{ab} = \Phi^{\star} (\lie_{\xi} \wt g )_{ab} = 2\wt{\Y}_{ab} = 2 \bcY_{ab}$ together with \eqref{equationsT2}, equations \eqref{elierictangn}-\eqref{ePcontractioneq} read
		\begin{align}
			2\lambda\bcr_a &= -\lie_n c^{(2)}_a - \big(2\bck_n + \tr_P\bU\big)c^{(2)}_a-\nablacero_a \mf{c}^{(2)} +\mc O^{(1)}_{ab}(\bcY) n^b,\label{equationr2fake}\\
			2\lambda\tr_P\bcY & = -2n(t^{(2)}) - 2\left(2\bck_n + \tr_P\bU\right)t^{(2)}+ 2\mf{c}^{(2)}\big(n(\elltwo)-\tr_P\bcY\big) \nonumber\\
			&\quad\, -4P\big(\bcr+\bs,\bm c^{(2)}\big) + 2\div_P\bm c^{(2)} + \tr_P\mc{O}^{(1)}(\bcY).\label{equationtrfake}
		\end{align}
		The contraction of \eqref{equationr2fake} with $n^a$ gives $$-2\lambda\bck_n  = \big(2\bck_n + \tr_P\bU\big)\mf{c}^{(2)} +\mc O^{(1)}_{ab}(\bcY) n^a n^b.$$ 
		
		Subtracting it from \eqref{cnn} yields 
		\begin{equation}
			\label{transportkappa2}
			2n\big(\bck^{(2)}-\mf c^{(2)}\big)+ \big(4\bck_n + \tr_P\bU\big)\big(\bck^{(2)}-\mf{c}^{(2)}\big) = 0.
		\end{equation}
		Since $\bck^{(2)}$ initially agrees with $\mf c^{(2)}$ it follows that $\bck^{(2)}=\mf c^{(2)}$ everywhere on $\mc H$. Next, subtracting \eqref{equationr2fake} from \eqref{cn}
		\begin{equation}
			\label{transport}
			2\lie_n \big(\bcr^{(2)}_a -c^{(2)}_a\big) + \big(4\bck_n + \tr_P\bU\big) \big({\bcr}^{(2)}_a-c^{(2)}_a \big) - 2P^{cd}\U_{ca}(\bcr^{(2)}_d-c^{(2)}_d)  = 0,
		\end{equation}
		which is a homogeneous transport equation for $\bcr^{(2)}_a-c^{(2)}_a$, and since both tensors agree on $\mc S$, they agree everywhere. Finally, subtracting \eqref{equationtrfake} from \eqref{trc} and using $\bcr^{(2)}=\bm c^{(2)}$ and $\bck^{(2)}=\mf{c}^{(2)}$ gives 
		\begin{equation}
			\label{diferenciatrazas}
			2n\big({\tr_P\bcY}{}^{(2)}-t^{(2)} \big) + \big(4\bck_n + \tr_P\bU\big)\big({\tr_P\bcY}{}^{(2)}-t^{(2)} \big) = 0.
		\end{equation} 
		Observe that the initial condition \eqref{incon} together with $c^{(2)}_a = \bcr^{(2)}_a$ implies that $\tr_P\bcY^{(2)} = t^{(2)}$ on $\mc S$ (see \eqref{trazas}), and thus by \eqref{diferenciatrazas} they also coincide on all $\mc H$. We have shown that the tensor $\bcY^{(2)}$ satisfies equations \eqref{xxixi}-\eqref{xtt} for $m=1$.\\
		
		In order to construct the rest of the $\bcY^{(k)}$'s we proceed analogously. Let us assume we have already constructed the sequence $\{\bcY,...,\bcY^{(k)}\}$ for some integer $k\ge 2$ such that equations \eqref{xconstraint}-\eqref{xtt} hold for every $m\le k-1$. We define the scalar $t^{(k+1)}$ and the one-form $\bm c^{(k+1)}$ by (cf. \eqref{xxixi}-\eqref{xxix})
		\begin{multicols}{2}
			\noindent
			\begin{equation}
				\label{xxixi3}
				0 = - t^{(k+1)} + \mc{O}^{(k)}(\bcY^{\le k}),
			\end{equation}
			\begin{equation}
				\label{xxix3}
				0 = c^{(k+1)}_a + \mc{O}_a^{(k)}(\bcY^{\le k}),
			\end{equation}
		\end{multicols}
		and the tensor $\bcY^{(k+1)}$ by integrating (cf. \eqref{xtt})
		\begin{equation}
			\label{xtt3}
			\begin{aligned}
				\hskip -0.5cm 2\lambda\bcY^{(k)}_{ab} &= -2\lie_n\bcY^{(k+1)}_{ab} - \left(2(k+1)\bck_n + \tr_P\bU\right)\bcY^{(k+1)}_{ab} - t^{(k+1)}\U_{ab} + 4P^{cd}\U_{c(a}\bcY^{(k+1)}_{b)d}  \\
				&\quad\,+4(\s-\bcr)_{(a} c^{(k+1)}_{b)} + 2\nablacero_{(a}c^{(k+1)}_{b)}-2\mf{c}^{(k+1)}\bcY_{ab} + \mc{O}^{(k)}_{ab}(\bcY^{\le k}),
			\end{aligned}
		\end{equation}
		where $\mf{c}^{(k+1)}\d - \bm c^{(k+1)}(n)$, with initial conditions given by $\iota^{\star}\big(\bcY^{(k+1)}(n,\cdot)\big) = \iota^{\star}\big(\bcY^{(k+1)}(\cdot,n)\big)=\iota^{\star}\big(\bm c^{(k+1)}\big)$, $\bcY^{(k+1)}(n,n) = -\mf{c}^{(k+1)}$ on $\mc S$ and
		\begin{equation}
			\label{initialcond2}
			\iota^{\star}\bcY^{(k+1)}_{AB} = \S^{(k+1)}_{AB} + \dfrac{1}{d-1}\big(t^{(k+1)}+2\bm c^{(k+1)}(\iota_{\star}\ell^{\sharp}) -\big(\elltwo-\elltwo_{\sharp}\big)\mf{c}^{(k+1)} \big) h_{AB}.
		\end{equation} 
		As in the case $k=1$ we define the tensor field $$\T^{(k+1)}_{ab} \d \dfrac{t^{(k+1)}-\mf{c}^{(k+1)}\elltwo}{d-1} \gamma_{ab} + 2c^{(k+1)}_{(a}\ell_{b)} + \mf{c}^{(2)}\ell_a\ell_b$$ 
		
		which satisfies $P^{ab}\T^{(k+1)}_{ab} = t^{(k+1)}$ and $\T^{(k+1)}_{ab}n^a = \T^{(k+1)}_{ba}n^a = c^{(k+1)}_b$. Then we construct an auxiliary semi-Riemannian manifold $(\wt{\mc M}_k,\wt{g}_k)$ using Theorem \ref{borel} with $\{\bcY,...,\bcY^{(k)},\bT^{(k+1)},0,...\}$ and denote its transverse expansion by $\wt{\bY}{}^{(i)} \d\frac{1}{2}\wt{\Phi}^{\star}\big(\lie_{\xi}^{(i)}\wt{g}_k\big)$, $i\ge 1$. Since $\bcY^{(i)}=\wt{\bY}{}^{(i)}$ for every $i=1,...,k$ and $\bT^{(k+1)}=\wt{\bY}{}^{(k+1)}$, the same argument as in the case $k=1$ proves that the Ricci tensor of $\wt{g}_k$ satisfies $\lie_{\xi}^{(i)}\wt{R}_{\mu\nu} = \lambda \mc{K}^{(i)}_{\mu\nu}$ for every $i=0,...,k-1$. Hence, by Proposition \ref{teo_bianchi} it then follows $n^b\wt{\mc R}{}^{(k+1)}_{ab} = \lambda n^b\mc{K}^{(k)}_{ab}$ and $P^{ab}\wt{\mc R}{}^{(k+1)}_{ab} = \lambda P^{ab}\mc{K}^{(k)}_{ab}$, and therefore (cf. \eqref{elierictangn}-\eqref{ePcontractioneq}) 
		\begin{align}
			2\lambda\bcr_a^{(k)} &= -\lie_n c^{(k+1)}_a - \big(2\bck_n + \tr_P\bU\big)c^{(k+1)}_a-\nablacero_a \mf{c}^{(k+1)} +\mc O^{(k)}_{ab}(\bcY^{\le k}) n^b,\label{equationrkfake}\\
			2\lambda\tr_P\bcY^{(k)} & = -2n(t^{(k+1)}) - 2\left(2\bck_n + \tr_P\bU\right)t^{(k+1)}+ 2\mf{c}^{(k+1)}\big(n(\elltwo)-\tr_P\bcY\big) \nonumber\\
			&\quad\, -4P\big(\bcr+\bs,\bm c^{(k+1)}\big) + 2\div_P\bm c^{(k+1)} + \tr_P\mc{O}^{(k)}(\bcY^{\le k}).\label{equationtrkfake}
		\end{align}
		Contracting \eqref{xtt3} with $n^b$ and $P^{ab}$ and combining the resulting equations with \eqref{equationrkfake}-\eqref{equationtrkfake} leads to transport equations for $\bck^{(k+1)}-\mf{c}^{(k+1)}$, $\bcr^{(k+1)}-\bm{c}^{(k+1)}$ and $\tr_P\bcY^{(k+1)}-t^{(k+1)}$ analogous to \eqref{transportkappa2}-\eqref{diferenciatrazas}. Since the initial conditions in the three cases agree, then $\bck^{(k+1)}=\mf c^{(k+1)}$, $\bcr^{(k+1)}=\bm{c}^{(k+1)}$ and $\tr_P\bcY^{(k+1)}=t^{(k+1)}$ everywhere on $\mc H$. Recall that this conclusion is independent on the auxiliary space $(\wt{\mc M}_k,\wt{g}_k)$ we have constructed. This proves that $\bcY^{(k+1)}$ satisfies \eqref{xtt} for $m=k$, and therefore by induction the whole sequence $\{\bcY^{(k)}\}$ constructed in this way satisfies the equations \eqref{xconstraint}-\eqref{xtt}. Moreover $\tr_P\bcY|_{\mc S} = \chi$, $\iota^{\star}\bcr = \bm\beta$, $\bck_n = f$ and $\mbox{tf }\big(\iota^{\star}\bcY^{(k)}\big) = \bS^{(k)}$ for every $k\ge 1$ (see \eqref{trazas} and \eqref{initialcond2}). Applying Theorem \ref{existencia} the result is established.
	\end{proof}
\end{teo}


\begin{rmk}
	The redundancy of equations \eqref{xxixi2}-\eqref{xxix2} with \eqref{cn}-\eqref{trc} (and similarly for higher orders) is a manifestation of the (ambient) contracted Bianchi identity. When the ambient manifold is given, this identity ensures that the objects coming from \eqref{cn}-\eqref{trc} and the ones coming from \eqref{xxixi2}-\eqref{xxix2} agree (see \eqref{B1}-\eqref{B2}). However, in Theorem \ref{teoexistencia} the ambient manifold is not given a priori, but it is being constructed, and thus analysing the compatibility between \eqref{cn}-\eqref{trc} and \eqref{xxixi2}-\eqref{xxix2} is a crucial point in the proof of the Theorem.
\end{rmk}

\begin{rmk}
	The existence of a function $f$ satisfying \eqref{equationf} is automatic whenever $\tr_P\bU\neq 0$ everywhere or $\bU = 0$ everywhere. It is only when $\tr_P\bU$ vanishes on a proper subset that \eqref{equationf} imposes restrictions on the function $f$. Equation \eqref{equationf} is the equivalent of the Raychaudhuri equation when expressed in terms of detached quantities.
	\end{rmk}

Next we write down two by-products of the proof of Theorem \ref{teoexistencia}.
\begin{lema}
	\label{lemaBianchi2}
Let $(\mc M,g)$ be a semi-Riemannian manifold and $\Phi:\H\hookrightarrow\mc M$ an embedded null hypersurface with rigging $\xi$ extended off $\Phi(\H)$ geodesically. Fix a natural number $m\ge 1$ and assume the Ricci tensor of $g$ satisfies $\mc{R}_{ab}\st{\H}{=} \lambda \gamma_{ab}$ (when $m=1$) and $\mc{R}^{(i)}_{ab}\st{\H}{=} \lambda \mc{K}_{ab}^{(i-1)}$ and $\xi^{\mu}R^{(i-1)}_{\mu\nu}\st{\H}{=} \lambda \xi^{\mu}\mc{K}_{\mu\nu}^{(i-2)}$ for $i=1,...,m$ (when $m\ge 2$). Let $t^{(m+1)}$ and $\bm c^{(m+1)}$ be defined by $$t^{(m+1)} \d \mc{O}^{(m)}(\bY^{\le m})-\delta^m_1\lambda\elltwo,\qquad c^{(m+1)}_a \d \delta_1^m \lambda\ell_a- \mc{O}_a^{(m)}(\bY^{\le m}),\qquad \mf{c}^{(m)}\d -\bm c^{(m)}(n),$$ where $\mc{O}^{(m)}$ and $\mc{O}_a^{(m)}$ are the same as in \eqref{ddotR}-\eqref{dotR}. Then the following equations hold
	\begin{align*}
	2\lambda\r_a^{(m)} &= -\lie_n c^{(m+1)}_a - \big(2m\kappa_n + \tr_P\bU\big)c^{(m+1)}_a-\nablacero_a \mf{c}^{(m+1)} + \mc{O}^{(m)}_{ab}n^b,\\
	2\lambda\tr_P\bY^{(m)} &= -2n\big(t^{(m+1)}\big) - 2\left((m+1)\kappa_n + \tr_P\bU\right)t^{(m+1)} + 2\mf{c}^{(m+1)}\big(n(\elltwo)-\tr_P\bY\big)\\
	&\quad\, -4P\big(\br+\bs,\bm c^{(m+1)}\big) + 2\div_P\bm c^{(m+1)} + P^{ab}\mc{O}^{(m)}_{ab}.
\end{align*}
\end{lema}

\begin{prop}
	\label{prop_firstorder}
	Let $\{\mc H,\bg,\bm\ell,\elltwo\}$ be null metric hypersurface data admitting a cross-section $\iota:\mc S\hookrightarrow\mc H$ with metric $h\d\iota^{\star}\bg$ and assume there exists a smooth function $f$ on $\mc H$ satisfying 
	\begin{equation*}
		-n(\tr_P\bU) + (\tr_P\bU)f - P^{ab}P^{cd}\U_{ac}\U_{bd}=0.
	\end{equation*} 
	Choose any real constant $\lambda$ and let $\chi$, $\bm\beta$ and $\bS$ be, respectively, a function, a one-form and a (0,2) symmetric tensor on $\mc S$ traceless w.r.t. $h$. Then, there exists a unique tensor $\bY$ on $\mc H$ satisfying $\tr_P\bY \st{\mc S}{=} \chi$, $\kappa_n \st{\mc S}{=} f$, $\iota^{\star}\br = \bm\beta$ and $\big(\iota^{\star}\bY\big)^{tf} = \bS$ such that $\{\mc H,\bg,\bm\ell,\elltwo,\bY\}$ is null hypersurface data satisfying $\R = \lambda\bg$.
\end{prop}

\subsection{An example: degenerate static Killing horizons with maximally symmetric cross sections}

In this subsection we analyze a particular case where the asymptotic expansion of the metric at a null hypersurface $\H$ can be explicitly computed in terms of certain data at $\H$. Given a degenerate Killing horizon, its so-called near horizon limit \cite{kunduri2009classification} is characterized by a function $F$, a one-form $\bm\omega$ and a metric $h$ on the cross-sections. In a recent work \cite{katona2024uniqueness} the authors show a uniqueness result for the $\mf n$-dimensional extremal Schwarzschild-de Sitter spacetime, which admits a degenerate Killing horizon with compact, maximally symmetric cross-sections. Under their assumptions the one-form $\bm\omega$ vanishes and thus the near horizon data simplifies to the function $F$ and a maximally symmetric metric $h$. By solving the $\Lambda$-vacuum equations order by order, the authors show that all the transverse derivatives of the metric at the horizon can be explicitly computed in terms of the data $(F,h)$ and that they agree either with those of extremal Schwarzschild-de Sitter or with its near horizon limit (Nariai).\\

In order to establish such a result, the authors in \cite{katona2024uniqueness} introduce Gaussian null coordinates $\{v,r,x^A\}$ in terms of which the metric takes the form $$g = \rho^2 F dv^2 + 2dvdr + 2\rho \omega_A dx^A dv + h_{AB} dx^A dx^B,$$ 

where $\eta\d \partial_v$ is the Killing field, $\xi\d \partial_{r}$ is the rigging (chosen to be geodesic, null and orthogonal to the cross-sections) and $\{r=0\}$ is the degenerate Killing horizon. The $\Lambda$-vacuum Einstein equations to first order (specifically $R_{vr}=\Lambda$ and $R_{AB}=\Lambda h_{AB}$) and the staticity of $\eta$ fix $F=\Lambda$ and $\omega_A=0$. In addition, the authors restrict $h$ to be maximally symmetric. \\

As proven in \cite{lilucietti}, there is a freedom that leaves the near horizon data invariant but modifies the transverse expansion. Such freedom can be fixed by requiring $h^{AB}\Y_{AB} = C$ for some $C\in\real$. By imposing the Einstein equations $R_{r A}^{(m)} = 0$, $R_{r v}^{(m+1)} = 0$ and $R_{AB}^{(m+1)} = 2\Lambda \Y_{AB}^{(m)}$ for all $m\ge 1$ the authors find that the full transverse expansion is given explicitly by means of
\begin{equation}
	\label{expansionLucietti}
	\begin{gathered}
\kappa_n = 0,\qquad \kappa^{(2)} = -\Lambda,\qquad \kappa^{(m+2)} = -\dfrac{\Lambda}{\mf{n}-1}\left(-C\right)^m\dfrac{(\mf{n}-2+m)!}{(\mf{n}-3)!},\qquad \r_A^{(m)} = 0,\\
 \Y_{AB} = \dfrac{(\mf{n}-3)C}{\Lambda}h_{AB},\qquad \Y^{(2)}_{AB} = \dfrac{(\mf{n}-3)C^2}{\Lambda}h_{AB},\qquad \Y_{AB}^{(m+1)} = 0 \quad (m\ge 1).
\end{gathered}
\end{equation}
When $C=0$ such expansion agrees with that of Nariai, and when $C\neq 0$ one can absorb $C$ by rescaling the coordinates $r$ and $v$ and the expansion \eqref{expansionLucietti} agrees with that of extremal Schwarzschild-de Sitter spacetime.\\

In this particular case the transverse expansion \eqref{expansionLucietti} is manifestly convergent and thus one can employ Theorem \ref{borel} with $\rho=1$ (see \eqref{borelsum}) to construct the spacetime $(\mc M,g)$, which clearly is either Schwarzschild-de Sitter (when $C\neq 0$) or Nariai (when $C=0$). In both cases, $(\mc M,g)$ solves exactly the $\Lambda$-vacuum equations.\\

There could be other situations (not necessarily Killing horizons) where one can find explicitly the transverse expansion but it may not converge (e.g. Fefferman-Graham metrics when the initial conformal class is not real analytic \cite{fefferman2012ambient} and the series does not converge). In these cases, Theorem \ref{borel} still ensures the existence of a smooth (and not analytic) ambient manifold $(\mc M,g)$. This space may or may not be a true solution (note that now all the freedom in selecting the function $\rho$ and the constants $\{\mu_k\}$ kicks in) of the $\Lambda$-vacuum equations in a neighbourhood of the hypersurface. However, if the transverse expansion has been constructed by imposing the Einstein equations $R_{vr} = \Lambda$, $R_{AB}=\Lambda h_{AB}$, $R_{r A}^{(m)}=0$, $R_{r v}^{(m+1)} = 0$ and $R_{AB}^{(m+1)} = 2\Lambda \Y_{AB}^{(m)}$ for all $m\ge 1$, Proposition \ref{teo_bianchi} guarantees that $(\mc M,g)$ solves the $\Lambda$-vacuum equations to infinite order on $\H$.

%% file: Killing2.tex
\section{Application to Killing horizons}

\label{sec_KH2}

The notion of a Killing horizon can be studied abstractly from many perspectives (see e.g. \cite{manzano2024embedded}, where the authors define the so-called abstract Killing horizons of order zero (AKH$_0$) and one (AKH$_1$)). In Section 6 of \cite{Mio3} we introduced a new definition of abstract Killing horizon from a detached point of view, namely the notion of abstract Killing horizon data (AKH data). The idea of the definition is that, once the data is embedded, AKH data is sufficient to determine the full transverse expansion of the metric at the horizon from the tower of derivatives of the ambient Ricci tensor at the horizon. Moreover, such expansion can be proven to be unique when the rigging $\xi$ is chosen suitably. This extends a recent work by O. Petersen and K. Kroencke \cite{oliver}. In the first part of this Section we review the basic definitions and results from \cite{Mio3}, and then we apply the existence results of Section \ref{sec_existence} to prove that given AKH data there exists an ambient manifold solving the $\Lambda$-vacuum equations to infinite order at the horizon and that such manifold admits a Killing vector which is null and tangent at the horizon.\\

A Killing horizon of $\eta$ is an embedded hypersurface $\Phi:\mc H\hookrightarrow\mc M$ to which $\eta$ is tangent, null and nowhere zero. We say that the Killing horizon is non-degenerate when its surface gravity satisfies $\wt\kappa\neq 0$ at some point, and we say it is degenerate when $\wt\kappa=0$ everywhere. Another well-known property of Killing horizons is that they are totally geodesic, i.e. $\bU=0$, which follows at once from $\Phi^{\star}\mc K[\eta] = 2\alpha\bU$. As a consequence, given any $X\in\X(\mc H)$, equations \eqref{connections} and \eqref{derivadannull} give
\begin{equation}
	\label{connectionskilling}
	\nabla_{\Phi_{\star}X}\eta = \nabla_{\Phi_{\star}X}(\alpha \nu) = \alpha\nabla_{\Phi_{\star}X} \nu + X(\alpha)\nu =  \big(\alpha (\bs-\br) + d\alpha\big)(X) \nu ,
\end{equation}
which motivates the definition of the one-form
\begin{equation}
	\label{def_omega}
	\bm\tau\d \alpha(\bs-\br)+d\alpha
\end{equation} 
satisfying the following properties \cite{Mio3}.
\begin{prop}
	\label{properties_omega}
	Let $\bm\tau$ be as in \eqref{def_omega} and $\kappa$ as in \eqref{kappasgeneral}. Then,\\
	\begin{minipage}{0.3\textwidth}
		\noindent
		\begin{enumerate}
			\item $\mc G_{(z,V)}\bm\tau=z\bm\tau$,
		\end{enumerate}
	\end{minipage}
	\begin{minipage}{0.3\textwidth}
		\noindent
		\begin{enumerate}
			\item[2.] $\bm\tau(n)=\kappa$,
		\end{enumerate}
	\end{minipage}
	\begin{minipage}{0.4\textwidth}
		\noindent
		\begin{enumerate}
			\item[3.] $\alpha\lie_n\bm\tau = n(\alpha)\bm\tau - \bm\tau(n) d\alpha$.
		\end{enumerate}
	\end{minipage}
\end{prop}

Particularizing equations \eqref{constraint} and \eqref{constraintn} to $\bU=0$,
\begin{align}
	\mc R_{ab}& = \accentset{\circ}{R}_{(ab)} -2\lie_n \Y_{ab} - 2\kappa_n\Y_{ab} + \nablacero_{(a}\left(\s_{b)}+2\r_{b)}\right) -2\r_a\r_b + 4\r_{(a}\s_{b)} - \s_a\s_b,\label{khconstraint}\\
	\mc R_{ab}n^a & = \lie_n(\s_b-\r_b) - \nablacero_b \kappa_n,\label{khconstraintn}
\end{align}
while, using that the tensors \eqref{hebrew} all vanish, equations \eqref{constraint22} and \eqref{constraintn2} become
\begin{align}
	\alpha\mc R_{ab} &= -2\kappa \Y_{ab} -2\alpha\nablacero_{(a}(\s-\r)_{b)} -4(\s-\r)_{(a}\nablacero_{b)}\alpha  -2\alpha(\s-\r)_a(\s-\r)_b \nonumber\\
	&\quad\,  - 2\nablacero_a\nablacero_b\alpha - \alpha\nablacero_{(a}\s_{b)}+\alpha\s_a\s_b  +\alpha \Rcero_{(ab)},\label{khconstraint2}\\
	\alpha\mc R_{ab}n^b &= -\nablacero_a\kappa.\label{khconstraintn2}
\end{align}
Similarly, identity \eqref{lierictang3} for $m\ge 1$ simplifies to
\begin{equation}
	\label{lierictangkh2}
	\begin{aligned}
		\alpha \mc{R}^{(m+1)}_{ab} & =  -2(m+1)\kappa\Y^{(m+1)}_{ab}  +4\alpha (\s-\r)_{(a} \r^{(m+1)}_{b)}+4\r^{(m+1)}_{(a}\nablacero_{b)}\alpha \\
		&\quad\,  + 2\alpha\nablacero_{(a}\r^{(m+1)}_{b)}-2\alpha\kappa^{(m+1)}\Y_{ab} + \alpha\mc{O}^{(m)}_{ab} + \mc{P}^{(m)}.
	\end{aligned}
\end{equation}

The abstract notion of Killing horizon introduced in \cite{Mio3} captures the minimum data on the horizon that allows for the determination of its full transverse expansion. Such data involves metric hypersurface data, the one-form $\bm\tau$ and the scalar $\alpha$. The first thing we require is $\bU=0$, since it guarantees that the pullback of the deformation tensor vanishes on $\Phi(\mc H)$. We must also impose that $\alpha$ is smooth and its zeroes have empty interior. Concerning the one-form $\bm\tau$, the abstract definition must ensure condition (iii) of Prop. \ref{properties_omega} as well as that the one-form $\alpha^{-1}(\bm\tau-d\alpha)$ extends smoothly to all $\mc H$, because for actual Killing horizons, the following equality holds (cf. \eqref{def_omega}) $$\alpha^{-1} ( \bm\tau - d \alpha) = \bs - \br,$$ 

and the right hand side is smooth everywhere on $\mc H$. This discussion motivates the following definition put forward in \cite{Mio3}.

\begin{defi}
	\label{defi_AKH}
	We say $\{\mc H,\bg,\bm\ell,\elltwo,\bm\tau,\alpha\}$ is abstract Killing horizon data (AKH data) provided that (i) $\{\mc H,\bg,\bm\ell,\elltwo\}$ is null metric hypersurface data satisfying $\bU\d\frac{1}{2}\lie_n\bg=0$, (ii) $\alpha$ is a smooth function such that the set $\{\alpha=0\}$ has empty interior and (iii) $\bm\tau$ is a one-form such that $\alpha\lie_n\bm\tau = n(\alpha)\bm\tau - \bm\tau(n) d\alpha$ and the one-form $\alpha^{-1}(\bm\tau-d\alpha)$ extends smoothly to all $\mc H$.
\end{defi}
Motivated by item (ii) of Prop. \ref{properties_omega} and $\alpha'=z\alpha$, the notion of gauge transformation can be extended to the context of AKH data as follows.
\begin{defi}
	\label{gaugeAKH}
	Let $\{\mc H,\bg,\bm\ell,\elltwo,\bm\tau,\alpha\}$ be AKH data and $(z,V)$ gauge parameters. We define the gauge-transformed data by $\{\mc H,\bg',\bm\ell',\elltwo{}',z\bm\tau,z\alpha\}$, where $\{\bg',\bm\ell',\elltwo{}'\}$ are given by \eqref{transgamma}-\eqref{transell2}.
\end{defi}
As already said, the condition $\bU=0$ of Definition \ref{defi_AKH} only guarantees that the pullback of the deformation tensor of $\eta$ vanishes on $\mc H$. To capture the full information about the deformation tensor we need to restrict ourselves to the embedded case.
\begin{defi}
	\label{EKHdata}
	Let $\k=\{\mc H,\bg,\bm\ell,\elltwo,\bm\tau,\alpha\}$ be AKH data and define $\bar\eta\d \alpha n$. We say that $\k$ is $(\Phi,\xi)$-embedded in $(\mc M,g)$ if (i) $\{\mc H,\bg,\bm\ell,\elltwo\}$ is $(\Phi,\xi)$-embedded in $(\mc M,g)$ as in Def. \ref{defi_embedded} and (ii) $\nabla_{\Phi_{\star}X}\Phi_{\star}\bar\eta = \bm\tau(X)\nu$ for every $X\in\X(\mc H)$. Moreover, we say that $\k$ is an $(\Phi,\xi)$-embedded Killing horizon data (EKH data) if, additionally, (iii) there exist an extension $\eta$ of $\Phi_{\star}\bar\eta$ such that its deformation tensor $\mc K[\eta]\d\lie_{\eta} g$ vanishes to all orders at $\Phi(\mc H)$. 
\end{defi}

\begin{rmk}
	\label{remarkembedding}
	As shown in \cite{Mio3}, if $\k=\{\mc H,\bg,\bm\ell,\elltwo,\bm\tau,\alpha\}$ is $(\Phi,\xi)$-embedded in $(\mc M,g)$, then $\mc{G}_{(z,V)}\k$ is $(\Phi,\xi')$-embedded in $(\mc M,g)$ with $\xi' = z(\xi+\Phi_{\star}V)$.
\end{rmk}

\begin{rmk}
	\label{notation}
The definition of AKH data allows us to define a smooth one-form $\bcr\d \bs - \alpha^{-1}(\bm\tau-d\alpha)$ and a scalar $\bck_n\d \alpha^{-1}(\bm\tau(n)-n(\alpha))$. In the embedded case, the one-form $\bcr$ and the function $\bck_n$ agree with $\br$ and $\kappa_n$, respectively, and hence we shall not distinguish between $\bcr$ and $\br$ anymore \cite{Mio3}. Observe also that this fact together with \eqref{kappasgeneral} implies that the surface gravity $\kappa$ is given by $\kappa = n(\alpha) + \alpha \kappa_n = \bm\tau(n)$.
\end{rmk}

In Lemma 6.13 of \cite{Mio3} we proved that given AKH data $\k=\{\mc H,\bg,\bm\ell,\elltwo,\bm\tau,\alpha\}$ satisfying $\bm\tau(n)\neq 0$ everywhere on $\mc H$, there exists a family of gauges (called $\eta$-gauges) satisfying $\bm\ell = \kappa^{-1}\bm\tau$ and $\elltwo=0$. The construction of such gauge can be achieved by a transformation of the form $\mc{G}_{(1,V)}$, with the vector $V$ uniquely defined by the conditions 
\begin{equation}
	\label{V}
	\bg(V,\cdot) = \kappa^{-1}\bm\tau-\bm\ell,\qquad \text{and}\qquad \bm\ell(V) = -\elltwo-\frac{1}{2}\kappa^{-2}P(\bm\tau,\bm\tau).
\end{equation}

Moreover, the remaining gauge freedom is the transformations $\mc{G}_{(z,0)}$. This gauge is particularly relevant because in the embedded case, when $\k$ is written in an $\eta$-gauge then any extension $\eta$ of the vector $\bar\eta \d \alpha n$ satisfying $\mc{K}[\eta](\xi,\cdot)= 0$ on $\Phi(\mc H)$ also satisfies
\begin{equation}
	\label{liexietaKH2}
	\lie_{\xi}\eta \st{\mc H}{=} (\alpha\kappa_n-\kappa) \xi \st{\H}{=} -n(\alpha)\xi .
\end{equation}

This guarantees in particular that the vector $X_{\eta}$ introduced in \eqref{liexieta} vanishes identically.\\

In \cite{Mio3} we introduced a natural notion of isometry in the context of AKH data, namely that two abstract Killing horizon data are isometric provided that they are related by a diffeomorphism and a gauge transformation. This notion is particularly relevant because it allowed us to prove a uniqueness result, namely that two isometric EKH data embedded in respective $\Lambda$-vacuum manifolds are necessarily isometric to infinite order at the horizons. This shows that every analytic $\Lambda$-vacuum manifold admitting a non-degenerate Killing horizon (possibly with bifurcation surfaces) is characterized near the horizon by its AKH data. In the last part of this paper we analyze the question of whether an AKH data gives rise to a $\Lambda$-vacuum manifold to infinite order admitting a Killing field with the given data as Killing horizon. A similar analysis can be applied to other field equations.\\

The idea of the Theorem is to construct explicitly the transverse expansion $\{\bcY^{(m)}\}$ from identities \eqref{khconstraint2} and \eqref{lierictangkh2} where we replace the Ricci tensor by $\lambda g$ and the deformation tensor by zero at all orders. A crucial point of the argument will rely on using that the transverse expansion constructed in this way has an appropriate behavior under gauge transformations of the AKH data. The following lemma is devoted to establishing that.
\begin{lema}
	\label{lemagauge}
Let $\k=\{\H,\bg,\bm\ell,\elltwo,\bm\tau,\alpha\}$ be AKH data satisfying that $\bm\tau(n)\neq 0$ is constant. Assume $\k$ is written in an $\eta$-gauge and define the tensors $\bm{c}\d \bs + \alpha^{-1}(d\alpha-\bm\tau)$, $\mf{c}\d -\bm{c}(n)$,
\begin{equation}
	\label{defYprima}
\begin{aligned}
	\bcY_{ab} &\d \dfrac{1}{2\kappa}\left(\alpha \Rcero_{(ab)}+\alpha\s_a\s_b - \alpha\nablacero_{(a}\s_{b)} - 2\nablacero_a\nablacero_b\alpha  -2\alpha(\s-c)_a(\s-c)_b \right.\\
	&\qquad\quad\left.-4(\s-c)_{(a}\nablacero_{b)}\alpha -2\alpha\nablacero_{(a}(\s-c)_{b)} -\alpha\lambda\gamma_{ab}\right),
\end{aligned}
\end{equation}
and for $k\ge 2$
\begin{equation}
	\label{Ykprima0}
	\begin{aligned}
		\bcY_{ab}^{(k)} &\d \dfrac{1}{2k\kappa}\left(\alpha \mc{O}^{(k-1)}(\bcY^{\le k-1})_{ab} -2\alpha \mf{c}^{(k)}\bcY_{ab} + 2\alpha\nablacero_{(a} c^{(k)}_{b)} + 4 c^{(k)}_{(a}\nablacero_{b)}\alpha\right.\\
		&\qquad\qquad\left. + 4\alpha(\s-c)_{(a} c_{b)}^{(k)} -2\alpha\lambda \bcY_{ab}^{(k-1)}\right),
	\end{aligned}
\end{equation}
where
	\begin{equation}
c_a^{(k)} \d \delta_2^k\lambda\ell_a - \mc{O}^{(k-1)}(\bcY^{\le k-1})_a,\qquad \mf{c}^{(k)}\d -\bm{c}^{(k)}(n).
	\end{equation}
Define the collection $\{\bcY^{(m)}{}'\}$ exactly in the same way as above but with all the terms in the right hand sides expressed in the gauge $\k'=\mc{G}_{(z,0)}\k$ with $z\in\mc{F}^{\star}(\H)$ arbitrary. Then, 
\begin{equation}
	\label{transYes0}
	\bcY_{ab}' = z\bcY_{ab} + \bm\ell \otimes_s dz,\qquad \bcY_{ab}^{(2)}{}' = z^2 \bcY_{ab}^{(2)}.
\end{equation}
Let $N\ge 2$ be an integer. If in addition $\tr_P\bcY^{(j)} = \mc{O}^{(j-1)}(\bcY^{\le j-1})$ for all $j=2,...,N$, then
\begin{equation}
	\label{transYes}
 \bcY_{ab}^{(k)}{}' = z^k \bcY_{ab}^{(k)}\quad \forall k =3,...,N+1.
\end{equation}
\begin{proof}
By definition, 
\begin{equation}
	\label{bcYprima}
	\begin{aligned}
		\bcY'_{ab} &\d \dfrac{1}{2\kappa}\left(\alpha' \Rcero_{(ab)}'+\alpha'\s_a'\s_b' - \alpha'\nablacero'_{(a}\s'_{b)} - 2\nablacero'_a\nablacero'_b\alpha'  -2\alpha'(\s'-c')_a(\s'-c')_b \right.\\
		&\qquad\quad\left.-4(\s'-c')_{(a}\nablacero'_{b)}\alpha' -2\alpha'\nablacero'_{(a}(\s'-c')_{b)} -\alpha'\lambda\gamma'_{ab}\right)
	\end{aligned}
\end{equation}
and
\begin{equation}
	\label{Ykprima}
	\begin{aligned}
		\bcY_{ab}^{(k)}{}' &\d \dfrac{1}{2k\kappa}\left(\alpha' \mc{O}_{ab}^{(k-1)}(\bcY'{}^{\le k-1}) -2\alpha' \mf{c}^{(k)}{}'\bcY_{ab}' + 2\alpha'\nablacero'_{(a} c_{b)}^{(k)}{}' + 4 c_{(a}^{(k)}{}'\nablacero'_{b)}\alpha'\right.\\
		&\qquad\qquad\left. + 4\alpha'(\s'-c')_{(a} c_{b)}^{(k)}{}' -2\alpha'\lambda \bcY_{ab}^{(k-1)}{}'\right),
	\end{aligned}
\end{equation}
where 
	\begin{equation}
		\label{cprima}
		c_a^{(k)}{}' \d \delta^k_2 \lambda\ell_a' - \mc{O}_a^{(k-1)}(\bcY'{}^{\le k-1}), \qquad \mf{c}^{(k)}{}'\d -\bm{c}^{(k)}{}'(n').
	\end{equation}
Recall that by Remark \ref{rmk_imp} there is no possible confusion in denoting $\mc{O}_{ab}^{(k-1)}(\bcY'{}^{\le k-1})$ and $\mc{O}_a^{(k-1)}(\bcY'{}^{\le k-1})$ by $\mc{O}_{ab}^{(k-1)}{}'$ and $\mc{O}_a^{(k-1)}{}'$, respectively. Moreover, under the hypothesis $\tr_P\bcY^{(j)} = \mc{O}^{(j-1)}(\bcY^{\le j-1})$ for all $j=2,...,N$ the transformations of Proposition \ref{propgaugeO} become
\begin{align}
\mc{O}_{ab}^{(k-1)}{}'&=	\mc{O}_{ab}^{(k-1)}(\bcY'{}^{\le k-1}) = z^{k-1}\mc{O}_{ab}^{(k-1)}(\bcY^{\le k-1}) + 2(k-1) z^{k-2} \mc{O}_{(a}^{(k-1)}(\bcY^{\le k-1})\nablacero_{b)}z ,\label{transOab2}\\
\mc{O}_a^{(k-1)}{}'&=	 \mc{O}_a^{(k-1)}(\bcY'{}^{\le k-1}) = z^{k-1}\mc{O}_a^{(k-1)}(\bcY^{\le k-1}).\label{transOa2}
\end{align}
Let us begin by showing that $\bcY_{ab}' = z\bcY_{ab} + \bm\ell \otimes_s dz$. We divide the computation into several pieces. Firstly, using $\alpha'=z\alpha$ and Corollary \ref{corRcero} the terms $\alpha \Rcero_{(ab)}+\alpha\s_a\s_b - \alpha\nablacero_{(a}\s_{b)}$ transform as (recall $\bU=0$)
\begin{align*}
	\alpha' \Rcero'_{(ab)}+\alpha'\s'_a\s'_b - \alpha'\nablacero'_{(a}\s'_{b)} = z\left(\alpha \Rcero_{(ab)}+\alpha\s_a\s_b - \alpha\nablacero_{(a}\s_{b)}\right).
\end{align*}
Secondly, from $\nablacero'=\nablacero + z^{-1} n\otimes\bm\ell\otimes_s dz$ it follows 
\begin{align*}
	\nablacero_a'\nablacero_b'\alpha' & = \nablacero'_a\nablacero_b(z\alpha) \\
	&= \nablacero_az \nablacero_b\alpha + z\nablacero_a'\nablacero_b\alpha +\nablacero_a\alpha\nablacero_b z + \alpha\nablacero_a'\nablacero_b z\\
	& = 2\nablacero_{(a}z\nablacero_{b)}\alpha +z\nablacero_a\nablacero_b \alpha +\alpha\nablacero_a\nablacero_b z -\big(n(\alpha)+z^{-1}\alpha n(z)\big)\ell_{(a}\nablacero_{b)}z.
\end{align*}
From $\bm\tau ' = z\bm\tau$ and $\alpha'=z\alpha$, the transformation of $\bs-\bm c$ is 
\begin{equation}
	\label{s-c}
	\bs'-\bm{c}' = \bs - \bm c -z^{-1} dz,
\end{equation}
from where it follows $$\big(\s'-c'\big)_a\big(\s'-c'\big)_b = \big(\s-c\big)_a\big(\s-c\big)_b -2z^{-1}(\s-c)_{(a}\nablacero_{b)}z + z^{-2}\nablacero_a z\, \nablacero_b z$$ 

and
\begin{align*}
	\nablacero'_{(a}(\s'-c')_{b)} & = \nablacero_{(a}\big(\s-c-z^{-1}dz\big)_{b)} + z^{-1}\big(\bm c-\bs + z^{-1}dz\big)(n) \ell_{(a}\nablacero_{b)}z\\
	&= \nablacero_{(a}\big(\s-c\big)_{b)} -z^{-1}\nablacero_a\nablacero_b z + z^{-2}\nablacero_a z \nablacero_b z + z^{-1}\big(z^{-1}n(z)-\mf{c}\big)\ell_{(a}\nablacero_{b)}z.
\end{align*}
Combining all the terms in the right hand side of \eqref{bcYprima}, $$\bcY'_{ab} = z\bcY_{ab} + \dfrac{1}{2\kappa}\big(2n(\alpha) + 2\alpha c\big)\ell_{(a}\nablacero_{b)}z = z\bcY_{ab} + \ell_{(a}\nablacero_{b)}z,$$

where we used $\kappa = \bm\tau(n) = \alpha \mf c + n(\alpha)$. Hence the first equation in \eqref{transYes0} follows. Next we compute the transformation of $\bcY^{(2)}$. Writing \eqref{Ykprima} for $k=2$ 
\begin{equation*}
	\bcY_{ab}^{(2)}{}' \d \dfrac{1}{4\kappa}\left(\alpha'\mc{O}_{ab}^{(1)}{}' - 2\alpha'\mf{c}^{(2)}{}'\bcY_{ab}' + 2\alpha'\nablacero_{(a}'c_{b)}^{(2)}{}' + 4c_{(a}^{(2)}{}'\nablacero_{b)}'\alpha' + 4\alpha'(\s'-c')_{(a}c_{b)}^{(2)}{}' - 2\alpha'\lambda\bcY'_{ab}\right),
\end{equation*}
where by \eqref{cprima} $c_a^{(2)}{}' \d \lambda\ell_a' - \mc{O}_a^{(1)}{}' = z\big(\lambda\ell_a-\mc{O}_a^{(1)}\big) = z c_a^{(2)}$ (see \eqref{tranfell} and \eqref{transOa2}) and hence $\mf{c}^{(2)}{}'=\mf{c}^{(2)}$. Inserting $\alpha' = z\alpha$, \eqref{transOab2}, $\bcY'_{ab} = z\bcY_{ab} + \ell_{(a}\nablacero_{b)}z$, \eqref{transnablcaero} and \eqref{s-c} it can be easily checked that $\bcY_{ab}^{(2)}{}' = z^2 \bcY_{ab}^{(2)}$. For arbitrary $3\le k\le N+1$ the process is analogous. Indeed, since $c_a^{(k)}{}' \d  -\mc{O}_a^{(k-1)}{}'$ equation \eqref{transOa2} gives $c_a^{(k)}{}' = z^{k-1}c_a^{(k)}$ and $\mf{c}^{(k)}{}' = z^{k-2}\mf{c}^{(k)}$. Inserting the transformations of $\alpha$, $\mc{O}^{(m)}_{ab}$ \eqref{transOab2}, $\bs-\bm c$ \eqref{s-c} and \eqref{induction} into \eqref{Ykprima} it follows $\bcY^{(k)}{}'=z^{k}\bcY^{(k)}$.
\end{proof}
\end{lema}

Now we are ready to show the main result of this section, namely that every AKH data with $\bm\tau(n)\neq 0$ constant gives rise to an ambient manifold solving the $\Lambda$-vacuum equations to infinite order and admitting a Killing field for which the initial data is a non-degenerate horizon. Our result is entirely covariant and does not make any assumptions regarding the dimension or topology of the null hypersurface.

\begin{teo}
	\label{teorema2}
Let $\wt\k=\{\mc H,\wt\bg,\wt{\bm\ell},\wt{\ell}{}^{(2)},\wt{\bm\tau},\wt\alpha\}$ be AKH data satisfying that $\wt{\bm\tau}(\wt n)$ is a non-zero constant. Then, there exists a smooth semi-Riemannian manifold $(\mc M,g)$, an embedding ${\Phi:\mc H\hookrightarrow\mc M}$, a rigging $\zeta$ of $\Phi(\mc H)$ and a smooth extension $\eta$ of $\bar\eta\d \wt{\alpha} \wt n$ off $\Phi(\mc H)$ such that (i) $\k$ is $(\Phi,\zeta)$-embedded in $(\mc M,g)$ (Def. \ref{EKHdata}), (ii) $\eta$ is a Killing vector of $g$ and (iii) $(\mc M,g)$ satisfies the $\Lambda$-vacuum equations to all orders on $\Phi(\mc H)$.
	\begin{proof}
The proof is divided into three main parts. Firstly, we define a suitable sequence of $(0,2)$ symmetric tensors $\{\bcY^{(k)}\}$ on $\H$ and we construct the ambient manifold $(\mc M,g)$ as in Theorem \ref{borel}. Secondly, we show that the ambient Ricci tensor of $(\mc M,g)$ satisfies the $\Lambda$-vacuum equations to infinite order on $\H$, i.e. $R^{(i)}_{\mu\nu}\st{\H}{=}\lambda \mc{K}^{(i-1)}_{\mu\nu}$ for every $i\ge 1$. Finally, we construct the Killing $\eta$ by extending $\bar\eta=\alpha n$ suitably.\\

\underline{1. Construction of the ambient space}\\

The idea of the first part of the proof is to construct the sequence $\{\bcY^{(k)}\}$ using identities \eqref{khconstraint2} and \eqref{lierictangkh2}, where the Ricci tensor is replaced by $\lambda g$ and the deformation tensor by zero at all orders. Observe that since $\kappa\d \wt{\bm\tau}(\wt n)\neq 0$ is constant it follows from \eqref{khconstraintn2} that $\R(\wt n,\cdot)=0$.\\

Let us transform the AKH data $\wt{\k}$ to the $\eta$-gauge constructed by the transformation $\mc{G}_{(1,V)}$ with $V$ defined by (compare with \eqref{V}) 
\begin{equation}
	\label{gauge2}
	\wt{\bg}(V,\cdot) = \kappa^{-1}\wt{\bm\tau} - \wt{\bm\ell},\qquad\text{and}\qquad \wt{\bm\ell}(V) =- \wt{\ell}{}^{(2)} - \frac{1}{2}\kappa^{-2}\wt{P}(\wt{\bm\tau},\wt{\bm\tau}).
\end{equation}
We denote the data in this $\eta$-gauge without the tilde and define the one-form $\bm c\d \bs +\alpha^{-1}(d\alpha -\bm\tau)$, the scalar $\mf{c} \d -\bm{c}(n) = \alpha^{-1}(\bm\tau(n)-n(\alpha))$ (recall that by Remark \ref{notation} both objects are smooth) and the tensor $\bcY$ by (cf. \eqref{khconstraint2})
\begin{equation}
	\label{defY}
	\begin{aligned}
\bcY_{ab} &\d \dfrac{1}{2\kappa}\left(\alpha \Rcero_{(ab)}+\alpha\s_a\s_b - \alpha\nablacero_{(a}\s_{b)} - 2\nablacero_a\nablacero_b\alpha  -2\alpha(\s-c)_a(\s-c)_b \right.\\
&\qquad\quad\left.-4(\s-c)_{(a}\nablacero_{b)}\alpha -2\alpha\nablacero_{(a}(\s-c)_{b)} -\alpha\lambda\gamma_{ab}\right).
\end{aligned}
\end{equation}
We also define the one-form $c^{(2)}_a$ by (cf. \eqref{xxix})
	\begin{equation}
		\label{dotRkh1}
c_a^{(2)}\d		\lambda \ell_a  - {\mc{O}}^{(1)}(\bcY)_a.
	\end{equation}
Similarly, for arbitrary $m\ge 2$ we define (cf. \eqref{lierictangkh2})
\begin{equation}
	\label{defYk}
	\begin{aligned}
		\bcY_{ab}^{(m)} &\d \dfrac{1}{2m\kappa}\left(\alpha \mc{O}^{(m-1)}(\bcY^{\le m-1})_{ab} -2\alpha \mf{c}^{(m)}\bcY_{ab} + 2\alpha\nablacero_{(a} c^{(m)}_{b)} + 4 c^{(m)}_{(a}\nablacero_{b)}\alpha\right.\\
		&\qquad\qquad\left. + 4\alpha(\s-c)_{(a} c_{b)}^{(m)} -2\alpha\lambda \bcY_{ab}^{(m-1)}\right)
	\end{aligned}
\end{equation}
and $\bm{c}^{(m+1)}$ by (cf. \eqref{xxix})
	\begin{equation}
c_a^{(m+1)} \d - \mc{O}^{(m)}(\bcY^{\le m})_a.
	\end{equation}
By Theorem \ref{borel} there exists an ambient space $(\mc M,g)$, an embedding $\Phi:\H\hookrightarrow\mc M$ and a rigging ${\xi}$ such that ${\bcY}^{(k)} = \frac{1}{2}{\Phi}^{\star}\big(\lie_{\xi}^{(k)}g\big)$ for every $k\ge 1$. By Remark \ref{remarkembedding}, item (i) of the Theorem follows with $\zeta = \xi - V$, with $V$ given by \eqref{gauge2}. Our objective now is to prove that $(\mc M,g)$ satisfies the $\Lambda$-vacuum equations at infinite order on $\Phi(\H)$ and that there exists an extension $\eta$ of $\bar\eta=\alpha n$ off $\Phi(\H)$ such that $\lie_{\eta}g = 0$ on $\mc M$. \\\\\\

\underline{2. $\Lambda$-vacuum equations}\\

Now we concentrate in proving that the $\Lambda$-vacuum equations hold at infinite order on $\Phi(\H)$, i.e. we want to show that for all $m\ge 1$
\begin{equation}
	\label{whatIwant}
	\mc{R}_{ab} = \lambda\gamma_{ab},\qquad	\ddot{\mc R}^{(m)} = 0,\qquad \dot{\mc R}_a^{(m)} = \delta_1^m\lambda\ell_a,\qquad  \mc{R}_{ab}^{(m+1)} = 2\lambda\Y_{ab}^{(m)}.
\end{equation}
By \eqref{constraint} and \eqref{ddotR}-\eqref{R} this is equivalent to check that the embedded expansion $\{\bY^{(m)}\}$ satisfies the following equations (recall $\elltwo=0$)
\begin{equation}
	\label{xconstrainte}
	\begin{aligned}
		\lambda\gamma_{ab}& = \accentset{\circ}{R}_{(ab)} -2\lie_n \Y_{ab} - (2\kappa_n+\tr_P\bU)\Y_{ab} + \nablacero_{(a}\left(\s_{b)}+2\r_{b)}\right)\\
		&\quad -2\r_a\r_b + 4\r_{(a}\s_{b)} - \s_a\s_b - (\tr_P\bY)\U_{ab} + 2P^{cd}\U_{d(a}\left(2\Y_{b)c}+\F_{b)c}\right),
	\end{aligned}
\end{equation}
\begin{multicols}{2}
	\noindent
	\begin{equation}
		\label{xxixie}
0 = - \tr_P\bY^{(m+1)} + \mc{O}^{(m)}(\bY^{\le m}),
	\end{equation}
	\begin{equation}
		\label{xxixe}
		\lambda \ell_a \delta_1^m = \r_a^{(m+1)} + \mc{O}^{(m)}(\bY^{\le m})_a,
	\end{equation}
\end{multicols}
\vspace{-0.4cm}
\begin{equation}
	\label{xtte}
	\begin{aligned}
		\hskip -0.5cm 2\lambda\Y^{(m)}_{ab} &= -2\lie_n\Y^{(m+1)}_{ab} - \left(2(m+1)\kappa_n + \tr_P\bU\right)\Y^{(m+1)}_{ab}- (\tr_P\bY^{(m+1)})\U_{ab}\\
		&\quad\,  + 4P^{cd}\U_{c(a}\Y^{(m+1)}_{b)d}  +4(\s-\r)_{(a} \r^{(m+1)}_{b)}+ 2\nablacero_{(a}\r^{(m+1)}_{b)}\\
		&\quad\, -2\kappa^{(m+1)}\Y_{ab} + \mc{O}^{(m)}_{ab}(\bY^{\le m}).
	\end{aligned}
\end{equation}
There are two main difficulties at this point. Firstly, we shall check that the tensors $\{\bm c^{(m)}\}$ agree with $\{\br^{(m)}\}$ at all orders, and secondly we need to compute Lie derivatives along $n$ of the expansion $\{\bY^{(m)}\}$. Unfortunately we cannot do this directly in the $\eta$-gauge because we have no information on the Lie derivative along $n$ of the various quantities. The reason is that $\bar{\eta}$ and $n$ are different in this gauge as they are related by the proportionality function $\alpha$. The obvious thing to try is to go into a gauge where $\alpha=1$. However, this requires some work because we are allowing $\alpha$ to have zeroes. This forces us to argue on a suitable open and dense subset and then come back to the original (and globally defined) $\eta$-gauge. Let us sketch the main lines of the argument:
\begin{enumerate}
	\item We restrict to the subset $\{\alpha\neq 0\}$ and rescale the rigging by $\xi'=\alpha^{-1}\xi$. This rescaling induces a transformation on the embedded expansion $\{\bY^{(m)}\}\mapsto \{\bY^{(m)}{}'\}$ (transformations \eqref{trans2}) as well as on the transverse derivatives of the ambient Ricci tensor $\{\ddot{\mc R}^{(m)},\dot{\mc R}_a^{(m)},\mc{R}_{ab}^{(m)}\}\mapsto\{\ddot{\mc R}^{(m)}{}',\dot{\mc R}_a^{(m)}{}',\mc{R}_{ab}^{(m)}{}'\}$ in \eqref{transR2}-\eqref{transRab2}. At the same time, we transform $\k$ with gauge parameters $(z=\alpha^{-1},0)$ (for consistency we shall denote the transformed data by a prime, $\k'$).
	\item We use \eqref{transYes0} to write down $\bY'$ and $\bY^{(2)}{}'$ explicitly. This allows us to check that $\br'=\bm c'$, $\lie_{n'}\bY' = \lie_{n'}\bY^{(2)}{}'=0$ and that $\mc{R}_{ab}' = \lambda\gamma'_{ab}$. Furthermore, applying Lemma \ref{lemaBianchi2} we will show $\br^{(2)}{}' = \bm{c}^{(2)}{}'$ and also $\dot{\mc R}_a^{(1)}{}' = \lambda \ell_a'$, $\ddot{\mc R}^{(1)}{}' = 0$ and $\mc{R}_{ab}^{(2)}{}' = 2\lambda\Y_{ab}'$.
	\item In order to continue with the argument inductively we need to determine $\{\bY^{(k\ge 3)}{}'\}$ explicitly. To do that we rely on Lemma \ref{lemagauge} that relates the unprimed quantities with the primed ones. This lemma has a key hypothesis, namely the validity of $$\tr_P\bcY^{(j)} = \mc{O}^{(j-1)}(\bcY^{\le j-1})\qquad \forall\, j=2,...,N.$$ 
	
	In order to check that this hypothesis holds we use the transformations \eqref{transR2}-\eqref{transRab2} between the primed and unprimed components of the Ricci tensor. By a similar application of Lemma \ref{lemaBianchi2} as in the case $k=2$ we will check that $\br^{(k)}{}' = \bm{c}^{(k)}{}'$, $\lie_{n'}\bY^{(k)}{}' = 0$ and that $\ddot{\mc R}^{(k-1)}{}'=0$, $\dot{\mc R}_a^{(k-1)}{}'=0$ and $\mc{R}_{ab}^{(k)}{}'=2\lambda\Y_{ab}^{(k-1)}{}'$.
	\item Finally, we come back to the unprimed quantities via transformations \eqref{transR2}-\eqref{transRab2} to show \eqref{xconstrainte}-\eqref{xtte}.
\end{enumerate}

Let $\H_0$ be the subset of $\H$ consisting of the points where $\alpha\neq 0$ and let $\k_0$ the restriction of the AKH data $\k$ to $\H_0$. Observe that by definition of AKH data, the closure of $\H_0$ is $\H$. Let us extend $\alpha$ off $\Phi(\H)$ by $\xi(\alpha)=0$ and let $\mc M_0$ be the subset of $\mc M$ where $\alpha\neq 0$. Clearly the closure of $\mc M_0$ is the whole $\mc M$. Let us now define another transverse vector field $\xi'$ on $\mc M_0$ by $\xi'\d \alpha^{-1}\xi$ (note that it is still geodesic). For such field the induced AKH data on $\H_0$, that we shall denote with a prime in accordance with the fact that the rigging is now $\xi'$, is written in the $\eta$-gauge satisfying $\alpha=1$, i.e. $\k'_0 = \mc{G}_{(\alpha^{-1},0)}\k_0$. Such gauge will be called ``adapted $\eta$-gauge'' and it is unique (because the only gauge transformation of the form $\mc{G}_{(z,0)}$ that leaves $\alpha$ invariant is the identity). Let us explore some properties of the adapted $\eta$-gauge. Firstly, by item (iii) in Definition \ref{defi_AKH} the condition $0=\lie_{n'}\bm\tau'=\kappa\lie_{n'}\bm\ell'$ implies $\bs'=0$ (see \eqref{lienell}), and thus all the Lie derivatives along $n'$ of the metric data vanish, i.e.
\begin{equation}
	\label{adapted1}
\lie_{n'}\bg'=0,\qquad \lie_{n'}\bm\ell'=0,\qquad \lie_{n'}\elltwo{}'=0.
\end{equation}
Secondly, $\mf{c}' = \bm\tau'(n')=\kappa$ is constant and $\bm\tau' = -\bm c'$, so $\lie_{n'}\bm c' = 0$. And finally, the tensor defined by $\Sigmacero' \d \lie_{n'}\nablacero'$ vanishes. Indeed, from \eqref{nablaell} and $\lie_{n'}\bF' = \frac{1}{2}\lie_{n'} d\bm\ell' = \frac{1}{2} d\lie_{n'}\bm\ell'=0$ as well as $\lie_{n'}\bm\ell'=0$ it follows (see Prop. \ref{propMarc} for $m=1$) $$0= \lie_{n'}\nablacero'_a\ell'_b = \nablacero_a'\lie_{n'}\ell'_b - \Sigmacero'{}^c{}_{ab}\ell'_c =- \Sigmacero'{}^c{}_{ab}\ell'_c,$$

and from \eqref{nablagamma} one has $$0= \lie_{n'}\nablacero'_a\gamma'_{bc} = \nablacero'_a \lie_{n'}\gamma'_{bc} - 2\Sigmacero'{}^d{}_{a(b}\gamma'_{c)d}=-2\Sigmacero'{}^d{}_{a(b}\gamma'_{c)d}.$$

Hence, $$0=\lie_{n'}\nablacero'_a\gamma'_{bc}+\lie_{n'}\nablacero'_b\gamma'_{ca}-\lie_{n'}\nablacero'_c\gamma'_{ab} = 2\Sigmacero'{}^d{}_{ab}\gamma'_{cd}.$$ 

Conditions $\Sigmacero'{}^c{}_{ab}\ell'_c=0$ and $\Sigmacero'{}^d{}_{ab}\gamma'_{cd}=0$ automatically imply $\Sigmacero'{}^d{}_{ab}=0$. Recalling identity \eqref{Yano} (which is valid for every vector and connection, and in particular for $n'$ and $\nablacero'$) we conclude $$\lie_{n'}\Rcero'_{ab} = \nablacero'_c\accentset{\circ}{\Sigma}'{}^c{}_{ab}-\nablacero'_b\accentset{\circ}{\Sigma}'{}^d{}_{ad} = 0.$$ 

Under the transformation $\xi' = \alpha^{-1}\xi$ the transverse expansion $\{\bY^{(m)}\}$ transforms as (see \eqref{trans}-\eqref{transkappam} with $z=\alpha^{-1}$ and recall that $\xi(\alpha)=0$)
\begin{multicols}{2}
	\noindent
\begin{align*}
\bY' &= \alpha^{-1}\bY -\alpha^{-2} d\alpha\otimes_s \bm\ell,\\
\br' &= \br - \dfrac{1}{2\alpha}\left(d\alpha + n(\alpha)\bm\ell\right),\\
\kappa_n' &= \alpha\kappa_n +  n(\alpha),
\end{align*}
\begin{align}
\bY^{(k)}{}' &= \alpha^{-k} \bY^{(k)} ,\label{trans2}\\
\br^{(k)}{}' &= \alpha^{-k+1}\br^{(k)}\quad (k\ge 2),\label{trans22}\\
\kappa^{(k)}{}' &= \alpha^{-k+2}\kappa^{(k)},
\end{align}
\end{multicols}
where the prime is to emphasize that $\bY^{(m)}{}' \d \frac{1}{2}\Phi^{\star}\lie_{\xi'}^{(m)}g$ on $\H_0$. Note that the transformation $\bY\mapsto \bY'$ is analogous to the gauge transformation \eqref{transY} with parameters $(\alpha^{-1},0)$, as it must be. In addition, under the change $\xi' = \alpha^{-1}\xi$ the derivatives of the Ricci tensor transform according to \eqref{transR}-\eqref{transRab} with $z=\alpha^{-1}$
\begin{align}
	\ddot{\mc R}^{(m)}{}' & = \alpha^{-m-1}\ddot{\mc R}^{(m)},\label{transR2}\\
	\dot{\mc R}^{(m)}_a{}' & = \alpha^{-m} \dot{\mc R}^{(m)}_a - (m-1)\alpha^{-m-1}\ddot{\mc R}^{(m-1)} \nablacero_a \alpha\label{transRa2},\\
	\mc{R}_{ab}^{(m+1)}{}' & = \alpha^{-m} \mc{R}_{ab}^{(m+1)} - 2m\alpha^{-m-1} \dot{\mc R}^{(m)}_{(a}\nablacero_{b)}\alpha +m(m-1) \alpha^{-m-2}\ddot{\mc R}^{(m-1)}\nablacero_a\alpha \, \nablacero_b \alpha.\label{transRab2}
\end{align}

Our aim now is to prove that the abstract transverse expansion defined in \eqref{defY} and \eqref{defYk} transform exactly as the embedded one in \eqref{trans2} under a gauge transformation with parameters $(z=\alpha^{-1}, V=0)$, as this will imply that the expressions for the tensors $\{\bY^{(m)}{}'\}$ agree with those of $\{\bcY^{(m)}{}'\}$ written in the adapted $\eta$-gauge. More specifically, we define the gauge-transformed tensors $\{\bcY^{(m)}{}'\}$ by the expressions \eqref{bcYprima} and \eqref{Ykprima}, where all the terms in the right hand sides are expressed in the gauge obtained by the transformation $\mc{G}_{(z,0)}$, and we want to show that, with these definitions, $\bcY'_{ab}$ and $\bcY_{ab}$ as well as $\bcY_{ab}^{(k)}{}'$ and $\bcY_{ab}^{(k)}$ are related by exactly the same expressions in \eqref{trans2} when $z$ is taken to be $z=\alpha^{-1}$, namely
\begin{multicols}{2}
	\noindent
\begin{align*}
	\bcY' &= \alpha^{-1}\bcY -\alpha^{-2} d\alpha\otimes_s \bm\ell,\\
	\bcr' &= \bcr - \dfrac{1}{2\alpha}\left(d\alpha + n(\alpha)\bm\ell\right),\\
	\bck_n' &= \alpha\bck_n +  n(\alpha),
\end{align*}
\begin{align}
	\bcY^{(k)}{}' &= \alpha^{-k} \bcY^{(k)} ,\label{transbcY}\\
	\bcr^{(k)}{}' &= \alpha^{-k+1}\bcr^{(k)}\quad (k\ge 2),\\
	\bck^{(k)}{}' &= \alpha^{-k+2}\bck^{(k)}.\label{transbck}
\end{align}
\end{multicols}
\vspace{-0.3cm}
As proven in Lemma \ref{lemagauge} the tensor $\bcY'$ satisfies $\bcY'_{ab} =  z\bcY_{ab} + \ell_{(a}\nablacero_{b)}z$, so putting $z=\alpha^{-1}$ yields $$\bcY'_{ab} = \alpha^{-1}\bcY_{ab} - \alpha^{-2}\ell_{(a}\nablacero_{b)}\alpha,$$ 

which is the first relation in \eqref{transbcY}. This proves that the tensor $\bY'$ on $\H_0$ has the same expression as \eqref{defY} but written in the adapted $\eta$-gauge, i.e. with $\alpha=1$
\begin{equation}
	\label{Yprima}
\Y'_{ab} = \dfrac{1}{2\kappa}\left(\Rcero'_{(ab)} - 2c'_ac'_b + 2\nablacero'_{(a}c'_{b)} - \lambda\gamma'_{ab}\right).
\end{equation}
By the properties of the adapted $\eta$-gauge described above it follows $\lie_{n'}\bY'=0$ on $\H_0$. Contracting \eqref{Yprima} with $n'{}^a$ and using $\mf{c}'=\kappa$, $\nablacero'_an'{}^b = 0$ (see \eqref{derivadannull}), $\Rcero'_{(ab)}n'{}^a=0$ (see \eqref{Rceron}) and $2n'{}^a\nablacero_{(a}' c'_{b)} = - \nablacero_b'\mf{c}'=0$ (see \eqref{nnablacerotheta2}) one also has $\br' = \bm c'$. In particular this implies that $\kappa_n'=\mf{c}'=\kappa$, so
\begin{equation}
	\label{equationwhY}
	\lambda\gamma'_{ab} = \Rcero'_{(ab)} -2\lie_{n'} \Y'_{ab} - 2{\kappa}'_n\Y'_{ab} + 2\nablacero'_{(a}\r'_{b)}-2\r'_a \r'_b.
\end{equation}
From Lemma \ref{lemagauge} the transformation of $\bcY^{(2)}$ is $\bcY_{ab}^{(2)}{}' = z^2 \bcY_{ab}^{(2)}$, which proves that the tensor $\bY^{(2)}{}'$ has the same expression as in \eqref{defYk} for $m=2$ but with all the data written in the adapted $\eta$-gauge, namely
\begin{equation}
	\label{defY2}
\Y_{ab}^{(2)}{}' = \dfrac{1}{4\kappa}\left(\mc{O}_{ab}^{(1)}{}' -2 \mf{c}^{(2)}{}'\Y'_{ab} + 2\nablacero'_{(a} c_{b)}^{(2)}{}'- 4\r'_{(a}c_{b)}^{(2)}{}'-2\lambda \Y'_{ab}\right),
\end{equation}
where recall $c_a^{(2)}{}' \d \lambda\ell_a' -\mc{O}^{(1)}_a{}'$. Since the tensors $\mc{O}_{ab}^{(1)}{}'$, $\mc{O}_a^{(1)}{}'$ and $\mc{O}^{(1)}{}'$ are constructed solely in terms of metric data and $\bY'$, it follows $\lie_{n'}\mc{O}_{ab}^{(1)}{}'=0$, $\lie_{n'}\mc{O}_a^{(1)}{}'=0$ and $\lie_{n'}\mc{O}^{(1)}{}'=0$, and thus $\lie_{n'}\bY^{(2)}{}'=0$ on $\H_0$ as well. Contracting \eqref{defY2} with $n'{}^b$, using $2n'{}^b\nablacero'_{(a}c_{b)}^{(2)}{}' = -\nablacero_a'\mf{c}^{(2)}$ (see \eqref{nnablacerotheta2}) and Lemma \ref{lemaBianchi2} with $m=1$, namely $$2\lambda\r'_a =  - 2\kappa'_n c_a^{(2)}{}'-\nablacero'_a c^{(2)}{}' + \mc{O}_{ab}^{(1)}{}' n'{}^b,$$ 

 gives $\r_a^{(2)}{}' = c_a^{(2)}{}' \d \lambda\ell_a' - \mc{O}_a^{(1)}{}'$. Contracting \eqref{defY2} with $P'{}^{ab}$ and using again Lemma \ref{lemaBianchi2} with $m=1$ and $\elltwo{}'=0$, i.e. $$2\lambda\tr_{P'}\bY' = -4\kappa \mc{O}^{(1)}{}' - 2\mf{c}^{(2)}{}'\tr_{P'}\bY ' - 4P'(\br',\bm{c}^{(2)}{}') + 2\div_{P'}\bm{c}^{(2)}{}' + P'{}^{ab}\mc O_{ab}^{(1)}{}',$$ 

one shows that $P'{}^{ab}\Y_{ab}^{(2)}{}' = \mc{O}^{(1)}{}'$. By combining $\lie_{n'} \bY^{(2)}{}'=0$ and \eqref{defY2} it follows
\begin{equation*}
	2\lambda \Y_{ab}' = -2 \lie_{n'} \Y_{ab}^{(2)}{}' -4\kappa'_n \Y_{ab}^{(2)}{}' - 4\r'_{(a}\r_{b)}^{(2)}{}'+ 2\nablacero'_{(a} \r_{b)}^{(2)}{}'-2 \kappa^{(2)}{}'\Y_{ab}'+\mc{O}_{ab}^{(1)}{}'.
\end{equation*}
So far we have shown that
\begin{equation}
	\label{orden1p}
	\mc{R}_{ab}^{(1)}{}' = \lambda\gamma_{ab}',\qquad \ddot{\mc R}^{(1)}{}' = 0,\qquad \dot{\mc R}_a^{(1)}{}' = \lambda\ell_a',\qquad \mc{R}_{ab}^{(2)}{}' = 2\lambda \Y_{ab}',
\end{equation}
which by \eqref{trans2}-\eqref{transRab2}, $\bm\ell = \alpha\bm\ell'$ and the gauge invariance of the constraint tensor and that of $\bg$ is equivalent to 
\begin{equation}
	\label{orden1}
	\mc{R}_{ab}^{(1)} = \lambda\gamma_{ab},\qquad \ddot{\mc R}^{(1)} = 0,\qquad \dot{\mc R}_a^{(1)} = \lambda\ell_a,\qquad \mc{R}_{ab}^{(2)} = 2\lambda \Y_{ab}.
\end{equation}
Let us show by induction that for every $m\ge 2$ 
\begin{equation}
	\label{induction}
	\bcY^{(k)}{}' = z^{k}\bcY^{(k)},\qquad \lie_{n'}\bY^{(k)}{}' = 0 \quad \forall\, k=2,...,m
\end{equation}
and also 
\begin{equation}
	\label{induction2}
	\ddot{\mc R}^{(j)}{}' = 0,\qquad \dot{\mc R}_a^{(j)}{}' = \delta_1^j\lambda\ell_a',\qquad  \mc{R}_{ab}^{(j+1)}{}' = 2\lambda\Y_{ab}^{(j)}{}' \quad \forall\, j=1,...,m-1.
\end{equation}
For $m=2$ it is clearly true because we have already shown $\bcY^{(2)}{}' = z^2 \bcY^{(2)}$, $\lie_{n'} \bY^{(2)}{}'=0$ and \eqref{orden1}. We assume \eqref{induction}-\eqref{induction2} hold for some $m\ge 2$ and we prove that they also hold for $m+1$. By \eqref{trans2}-\eqref{transRab2} equations \eqref{induction2} are equivalent to
\begin{equation}
	\label{induction3}
\ddot{\mc R}^{(j)} = 0,\qquad \dot{\mc R}_a^{(j)} = \delta^j_1 \lambda\ell_a,\qquad \mc{R}_{ab}^{(j+1)} = 2\lambda\Y_{ab}^{(j)} \quad \forall\, j=1,...,m-1.
\end{equation}
In particular, since $\ddot{\mc R}^{(j)}=0$ it follows $\tr_{P}\bY^{(j+1)} = \mc{O}^{(j)}$ $\forall j=1,...,m-1$ and hence by Lemma \ref{lemagauge} one concludes $\bcY^{(m+1)}{}'=z^{m+1}\bcY^{(m+1)}$, so the expression of the tensor $\bY^{(m+1)}{}'$ agrees with \eqref{defYk} in the adapted $\eta$-gauge, namely
\begin{equation}
	\label{ykprime}
		\Y_{ab}^{(m+1)}{}' = \dfrac{1}{2(m+1)\kappa}\left( \mc{O}_{ab}^{(m)}{}' -2 \mf{c}^{(m+1)}{}'\bcY_{ab}' + 2\nablacero'_{(a} c_{b)}^{(m+1)}{}'  - 4c_{(a}' c_{b)}^{(m+1)}{}' -2\lambda \bcY_{ab}^{(m)}{}'\right).
\end{equation}
A similar argument as before based on Lemma \ref{lemaBianchi2} proves $\lie_{n'}\bY^{(m+1)}{}'=0$, $\r_a^{(m+1)}{}' = c_a^{(m+1)}{}'$ and $\tr_P\bY^{(m+1)}{}' = \mc{O}^{(m)}{}'$, so $$0 = \r_a^{(m+1)}{}' + \mc{O}_a^{(m)}{}', \qquad 0 = -\tr_P\bY^{(m+1)}{}' + \mc{O}^{(m)}{}'.$$ 

Moreover, combining $\lie_{n'}\bY^{(m+1)}{}'=0$ with \eqref{ykprime} gives $$2\lambda \Y_{ab}^{(m)}{}' = -2\lie_{n'}\bY_{ab}^{(m+1)}{}' -2(m+1)\kappa_n' \Y_{ab}^{(m+1)}{}' -2 \kappa^{(m+1)}{}'\Y_{ab}' + 2\nablacero'_{(a} \r_{b)}^{(m+1)}{}'  - 4\r_{(a}' \r_{b)}^{(m+1)}{}'+\mc{O}_{ab}^{(m)}{}'.$$ 

This proves $$\ddot{\mc R}^{(m)}{}' = 0,\qquad \dot{\mc R}_a^{(m)}{}' = 0,\qquad  \mc{R}_{ab}^{(m+1)}{}' = 2\lambda\Y_{ab}^{(m)}{}',$$ 

and since $\bcY^{(m+1)}{}'=z^{m+1}\bcY^{(m+1)}$ and $\lie_{n'}\bY^{(m+1)}{}'=0$ this closes the induction argument. Hence we have proved that for all $m\geq 1$ 
\begin{equation}
	\label{adaped2}
	\lie_{n'}\bY^{(m)}{}'=0 \quad \forall m\ge 1,
\end{equation}
and by \eqref{trans2}-\eqref{transRab2},
\begin{equation}
\mc{R}_{ab} = \lambda\gamma_{ab},\qquad	\ddot{\mc R}^{(m)} = 0,\qquad \dot{\mc R}_a^{(m)} = \delta_1^m\lambda\ell_a,\qquad  \mc{R}_{ab}^{(m+1)} = 2\lambda\Y_{ab}^{(m)},
\end{equation}
so equations \eqref{whatIwant} are established on $\H_0$, and by continuity on the full $\H$. Therefore $(\mc M,g)$ solves the $\Lambda$-vacuum equations to infinite order on $\H$. This proves item (iii) of the theorem.\\

\underline{3. Construction of the Killing field}\\

In order to finish the proof it only remains to show that $(\mc M,g)$ admits a Killing field that is null and tangent to $\Phi(\H)$. Let us now extend $\bar\eta = \alpha n$ off $\Phi(\H)$ by means of $\lie_{\xi}\eta = \beta\xi$, where $\beta$ is the function on $\mc M$ that solves $\xi(\beta)=0$ with initial condition $\beta = -n(\alpha)$ on $\Phi(\mc H)$. Equation \eqref{liexietaKH2} (in fact combined with Lemma 5.7 in \cite{Mio3}) suggests that this is indeed a natural candidate to be a Killing. Our aim is to prove that the deformation tensor of $\eta$ vanishes on $\mc M_0$, and thus by continuity on all $\mc M$. Recall that the construction of the metric $g$ in the proof of Theorem \ref{borel} (see \eqref{metric0}) consists of choosing local coordinates $\{r,u,x^A\}$, where $n=\partial_u$ and $\xi=\partial_r$, and then using Borel's lemma (see Lemma \ref{Borellemma}) to construct the coefficients $f_A$, $h_A$, $f$ and $H_{AB}$ in such a way that the transverse expansion at $\Phi(\H)$ agrees with $\{\bcY^{(k)}\}$. Each of the coefficients is given by a sum like \eqref{borelsum}, and from the properties of the sequence $\{\mu_k\}$ it follows that for any given $r>0$ the sum has only finitely many terms. This will allows us to interchange the sum and a derivative when necessary. It is easy to check that the vector $\eta$ in these coordinates is given by $\eta = \alpha\partial_u + \beta r\partial_r= \alpha\partial_u - r\partial_u\alpha \ \partial_r$. In these coordinates the computation of $\lie_{\eta}g$ is somewhat long and requires using the equations $$\lie_{\alpha n}\bY = n(\alpha)\bY + \alpha d\big(n(\alpha)\big)\otimes_s\bm\ell,\qquad \lie_{\alpha n}\bY^{(k)} = k n(\alpha)\bY^{(k)} \quad (k\ge 2)$$

on $\H_0$, which follow after transforming $\lie_{n'}\bY^{(m)}{}'=0$ $\forall m\ge 1$. Fortunately there exists a quicker strategy. If instead of working directly with the coordinate system $\{r,u,x^A\}$ one defines the new coordinates $\{r'\d \alpha r,u' \d \int\alpha^{-1} du,x^A\}$ on $\mc M_0$, it follows 
\begin{multicols}{2}
\noindent
\begin{align*}
\partial_u &= \alpha^{-1}\partial_{u'} + r\partial_u\alpha\ \partial_{r'} ,\\
\partial_r &=\alpha\partial_{r'},
\end{align*}
\begin{align*}
du &= \alpha du',\\
dr &= \alpha^{-1}dr' -r\partial_u\alpha\ du'-\alpha^{-1}r\partial_{x^A}\alpha\ dx^A,
\end{align*}
\end{multicols} 
and hence $$\eta = \alpha\partial_u -r\partial_u\alpha\ \partial_r = \alpha \big(\alpha^{-1}\partial_{u'}+ r\partial_u\alpha\ \partial_{r'}\big) - \alpha  r \partial_u\alpha\ \partial_{r'} = \partial_{u'}.$$

Moreover, transforming the metric $g$ in \eqref{metric0} yields (recall $h=0$ because $\elltwo=0$)
\begin{equation}
	\label{metric02}
	g = 2 dr' du'  +2f'_A dr' dx^A + 2 h'_A du' dx^A +f' du'{}^2 + H'_{AB}dx^A dx^B,
\end{equation} 
where $f'_A$, $h'_A$, $f'$ and $H'_{AB}$ are given by 
\begin{multicols}{2}
	\noindent
\begin{align*}
f'_A &= \alpha^{-1}f_A,\\
f' &= \alpha^2 f -2\alpha  r\partial_u\alpha,
\end{align*}
\begin{align*}
h_A' &= \alpha h_A -r\partial_u\alpha\ f_A-r\partial_{x^A}\alpha,\\
H_{AB}' &= H_{AB}-2\alpha^{-1}rf_{(A} \partial_{x^B)}\alpha.
\end{align*}
\end{multicols}
\vspace{-0.3cm}
By taking into account \eqref{order0}-\eqref{orderk} this means that 
\begin{equation*}
 f'|_{r'=0} = 0, \qquad f'_A|_{r'=0} = \alpha^{-1}\ell_A=\ell_A',\qquad h'_A|_{r'=0} = 0,\qquad H'_{AB}|_{r'=0} = \gamma_{AB} = \gamma'_{AB},
\end{equation*}
as well as
\begin{equation*}
	\begin{gathered}
		\partial_{r'}^{(k)}f'|_{r'=0} = -2\alpha^{-k+2}\bck^{(k)}-2\delta^k_1\partial_u\alpha=-2\bck'{}^{(k)},\qquad \partial_{r'}^{(k)} f'_A|_{r'=0} = 0,\\
\partial_{r'}^{(k)} h'_A|_{r'=0} = 2\alpha^{-k+1}\bcr^{(k)}_A -\delta_1^k \alpha^{-1}\big(n(\alpha)\ell_A+\partial_{x^A}(\alpha)\big) = 2\bcr^{(k)}{}',\\
		\partial_{r'}^{(k)} H'_{AB}|_{r'=0} = 2\alpha^{-k}\bcY^{(k)}_{AB} - 2\delta_k^1 \alpha^{-2} \ell_{(A}\partial_{x^B )}(\alpha) = 2\bcY^{(k)}{}',
	\end{gathered}
\end{equation*}
where the last equalities involving $\bck^{(k)}{}'$, $\bcr^{(k)}{}'$ and $\bcY^{(k)}{}'$ follow from \eqref{transbcY}-\eqref{transbck}. This proves that the coefficients $f'_A$, $h'_A$, $f'$ and $H'_{AB}$ are the same as the ones constructed as in Theorem \ref{borel} but in terms of $\{\bcY^{(m)}{}'\}$ instead of $\{\bcY^{(m)}\}$ (because $\xi' = \partial_{r'}$). Hence, after commuting $\lie_{\eta}$ with the sum in \eqref{borelsum}, the computation of $\lie_{\eta} g$ only requires calculating Lie derivatives of the metric data and of the expansion with respect to $\partial_{u'}|_{\H_0} = n'$, and thus all of them will automatically vanish (see \eqref{adapted1} and \eqref{adaped2}). This proves that $\eta$ is a Killing vector of $g$ on $\mc M_0$, and thus on $\mc M$.
	\end{proof}
\end{teo}

\begin{rmk}
When the function $\wt\alpha$ in the data of Theorem \ref{teorema2} has no zeroes the proof can be made much simpler because one can work directly in the adapted $\eta$-gauge from the beginning. 
\end{rmk}

%% file: Ack.tex
\section*{Acknowledgements}
This work has been supported by Projects PID2021-122938NB-I00 (Spanish Ministerio de Ciencia e Innovación and FEDER ``A way of making Europe''). M. Mars acknowledges financial support under projects SA097P24 (JCyL) and RED2022-134301-T funded by MCIN/AEI/10.13039/ 501100011033. G. Sánchez-Pérez also acknowledges support of the PhD. grant FPU20/03751 from Spanish Ministerio de Universidades.